\documentclass[aps,prx,twocolumn,footinbib,superscriptaddress,floatfix]{revtex4-2}
\usepackage{graphicx}
\usepackage{indentfirst}
\usepackage{physics}
\usepackage{braket}
\usepackage{float}
\usepackage{amsmath}
\usepackage{amssymb}
\usepackage{mathtools}
\usepackage{epstopdf}
\usepackage{footnote}
\usepackage{CJK}
\usepackage{esint}
\usepackage{comment}
\usepackage{color}
\usepackage[T1]{fontenc}
\usepackage{subfigure}
\usepackage{amsfonts}
\usepackage{footmisc}
\usepackage{scrextend}
\usepackage{multirow}
\usepackage[hyperfootnotes=false]{hyperref}
\usepackage[acronym]{glossaries}
\usepackage[english]{babel}
\usepackage{url}
\usepackage{bm}
\usepackage{hyperref}
\usepackage{algpseudocode}
\usepackage{soul}
\usepackage{xcolor}
\definecolor{darkblue}{rgb}{0,0,0.5}
\hypersetup{
colorlinks=true,
linkcolor=black,
filecolor=blue,
citecolor=darkblue,  
urlcolor=black,
}

\usepackage{stmaryrd}
\usepackage{trimclip}
\usepackage[normalem]{ulem}

\makeatletter
\DeclareRobustCommand{\shortto}{%
  \mathrel{\mathpalette\short@to\relax}%
}

\newcommand{\short@to}[2]{%
  \mkern2mu
  \clipbox{{.5\width} 0 0 0}{$\m@th#1\vphantom{+}{\shortrightarrow}$}%
  }
\makeatother

\newtheorem{theorem}{Theorem}

\newtheorem{lemma}[theorem]{Lemma}

\newcommand{\calA}{{\cal A}}

\newcommand{\calE}{{\cal E}}
\newcommand{\calF}{{\cal F}}

\newcommand{\calL}{{\cal L}}
\newcommand{\calN}{{\cal N}}

\newcommand{\calU}{{\cal U}}

\newcommand{\1}{^{(1)}}

\newcommand{\state}[1]{\ketbra{#1}{#1}}

\newcommand{\bI}{\boldsymbol I}

\newcommand{\QZ}[1]{{{\textcolor{black}{#1}}}}

\def\be{\begin{equation}}
\def\ee{\end{equation}}
\def\ba{\begin{eqnarray}}
\def\ea{\end{eqnarray}}

\begin{document}
\title{Dynamical transition in controllable quantum neural networks with large depth}

\date{\today}

\author{Bingzhi Zhang}
\thanks{These two authors contributed equally.}
\affiliation{Department of Physics and Astronomy, University of Southern California, Los Angeles, CA 90089, USA}
\affiliation{Ming Hsieh Department of Electrical and Computer Engineering, University of Southern California, Los Angeles, CA 90089, USA}

\author{Junyu Liu}
\thanks{These two authors contributed equally.}
\affiliation{Pritzker School of Molecular Engineering, The University of Chicago, Chicago, IL 60637, USA}
\affiliation{Department of Computer Science, The University of Chicago, Chicago, IL 60637, USA}
\affiliation{Kadanoff Center for Theoretical Physics, The University of Chicago, Chicago, IL 60637, USA}

\author{Xiao-Chuan Wu}
\affiliation{Kadanoff Center for Theoretical Physics, The University of Chicago, Chicago, IL 60637, USA}

\author{Liang Jiang}
\affiliation{Pritzker School of Molecular Engineering, The University of Chicago, Chicago, IL 60637, USA}

\author{Quntao Zhuang}
\email{qzhuang@usc.edu}
\affiliation{Ming Hsieh Department of Electrical and Computer Engineering,
University of Southern California, Los Angeles, CA 90089, USA}
\affiliation{Department of Physics and Astronomy, University of Southern California, Los Angeles, CA 90089, USA}

\begin{abstract}
Understanding the training dynamics of quantum neural networks is a fundamental task in quantum information science with wide impact in physics, chemistry and machine learning. In this work, we show that the late-time training dynamics of quantum neural networks with a quadratic loss function can be described by the generalized Lotka-Volterra equations, which lead to a transcritical bifurcation transition in the dynamics. When the targeted value of loss function crosses the minimum achievable value from above to below, the dynamics evolve from a {\em frozen-kernel dynamics} to a {\em frozen-error dynamics}, showing a duality between the quantum neural tangent kernel and the total error. In both regions, the convergence towards the fixed point is exponential, while at the critical point becomes polynomial. We provide a non-perturbative analytical theory to explain the transition via a restricted Haar ensemble at late time, when the output state approaches the steady state. Via mapping the Hessian to an effective Hamiltonian, we also identify a linearly vanishing gap at the transition point.  Compared with the linear loss function, we show that a quadratic loss function within the {\em frozen-error dynamics} enables a speedup in the training convergence.
The theory findings are verified experimentally on IBM quantum devices.

\end{abstract}


\maketitle

\section{Introduction}

As a paradigm of near-term quantum computing, variational quantum algorithms~\cite{peruzzo2014variational,farhi2014quantum,mcclean2016theory,mcclean2018barren,mcardle2020quantum,cerezo2021variational} have been widely applied to chemistry~\cite{peruzzo2014variational,kandala2017hardware}, optimization~\cite{farhi2014quantum,ebadi2022quantum}, quantum simulation~\cite{yuan2019theory,yao2021}, condensed matter physics~\cite{cong2019quantum}, communication~\cite{zhang2022hybrid,ur2022variational},  sensing~\cite{zhuang2019,xia2021} and machine learning~\cite{wittek2014quantum,schuld2015introduction,biamonte2017quantum,farhi2018classification,dunjko2018machine,schuld2019quantum,havlivcek2019supervised,abbas2021power}. Adopting layers of gates and stochastic gradient descent, they are regarded as `quantum neural networks' (QNNs), analog to classical neural networks that are crucial to machine learning. Concepts and methods related to variational quantum algorithms are also beneficial for quantum error correction and quantum control~\cite{ni2023beating,sivak2023real}, bridging near-term applications with the fault-tolerant era.

Despite the progress in applications, theoretical understanding of the training dynamics of QNN is limited, hindering the optimal design of quantum architectures and the theoretical study of quantum advantage in such applications. Previous works adopt tools from quantum information scrambling for empirical study of QNN training~\cite{shen2020information,garcia2022quantifying}. Recently, the Quantum Neural Tangent Kernel (QNTK) theory presents a potential theoretical framework for an analytical understanding of variational quantum algorithms, at least within certain limits \cite{liu2022representation,liu2023analytic,liu2022laziness,wang2022symmetric,yu2023expressibility}, revealing deep connections to their classical machine learning counterparts \cite{lee2017deep,jacot2018neural,lee2019wide,sohl2020infinite,yang2020feature,yaida2020non,arora2019exact,dyer2019asymptotics,halverson2021neural,roberts2021ai,roberts2021principles}. 
However, the theory of QNTK relies on the assumption of sufficiently random quantum circuit set-ups known as unitary $k$-designs~\cite{roberts2017chaos,Cotler:2017jue,Liu:2018hlr,Liu:2020sqb} that is only true at random initialization, preventing the theory from describing the more important late-time training dynamics. Similar limitations also exist for other theoretical works~\cite{you2022convergence,you2021exponentially,anschuetz2021critical,mcclean2018barren,cerezo2021cost}.


\begin{figure}[t]
    \centering
    \includegraphics[width=0.5\textwidth]{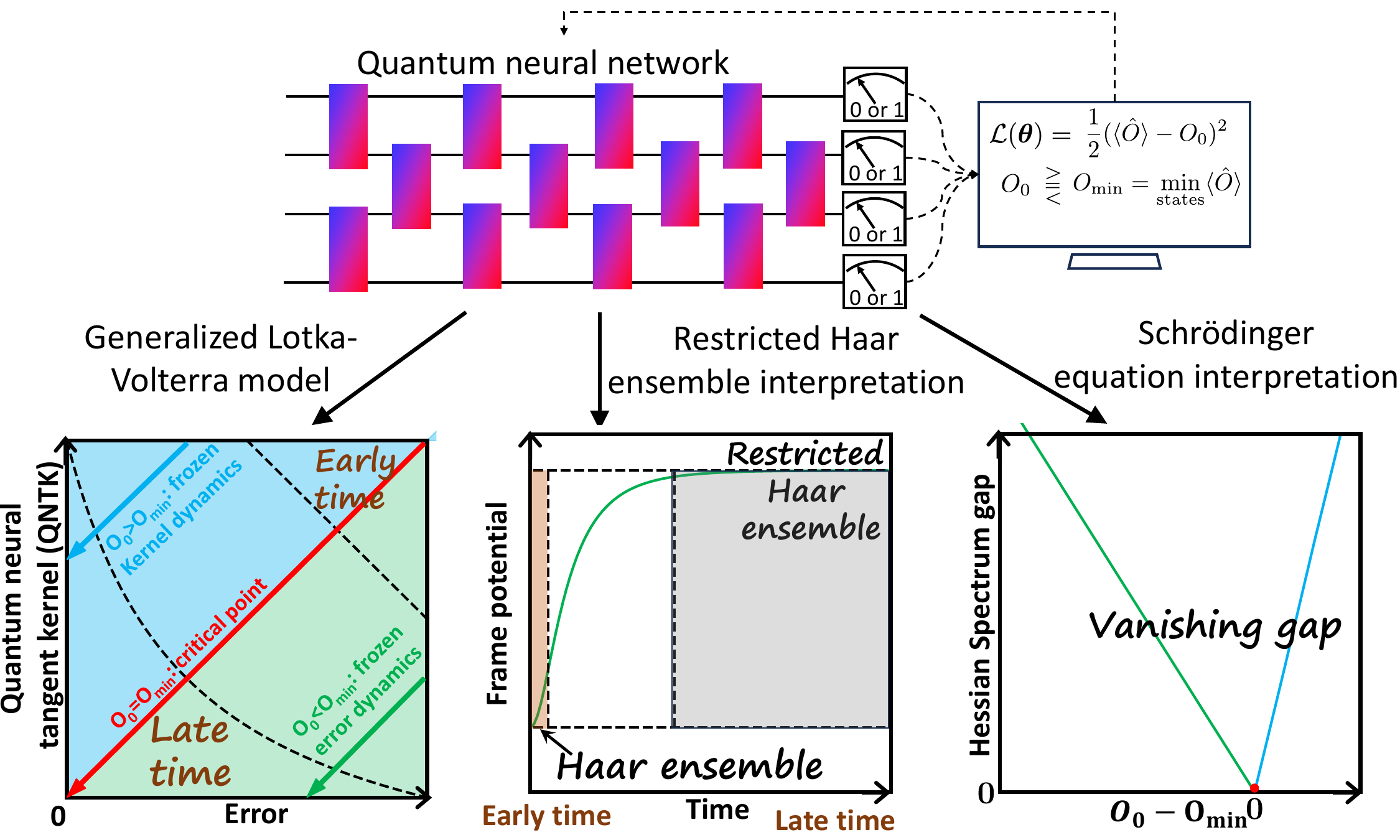}
    \caption{{\bf Illustration of setup and main results of this work.} We study the training dynamics of quantum neural networks with loss function $\calL(\bm \theta)=(\braket{\hat{O}} - O_0)^2/2$, and identify a dynamical transition. We derive a first-principle generalized Lotka-Volterra model to characterize it, and also provide interpretations from random unitary ensemble and Schr\"{o}dinger equation.}
    \label{fig:concept}
\end{figure}


In this work, we go beyond QNTK theory and identify a dynamical transition in the training of QNNs with a quadratic loss function, when the target loss function value $O_0$ cross the minimum achievable value (ground state energy $O_{\rm min}$).  We show that the training dynamics of deep QNNs is governed by the generalized Lotka-Volterra (LV) equations describing a competitive duality between the quantum neural tangent kernel and the total error. The LV equations can be analytically solved and the dynamics is determined by the value of a conserved quantity. When the target value crosses $O_{\rm min}$, the conserved quantity changes sign and induces a transcritical bifurcation transition.
As depicted in Fig.~\ref{fig:concept}, in the {\em frozen-kernel dynamics} where $O_0>O_{\rm min}$ is in the bulk of spectrum, the kernel is approaching a constant while the error decays exponentially with training steps; At the {\em critical point} when $O_0=O_{\rm min}$ exactly, both the kernel and the error decay polynomially; In the {\em frozen-error dynamics} when $O_0<O_{\rm min}$ is unachievable, the output from QNN still converges to the ground state leaving the error approaching a constant $O_{\rm min}-O_0$, while the kernel experiences an exponential decay. We provide a non-perturbative analytical theory to explain the dynamical transition via a restricted Haar ensemble at late time, when the QNN output state approaches the steady state. We also identify a vanishing Hessian gap at the transition point,  \QZ{which corresponds to Hamiltonian gap closing in the imaginary-time Schr\"odinger equation interpretation.} While our theory analyses assume the large-depth limit, the dynamical transition is also numerically identified in QNNs with limited depths.   Compared to the exponential decay of linear loss function with a non-tunable exponent, we identify convergence speed-up via tuning the quadratic loss function to be within the {\em frozen-error dynamics}.  The theory findings are experimentally verified on IBM quantum devices. Our results imply that designing the loss function properly is important to achieve fast convergence. 

\begin{figure*}[t]
    \centering
    \includegraphics[width=0.95\textwidth]{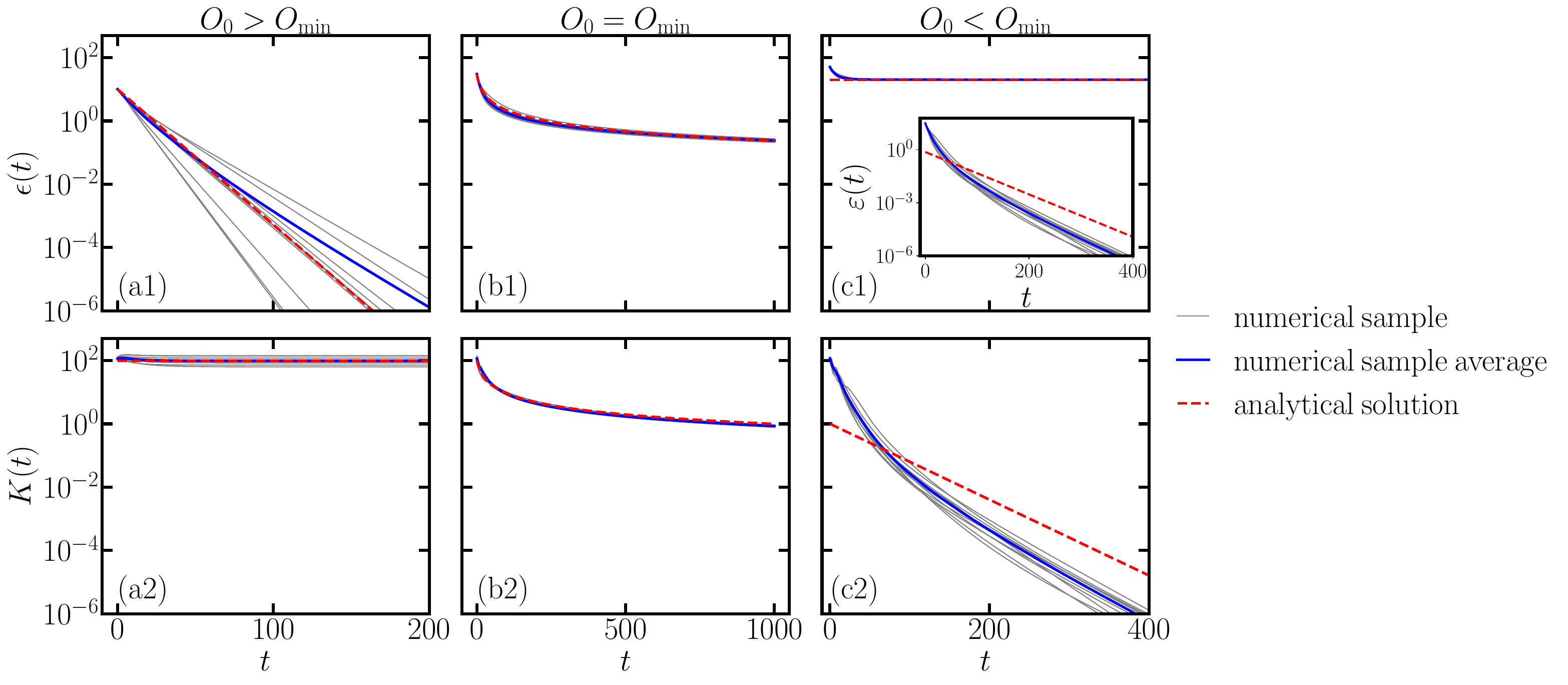}
    \caption{{\bf Dynamics in QNN in the example of XXZ model.}  
    The top and bottom panel shows the dynamics of total error ${\epsilon}(t)$ and QNTK ${K}(t)$ with respect to the three cases $O_0 \gtreqless O_{\rm min}$. Blue solid curves represent numerical ensemble average result. Red dashed curves in panels represents theoretical predictions on the dynamics of total error in Eq.~\eqref{eq:eps_t_frozen_exp},~\eqref{eq:K_poly_asym_main},~\eqref{eq:K_nonfrozen_exp_main} (from left to right). Grey solid lines show the dynamics for each random sample. The inset in (c1) shows the exponential decay of residual error $\varepsilon(t)$. Here random Pauli ansatz (RPA) consists of $L=768$ variational parameters ($D=L$ for RPA) on $n=8$ qubits, and the parameter in XXZ model is $J=2$. }
    \label{fig:main_results}
\end{figure*}


We begin by first introducing the model of the QNN and the necessary quantities. Then, we uncover the dynamical transition phenomena as a bifurcation transition in LV model. The unitary ensemble theory is then developed to support assumptions in obtaining the LV model. Afterwards, we characterize the transition with tools from statistical physics. After finishing the theory, we provide numerical extensions and discuss the potential training speed-up brought by our results. Finally, we confirm the results in experiments.

\section{Training dynamics of quantum neural networks}

A ${D}$-depth QNN is composed of ${D}$ layers of parameterized quantum circuits, realizing a unitary transform 
$\hat{U}(\bm \theta)$ on $n$ qubits, with ${L}$ variational parameters $\bm \theta = (\theta_1,\dots,\theta_L)$. The gate configuration of each layer varies between different circuit ansatz (see Sec.~\ref{app:methods} for examples). When inputting a trivial state $\ket{0}^{\otimes n}$, the final output state of the neural network $\ket{\psi(\bm \theta)}=\hat{U}(\bm \theta)\ket{0}^{\otimes n}$, from which one can measure a Hermitian observable $\hat{O}$ leading to expectation value $\braket{\hat{O}}=\braket{\psi(\bm \theta)|\hat{O}|\psi(\bm \theta)}$.
To optimize the expectation of an observable $\hat{O}$ towards the target value $O_0$, a general choice of loss function is in a quadratic form,
\begin{align}
    \mathcal{L}(\bm \theta) = \frac{1}{2}\left(\braket{\hat{O}} - O_0\right)^2 \equiv \frac{1}{2}\epsilon(\bm \theta)^2,
    \label{loss_function}
\end{align}
where the total error
$
\epsilon(\bm \theta)=\braket{\hat{O}} - O_0.
$
Suppose observable $\hat{O}$ has possible values in the range of $[O_{\rm min},O_{\rm max}]$. Without further specification, $O_{\rm min}$ and $O_{\rm max}$ refer to the minimum and maximum eigenvalue of $\hat{O}$.
Now due to symmetry of maximum and minimum in optimization problems, we assume $O_0<O_{\rm max}$ is true. 

A QNN goes through training to minimize the loss function.
In each training step, every variational parameter is updated by the gradient descent
\begin{align}
    \delta \theta_\ell(t) \equiv \theta_\ell(t+1) - \theta_\ell(t) = -\eta\frac{\partial \mathcal{L(\bm \theta)}}{\partial \theta_\ell} = -\eta \epsilon (\bm \theta)\frac{\partial \epsilon (\bm \theta)}{\partial \theta_\ell},
    \label{dtheta_grad}
\end{align}
where $\eta$ is the fixed learning rate and $t$ is the discrete number of time steps in the training. 
With the update of parameters $\bm \theta$, quantities depending on $\bm \theta$ also acquire new values in each training step. For simplicity of notion, we denote their dependence on $t$ explicitly omitting $\bm \theta$, e.g. $\epsilon(t)\equiv \epsilon(\bm \theta(t))$. To study the convergence, we separate the error into two parts,  
$
\epsilon (t)\equiv \varepsilon(t) +R
$
consists of a constant remaining term
$
R=\lim_{t\to \infty}\epsilon(t)
$
and a vanishing residual error
$
\varepsilon(t).
$

When $\eta \ll 1$ is small, the total error is updated as
\begin{align}
    \delta \epsilon(t) &\simeq \sum_{\ell}\frac{\partial\epsilon (\bm \theta)}{\partial\theta_\ell} \delta \theta_\ell + \frac{1}{2}\sum_{\ell_1, \ell_2}\frac{\partial^2\epsilon (\bm \theta)}{
    \partial\theta_{\ell_1}\partial\theta_{\ell_2}
    }\delta \theta_{\ell_1}\delta \theta_{\ell_2}\\
    &= -\eta \epsilon (t) K (t) + \frac{1}{2}\eta^2\epsilon (t)^2 \mu(t),
    \label{eq:delta_eps}
\end{align}
where the QNTK $K$ and dQNTK $\mu$ are defined as~\cite{liu2023analytic}
\begin{align}
     K(t) &\equiv \sum_\ell  \left.\left(\frac{\partial\epsilon(\bm \theta)}{\partial\theta_\ell}\right)^2 \right\rvert_{\bm \theta=\bm \theta(t)},\label{eq:K_def} \\
     \mu(t) &\equiv \sum_{\ell_1, \ell_2} \left. \frac{\partial^2\epsilon(\bm \theta)}{
    \partial\theta_{\ell_1}\partial\theta_{\ell_2}
    } \frac{\partial\epsilon(\bm \theta)}{\partial\theta_{\ell_1}}\frac{\partial\epsilon(\bm \theta)}{\partial\theta_{\ell_2}} \right\rvert_{\bm \theta=\bm \theta(t)} \label{eq:mu_def}.
\end{align}
In the dynamics of $\epsilon(t)$, as $\eta \ll 1$, we focus on the first order of $\eta$ in Eq.~\eqref{eq:delta_eps} as
\begin{align}
    \delta \epsilon(t) = -\eta \epsilon(t) K(t) + \mathcal{O}(\eta^2).
    \label{epsilon_dynamics}
\end{align}
To characterize the dynamics of $\epsilon(t)$, it is necessary and sufficient to understand the dynamics of QNTK $K(t)$.
Towards this end, we derive a first-order difference equation for QNTK $K(t)$ as (see details in App.~\ref{app:diff_eq})
\begin{align}
    \delta K(t) = -2\eta \epsilon(t) \mu(t) + O(\eta^2).
    \label{eq:K_de1}
\end{align}
Combining Eq.~\eqref{epsilon_dynamics} and Eq.~\eqref{eq:K_de1}, we aim to develop the dynamical model in training QNNs.

\section{Dynamical transition}
Our major finding is that when the circuit is deep and controllable, the QNN dynamics exhibit a dynamical transition at $O_{\rm min}$ (and $O_{\rm max}$ similarly) as we depict in Fig.~\ref{fig:main_results}, where a QNN with random Pauli ansatz (RPA) is utilized to optimize the XXZ model Hamiltonian (see Sec.~\ref{app:methods} for details of the circuit and observable).

{\em Frozen-kernel dynamics}: When $O_0> O_{\rm min}$, the total error decays exponentially and the energy converges towards $O_0$, as shown in Fig.~\ref{fig:main_results}(a1). This is triggered by the frozen QNTK as shown in Fig.~\ref{fig:main_results}(a2). Each individual random sample (gray) has slightly different value of frozen QNTK due to initialization, while all possess the exponential convergence. Our theory prediction (red dashed) agrees with the actual average (blue solid) for both the ensemble averaged QNTK $\overline{K}$ and the error, while deviations due to early time dynamics can be seen (see Sec.~\ref{app:methods} for details).

{\em Critical point}: When targeting right at the GS energy $O_0=O_{\rm min}$, both the total error and QNTK decay as $1/t$, independent of system dimension $d$. As shown in Fig.~\ref{fig:main_results}(b2), the QNTK ensemble average (blue solid) agrees very well with the theory prediction shown as the red dashed line. Due to initial time discrepancy in QNTK that is beyond our late time theory, the actual error dynamics has a constant deviation from the theory prediction (red dashed), however still has the $1/t$ late time scaling, as shown in Fig.~\ref{fig:main_results}(b1).

{\em Frozen-error dynamics}: When targeting below the GS energy $O_0< O_{\rm min}$, the total error converges to a constant $R=O_{\rm min}-O_0>0$ exponentially, as shown in Fig.~\ref{fig:main_results}(c1). The inset shows the exponential convergence via the residual error $\varepsilon=\epsilon-R$. In this case, the QNTK also decays exponentially with the training steps, as shown in Fig.~\ref{fig:main_results}(c2). Deviation between the theory (red dashed) and numerical results (blue solid) can be seen due to early time dynamics beyond our theory.


\begin{figure*}
    \centering
    \includegraphics[width=0.85\textwidth]{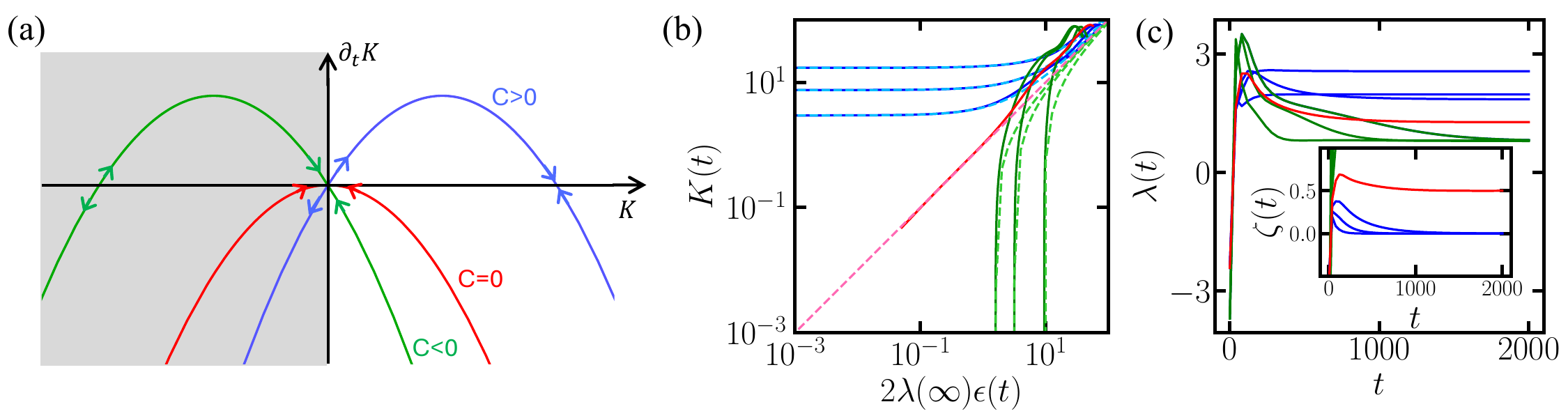}
    \caption{{\bf Classical dynamics interpretation of total error and QNTK dynamics.} (a) The RHS of Eq.~\eqref{eq:Kdiff} showing a bifurcation. The gray region is nonphysical as $K\ge 0$. In the physical region ($K\ge 0$), we have a single stable fixed point $K=C$ when $C >0$, corresponding to the frozen-kernel dynamics (blue in (b)); and a single stable fixed point $K=0$ when $C \le 0$, corresponding to the frozen error dynamics and critical point (green and red in (b)) separately. (b) Trajectories of $(2\lambda(\infty) \epsilon(t), K(t))$ in dynamics of QNN with different $O_0 \gtreqless O_{\rm min}$, plotted in solid blue, red and green. Dashed curves show the trajectory from Eq.~\eqref{eq:LV_conserve}. The arrows denote the flow of time in QNN optimization. Logarithmic scale is taken to focus on the late-time comparison. (c) The dynamics of corresponding $\lambda(t) = \mu(t)/K(t)$. The inset shows the dynamics of $\zeta(t) = \epsilon(t)\mu(t)/K(t)^2$. The observable is XXZ model with $J=2$, and QNN is a $n=6$-qubit RPA with ${L}=192$ parameters (for RPA ${D}={L}$). The legend in (b) is also shared with (c) and its inset.}
    \label{fig:LVmodel}
\end{figure*}

\section{Generalized Lotka-Volterra model: bifurcation }
In this section, we reveal the nature of the transition as a transcritical bifurcation of an effective nonlinear dynamical equation.
With large depth ${D}\gg1$ and full control, QNNs are commonly modeled as a random unitary~\cite{liu2023analytic,mcclean2018barren,cerezo2021cost}. However, at late time, the convergence of QNN training imposes constraints on the QNN unitary. As we will detail in `Unitary ensemble theory' section, assuming that the late-time QNN is typical among random ensemble of unitaries under the convergence constraint, we can show that the relative dQNTK---the ratio of dQNTK and QNTK 
\be 
\lambda(t) = \mu(t)/K(t)
\label{eq:lambda_def}
\ee
converges towards a constant dependent on the number of parameters $L$ and Hilbert space dimension of the system $d=2^n$. Under the assumption that $\lambda(t)=\lambda$ being a constant and taking the continuous limit, Eqs.~\eqref{epsilon_dynamics} and~\eqref{eq:K_de1} lead to a coupled set of equations,
\begin{align}
\label{eq:LV-model-main}
\left\{ \begin{array}{ll}
         \partial_t \epsilon(t) & = -\eta \epsilon(t) K(t)\\
        \partial_t K(t) &= -2\eta  \lambda \epsilon(t) K(t) \end{array} \right. 
\end{align}
to the leading order in $\eta$.
This is the generalized Lotka-Volterra equation developed in modeling nonlinear population dynamics of species competing for some common resource~\cite{hofbauer1998evolutionary}. 
The two `species' represented by $K$ and $\epsilon$ are in direct competition as the interaction terms are negative. As Eqs.~\eqref{eq:LV-model-main} have zero intrinsic birth/death rate, there is no stable attractor where all species $K(t)$ and $\epsilon(t)$ are positive, as sketched in Fig.~\ref{fig:concept} bottom left, where $2\lambda \epsilon$ and $K$ are the x and y axis. From Eqs.~\eqref{eq:LV-model-main}, we can identify the conserved quantity at late time
\begin{align}
    C =K(t) - 2\lambda \epsilon(t) = {\rm const}.
    \label{eq:LV_conserve}
\end{align}
Each trajectory of $(2\lambda \epsilon(t), K(t))$ governed by Eqs.~\eqref{eq:LV-model-main} is thus a straight line quantified by the conserved quantity $C$. We verify the trajectory from conserved quantity of Eq.~\eqref{eq:LV_conserve} in Fig.~\ref{fig:LVmodel}(b), where good agreement between QNN dynamics (solid) and generalized LV dynamics (dashed) can be identified. The conservation law in Eq.~\eqref{eq:LV_conserve} indicates that a classical Hamiltonian description of LV dynamics is possible, via mapping the scaled error and kernel to the canonical position and momentum (see Sec.~\ref{app:methods}). Therefore, the position-momentum duality in Hamiltonian formulation implies an {\it error-kernel} duality between $K$ and $\epsilon$. 

Thanks to the conserved quantity $C$, we can reduce the coupled differential equations of LV model in Eqs.~\eqref{eq:LV-model-main} to a single variable differential equation with the kernel or the error alone, e.g.,
\begin{align}
    \partial_t K(t) = \eta(C-K(t)) K(t). 
    \label{eq:Kdiff}
\end{align}
This is a canonical example of a transcritical bifurcation, with two fixed points $K=C$ and $K=0$~\cite{strogatz2018nonlinear}. To see this, we plot the RHS of Eq.~\eqref{eq:Kdiff} in Fig.~\ref{fig:LVmodel}(a). When $C>0$ (blue curve), via the sign of $\partial_t K$, we can see that only $K = C$ (therefore $\epsilon=0$) is stable, corresponding to the frozen-kernel dynamics. On the other hand, when $C<0$ (green curve), $K= 0$ (therefore $2\lambda \epsilon=-C>0$) is the only stable fixed point, corresponding to the frozen-error dynamics. Specifically, for $C = 0$ (red curve), the two candidates collide and $K = 0$ (therefore $\epsilon=0$) becomes the bifurcation point. As the fixed points collide and their stability exchange through the bifurcation point $(K, C) =(0, 0)$, the transition is identified as the transcritical bifurcation.


Overall, we see that the two dynamics (and the critical point) of the QNN dynamics has a one-to-one correspondence to the two families of fixed points (and their common fixed point) of the generalized LV equation. 
The conserved quantity $C=K(t)-2\lambda \epsilon(t)=(K(t)^2-2\epsilon(t) \mu(t))/K(t)$. Since $K(t) > 0$ at any finite time, the sign of constant is determined by the dynamical index defined as
\be 
\zeta =\epsilon(t) \mu(t)/K(t)^2.
\label{eq:zeta_def}
\ee 
If $\zeta \gtreqless 1/2$, we have $C \lesseqgtr 0$, determining the bifurcation dynamics.

Indeed, the analytical closed-form solution (see Sec.~\ref{app:methods}) to the LV dynamics of Eqs.~\eqref{eq:LV-model-main} supports the following theorem at the $t\gg 1$ late time limit.

\begin{theorem}
\label{main_theorem}
Assuming relative dQNTK $\lambda=\mu(t)/K(t)$ being a constant at late time, the QNN dynamics is governed by the generalized Lotka-Volterra equation in Eq.~\eqref{eq:LV-model-main} and possesses a bifurcation to two different branches of dynamics, depending on the value of a conserved quantity $C=K(t)-2\lambda \epsilon(t)=(1-2\zeta)K(t)$ or equivalently the dynamical index $\zeta =\epsilon(t) \mu(t)/K(t)^2$.
\begin{enumerate}
    \item 
When $\zeta<1/2$ thus $C>0$, we have the `frozen-kernel dynamics' (c.f. \cite{liu2023analytic}),
where the QNTK $K(t) = C$ is frozen and
\begin{align}
    \epsilon(t)\propto e^{-\eta C t},
    \label{eq:eps_t_frozen_exp}
\end{align}
\item 
When $\zeta=1/2$ thus $C=0$, we have the `critical point', where both the QNTK and total error decay polynomially, 
\begin{align}
K(t)=2\lambda \epsilon(t)= 1/(\eta t+c),
\label{eq:K_poly_asym_main}
\end{align}
where $c$ is a constant.

\item 
When $\zeta>1/2$ thus $C<0$, we have the `frozen-error dynamics', where the total error $\epsilon(t)=R$ is frozen and both the kernel and the residual error decay exponentially
\begin{align}
K(t)=2\lambda \varepsilon(t) \propto e^{-2\eta \lambda Rt}.
\label{eq:K_nonfrozen_exp_main}
\end{align}

\end{enumerate}

\end{theorem}

The bifurcation can be connected to $O_0 \gtreqless O_{\rm min}$ intuitively. When $O_0< O_{\rm min}$, it is clear that $R>0$ and we expect dynamical index $\zeta>1/2$ and $C<0$ so that it is the `\textit{frozen-error dynamics}'. When $O_0>O_{\rm min}$, we know the total error will decay to zero eventually, and therefore we can correspond this branch to the `\textit{frozen-kernel dynamics}', where dynamical index $\zeta<1/2$ and $C>0$. The case $O_0=O_{\rm min}$ is therefore the critical point. In Fig.~\ref{fig:LVmodel}(c) inset, we indeed see the dynamical index $\zeta \to 0, 1/2, +\infty$ when $O_0 \gtreqless O_{\rm min}$.
In our later theory analyses, we will make this connection rigorous between $O_0 \gtreqless O_{\rm min}$, the dynamical index $\zeta \gtreqless 1/2$ and the bifurcation transition.

\section{Unitary ensemble theory}
In this section, we provide analytical results to resolve two missing pieces of the LV model---the assumption that the relative dQNTK $\lambda$ in Eq.~\eqref{eq:lambda_def} is a constant at late time and the connection between the dynamical index $\zeta\gtreqless 1/2$ in Eq.~\eqref{eq:zeta_def} and the $O_0 \gtreqless O_{\rm min}$ cases. Our analyses will rely on large depth $D\gg1$ (equivalently $L\gg1$), which allow us to model each realization of the QNN $\hat{U}(\bm \theta)$ as a sample from an ensemble of unitaries and consider ensemble averaged values to represent the typical case,
$
    \overline{\zeta} = {\overline{\epsilon \mu}}/{\overline{K}^2},
    \overline{\lambda} = {\overline{\mu}}/{\overline{K}}
$.
Note that we take the ratio between averaged quantity via considering the sign of $\overline{C}$. The ordering of ensemble averages has negligible effects (see App.~\ref{app:rh_result}).


As the QNN is initialized randomly at the beginning, the unitary $\hat{U}(\bm \theta)$ being implemented can be regarded as typical ones satisfying Haar random distribution~\cite{liu2023analytic,mcclean2018barren,cerezo2021cost}, regardless of the circuit ansatz. While this is a good approximation at initial time, we notice that at late time, the QNN $\hat{U}(\bm \theta)$ is constrained in the sense that it maps the initial trivial state (e.g. product of $\ket{0}$) towards a {\em single} quantum state, regardless of whether the quantum state is the unique optimum or not. Therefore, the late-time dynamics are always restricted due to convergence, which we model as the restricted Haar ensemble with a block diagonal form,
\begin{align}
    \calE_{\rm RH} = \{{U}|{U} = 
    \begin{pmatrix}
    1 & {\bm 0} \\
    {\bm 0} & V
\end{pmatrix}
\},
\label{eq:restrict_U}
\end{align}
where $V$ is a unitary with dimension $d-1$ following a sub-system Haar random distribution (only $4$-design is necessary). Here we have set the basis of the first column and row to represent mapping from the initial state to the final converged state. At late time, QNN converges to a restricted Haar ensemble determined by the converged state. When the converged state is unique, frame potential~\cite{roberts2017chaos} of the ensemble can be evaluated by considering different training trajectories, which confirms the ansatz in Eq.~\eqref{eq:restrict_U}, as shown in Fig.~\ref{fig:concept} and App.~\ref{app:k_design}.

The ensemble average for a general traceless operator is challenging to analytically obtain. To gain insights to QNN training, we consider a much simpler problem of state preparation, where $\hat{O}=\state{\Phi}$ is a projector. In this case, we are interested in target values $O_0$ near the maximum loss function $O_{\rm max}=1$.
Under such restricted Haar ensemble, we have 
\begin{lemma}
\label{lemma:RH_ratio}

When the circuit satisfies the restricted Haar random (restricted 4-design) ensemble and ${D}\gg 1$ (therefore ${L}\gg1$), in state preparation tasks the relative dQNTK $\overline{\lambda_\infty}$ goes to an ${L},d$ dependent constant. When $O_0<O_{\rm max}$, the dynamical index $\overline{\zeta_\infty} = 0$; when $O_0=O_{\rm max}$, the dynamical index
$
\overline{\zeta_\infty} =  1/2;  
$
when $O_0>O_{\rm max}$, the dynamical index $\zeta$ diverges to $+\infty$.
\label{lemma_lambda}
\end{lemma}
This lemma derives from Theorem~\ref{theorem:RH} in Sec.~\ref{app:methods}.

While our results are general, in our numerical study that verifies the analytical results, we adopt the random Pauli ansatz (RPA)~\cite{liu2023analytic} as an example (see Sec.~\ref{app:methods}).
Due to symmetry between maximum and minimum in optimization, this restricted Haar ensemble therefore fully explains the branches of dynamics in Theorem~\ref{main_theorem} quantitatively and the assumption of $\lambda$ approaches a constant qualitatively. 
From asymptotic analyses of the restricted Haar ensemble in App.~\ref{app:rh_result}, we also have both $\lambda, C \propto {L}/d$, thus the exponential decay in LV has exponent $\propto \eta {L}t/d$. 
Indeed, in a computation, $\eta {L}t$ describes the resource---when number of parameters ${L}$ is larger, one needs to compute and update more parameters, while taking fewer steps $t$ to converge.

As we show in Sec.~\ref{app:methods}, Haar ensemble fails to capture the $\zeta$ dynamics nor the bifurcation transition. Only in the case of \textit{frozen-kernel dynamics}, as the kernel does not change much during the dynamics, the Haar predictions roughly agree with the actual kernel, as shown in Fig.~\ref{fig:main_results}(a2) (see Sec.~\ref{app:methods}).

\section{Schr\"odinger equation interpretation}
Besides the LV dynamics, we can also connect the transition to the gap closing of the Hessian, via interpreting the training dynamics around the extremum as imaginary Schrodinger evolution as we detail below. The gradient descent dynamics in Eq.~\eqref{dtheta_grad} leads to the time evolution of the quantum state $\ket{\psi(\bm \theta)}$, where $\bm \theta$ are the variational parameters. In the late time limit, omitting \QZ{the} $t$ dependence in \QZ{our} notation, we can expand the shifts $\delta\theta_{\ell}$ around the extremum $\bm \theta^*$  to second order as
\begin{align}
    \delta \bm \theta \simeq -\eta M(\bm \theta)|_{\bm \theta = \bm \theta^*}\left(\bm \theta - \bm \theta^*\right),
    \label{delta_theta_M}
\end{align}
where the first-order term vanishes due to convergence and the Hessian matrix $M(\bm \theta)$ is
\begin{align}
{{M}_{\ell_1 \ell_2 }}(\bm \theta)=\left( \frac{{{\partial }^{2}} \calL (\bm \theta) }{\partial {{\theta }_{\ell_1 }}\partial {{\theta }_{\ell_2 }}} \right)=\frac{{{\partial }}\epsilon (\bm \theta) }{\partial {{\theta }_{\ell_1 }}} \frac{{{\partial }}\epsilon (\bm \theta) }{\partial {{\theta }_{\ell_2 }}}+\epsilon(\bm \theta) \frac{{{\partial }^{2}}\epsilon (\bm \theta) }{\partial {{\theta }_{\ell_1 }}\partial {{\theta }_{\ell_2 }}}.
\label{eq:hessian}
\end{align}
We can then model a difference equation for the unnormalized ``differential state'' $\ket{\Psi(\bm \theta)} \equiv \ket{\psi(\bm \theta)} - \ket{\psi(\bm \theta^*)}$ as
\begin{align}
    \delta \ket{\Psi(\bm \theta)} = -\eta H_\infty(\bm \theta) \ket{\Psi(\bm \theta)},
    \label{eq:sch_eq}
\end{align}
where $H_\infty(\bm \theta) \sim M(\bm \theta)$ is similar to the Hessian matrix (see App.~\ref{app:Schrondinger}). The difference  equation can be interpreted as an imaginary time Schr\"odinger equation, and we identify a transition with gap closing of $H_\infty$ (equivalently $M(\bm \theta)$) driven by $O_0$ at the infinite time limit.

\begin{figure}[t]
    \centering
    \includegraphics[width=0.45\textwidth]{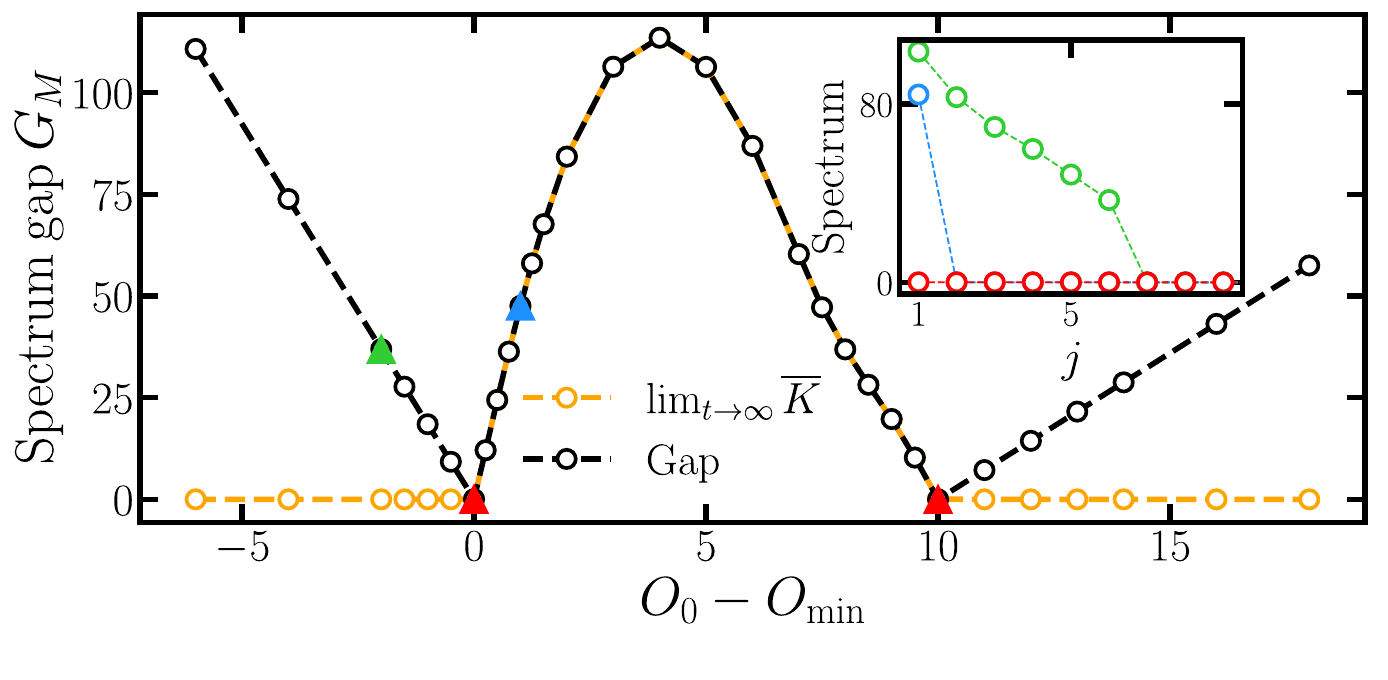}
    \caption{{\bf Spectrum gap of the effective Hamiltonian in Schr\"odinger interpretation of QNN in the example of XXZ model.} The spectrum gap of Hessian matrix of the effective Shr\"{o}dinger dynamics in $t\to \infty$ (black). The gapless transition point corresponds to $O_0 = O_{\rm min}, O_{\rm max}$ (red triangles). The orange line represents the QNTK $\lim_{t\to \infty}\overline{K}$. Inset shows the Hessian spectrum of the largest $10$ eigenvalues for the three cases $O_0 \lesseqgtr O_{\rm min}$ marked by triangles. The RPA consists of $D = 64$ layers (equivalently $L = 64$ parameters) on $n = 2$ qubits. The parameters in XXZ model is $J = 2$.
    }
    \label{fig:main_results_CMT}
\end{figure}


To provide insight into the transition, \QZ{we explore} the behaviors of the gap of Hessian matrix. We consider the Hessian eigenvalues at the late time limit of $t\to\infty$ and large circuit depth in Fig.~\ref{fig:main_results_CMT}. For {\it frozen-kernel dynamics} of $O_{\rm min}< O_0<O_{\rm max}$, Hessian matrix in Eq.~\eqref{eq:hessian} becomes a rank-one matrix with only one nonzero eigenvalue as $\epsilon(\bm \theta) \to 0$ (see blue in the \QZ{inset}), which equals the kernel and is verified by the orange and black curve in \QZ{Fig.~\ref{fig:main_results_CMT}}. While for {\it frozen-error dynamics} with $O_0 < O_{\rm min}$ (or $O_0 < O_{\rm max}$), due to non-vanishing $\epsilon (\bm \theta)$, the Hessian has multiple nonzero eigenvalues (see green in \QZ{the inset}). Overall, gap closing is observed at the critical point.
\QZ{Such a transition at a finite system size resembles that for non-Hermitian dynamical systems~\cite{berry2004physics,rotter2015review,el2018non}. More discussions on the statistical physics interpretation and the closing of the gap under different number of parameters can be found in App.~\ref{app:Schrondinger}.}

\begin{figure}[b]
    \centering
    \includegraphics[width=0.45\textwidth]{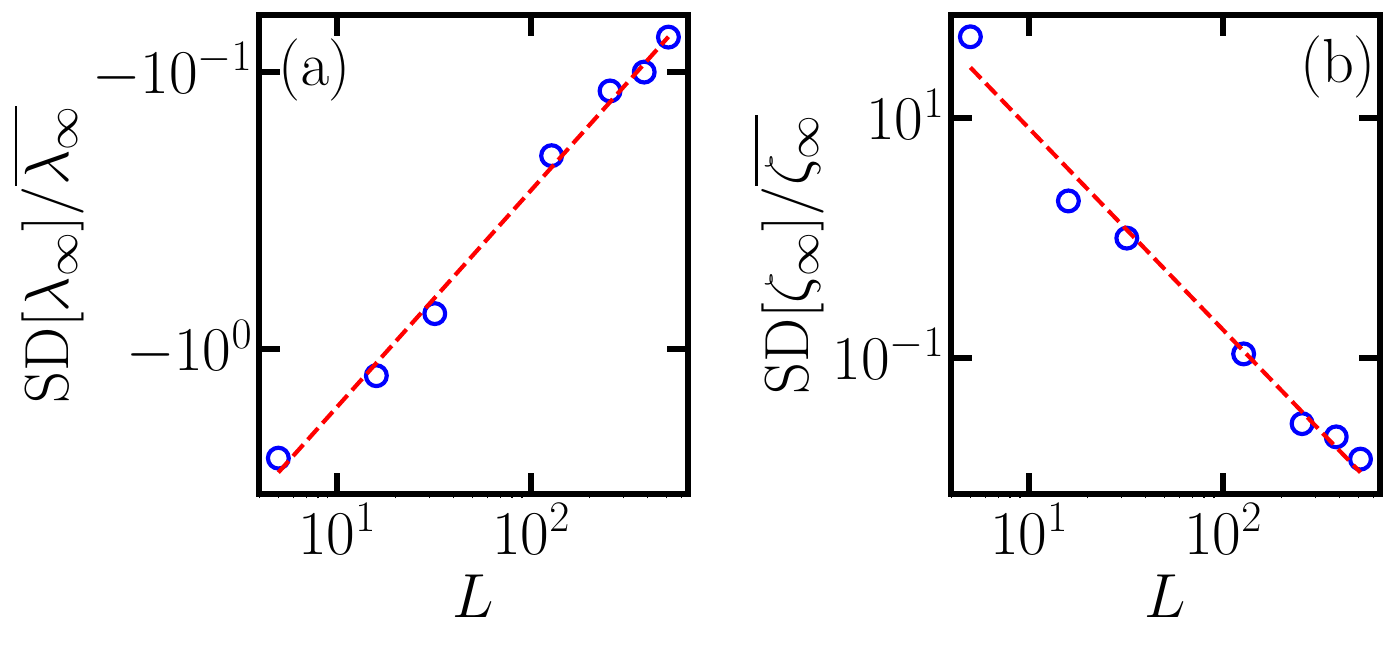}
    \caption{
    {\bf Late-time sample fluctuations.} 
    The standard deviations normalized by mean for the relative dQNTK $\lambda_\infty$ (a) and dynamical index $\zeta_\infty$ (b) are plotted versus the number of parameters $L$. Red dashed lines represent power-law fitting results. Here the RPA is applied on $n=5$ qubits with different $L$ parameters (via tuning number of layers $D$). The observable is a state projector and the target value is $O_0 = 1$.}
    \label{fig:main_results_flucs}
\end{figure}

\begin{figure*}
    \centering
    \includegraphics[width=0.95\textwidth]{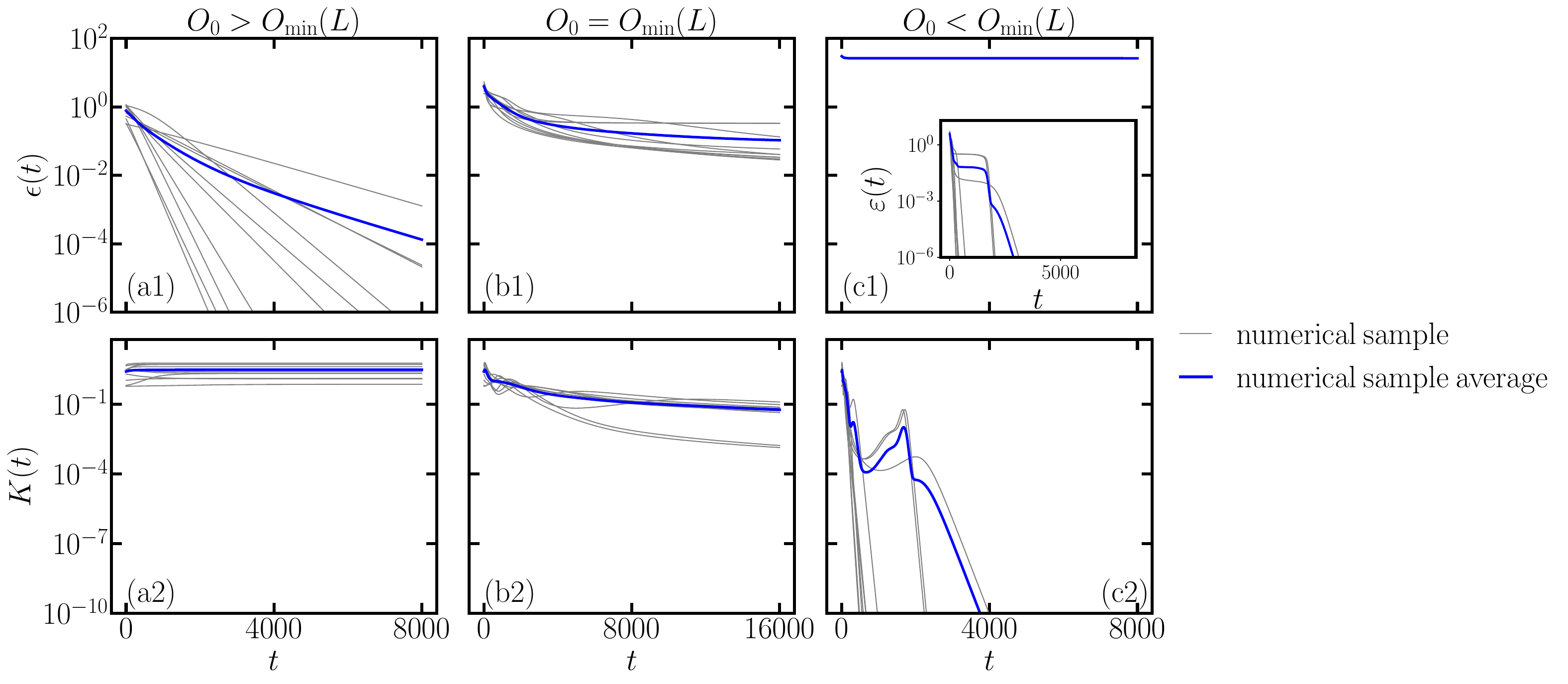}
    \caption{{\bf Dynamics in limited-depth QNNs in the example of the XXZ model.} All notations share the same meaning as in Fig.~\ref{fig:main_results}. The critical point $O_{\rm min}(L)$ for such QNNs depends on $L$ and has sample fluctuations. Here random Pulai ansatz (RPA) consists of $L=6$ variational parameters ($D=L$ for RPA) on $n=6$ qubits, and the parameter in XXZ model is $J=2$.}
    \label{fig:main_results_finite}
\end{figure*}

\section{Dynamics of limited-depth QNN}
We have so far focused on controllable QNNs with a universal gate set and large depth. In particular, for a general observable $\hat{O}$, reaching $O_{\rm min}$ may require a circuit of exponential depth (in the number of qubits)~\cite{kiani2020learning}. In addition, Lemma~\ref{lemma:RH_ratio} requires the (restricted) 4-design that involves a polynomial circuit depth. However, we point out that these depth requirements may be only necessary for the theory derivations and not necessary for the transition phenomenon. Indeed, it is an interesting question whether the transition still exists when the circuit is not controllable---either the ansatz is not universal~\cite{wiersema2020exploring,larocca2022diagnosing} or the depth is limited. Here we provide some results to the limited depth region of the QNNs under study. In this case, the circuit depth $L$ is limited such that the QNN's minimum achievable value of the observable $O_{\rm min}(L)$ deviates from the ground state energy $O_{\rm min}$, which is visible in Fig.~\ref{fig:main_results_CMT}(b). Such a scenario is often referred to as underparameterization.

We first consider the relative dQNTK $\overline{\lambda_\infty}$ and dynamical index $\overline{\zeta_\infty}$ versus the depth. In Lemma~\ref{lemma:RH_ratio}, we provide a justification of both quantities being constants for QNNs with a large depth $D$ to approach the restricted 4-design. 
In Fig.~\ref{fig:main_results_flucs}, we present a numerical example for the target $O_0 = 1$ in state-preparation tasks. The relative sample fluctuations, defined as the standard deviation compared to its mean, decay in a power-law scaling with $L$, and thus vanish in the asymptotic limit of $L\gg 1$. The mean values $\overline{\lambda_\infty} \propto -L$ and $\overline{\zeta_\infty} \to 1/2$ are shown in Fig.~\ref{fig:rstHaar} (b),(c) in Sec.~\ref{app:methods}.
The decay of fluctuation suggests that the ensemble-average results in Lemma.~\ref{lemma_lambda} can represent the typical samples. Note that changing the order of ensemble average for $\lambda_\infty$ (see Eq.~\eqref{eq:lambda_def}) and $\zeta_\infty$ (see Eq.~\eqref{eq:zeta_def}) has negligible effects (see App.~\ref{app:rh_result}). Similar results for other observables, e.g. XXZ model, are shown in App.~\ref{app:rh_result}. The speed of convergence roughly agrees with the 4-design requirement of Lemma~\ref{lemma:RH_ratio}. However, we emphasize that sample fluctuation being small is only a sufficient but not necessary condition for the dynamical transition, as we show in the below example.

To our surprise, in Fig.~\ref{fig:main_results_finite}, we find that the dynamical transition induced by the target value $O_0$ persists for a QNN with depth $D=L=n$ equaling the number of qubits, much less than what the theory requires. The results align with the dynamics of the controllable QNN presented in Fig.~\ref{fig:main_results}. We numerically find that the critical values for limited-depth QNNs, denoted as $O_{\rm min}(L)$, can deviate from the true ground state energy $O_{\rm min}$ of a given observable $\hat{O}$. The critical value for a QNN with $L\ll d$ will not only depend on depth due to limited expressivity, but also fluctuates due to different initializations.
We suspect this may be caused by the training converging to different local minimum traps~\cite{anschuetz2022quantum, you2021exponentially}. The deviation of the critical point $O_{\rm min}(L)$ from $O_{\rm min}$ indicates that the exponential depth for the convergence to $O_{\rm min}$ is not necessary for the dynamical transition to persist.
Moreover, despite the example being also not within the applicability of Lemma~\ref{lemma_lambda}, the relative dQNTK $\lambda$ still converges to a constant at late time as we show in SI. However, large sample fluctuation persists in this example due to $D=L=n$ being shallow, violating the unitary design assumption in Lemma~\ref{lemma_lambda}. However, we point out that as long as $\lambda$ has small time fluctuation at late time, its dynamics still follows the eneralized LV equation discussed in Eq.~\eqref{eq:LV-model-main}. The above results indicate that the depth requirement of the transition may be much less than that for overparametrization~\cite{larocca2023theory}.

\section{Speeding up the convergence}

While the transition in training dynamics is interesting, the crucial question in practical applications is about how to speed up the training convergence of QNNs. Typically, two types of loss functions are adopted in optimization problems, the quadratic loss function in Eq.~\eqref{loss_function} that we have focused on, and the linear loss function 
\begin{align}
    \calL(\bm \theta) = \braket{\hat{O}}.
    \label{loss_function_linear}
\end{align}
While the linear loss function is widely used in variational quantum eigensolver~\cite{kandala2017hardware,wiersema2020exploring}, we note that unlike the versatile quadratic loss function that has a tunable target value, a linear function does not allow preparing excited states above the ground state energy nor can it be utilized to data classification and regression. Moreover, for the case of solving the ground state, we show that adopting the quadratic loss function and choosing a target value well below the achievable minimum can speed up the convergence compared to the linear loss function case. Interestingly, `shooting for the star' will allow a faster convergence.

To begin with, we extend our theory framework to characterize the training dynamics of deep controllable QNNs with a linear loss function. To study its convergence, we further consider its residual error $\varepsilon(\bm \theta) = \braket{\hat{O}} - O_{\rm min}$. Via a similar approach (see details in App.~\ref{app:linear_loss_detail}), we have the dynamical equations for the error $\varepsilon(t)$ as
\begin{align}
    \delta \varepsilon(t) = -\eta K(t) + \mathcal{O}(\eta^2),
    \label{epsilon_dynamics_linear}
\end{align}
where $K(t)$ is still the QNTK defined in Eq.~\eqref{eq:K_def}. The dynamical equation for QNTK $K(t)$ becomes
\begin{align}
    \delta K(t) = -2\eta \mu(t) + \mathcal{O}(\eta^2),
    \label{K_dynamics_linear}
\end{align}
with $\mu(t)$ being dQNTK defined in Eq.~\eqref{eq:mu_def}.
One may notice that the only difference compared to Eqs.~\eqref{epsilon_dynamics} and~\eqref{eq:K_de1} is the missing of $\epsilon(t)= \varepsilon(t)$ on RHS due to a linear loss.

\begin{figure}[b]
    \centering
    \includegraphics[width=0.45\textwidth]{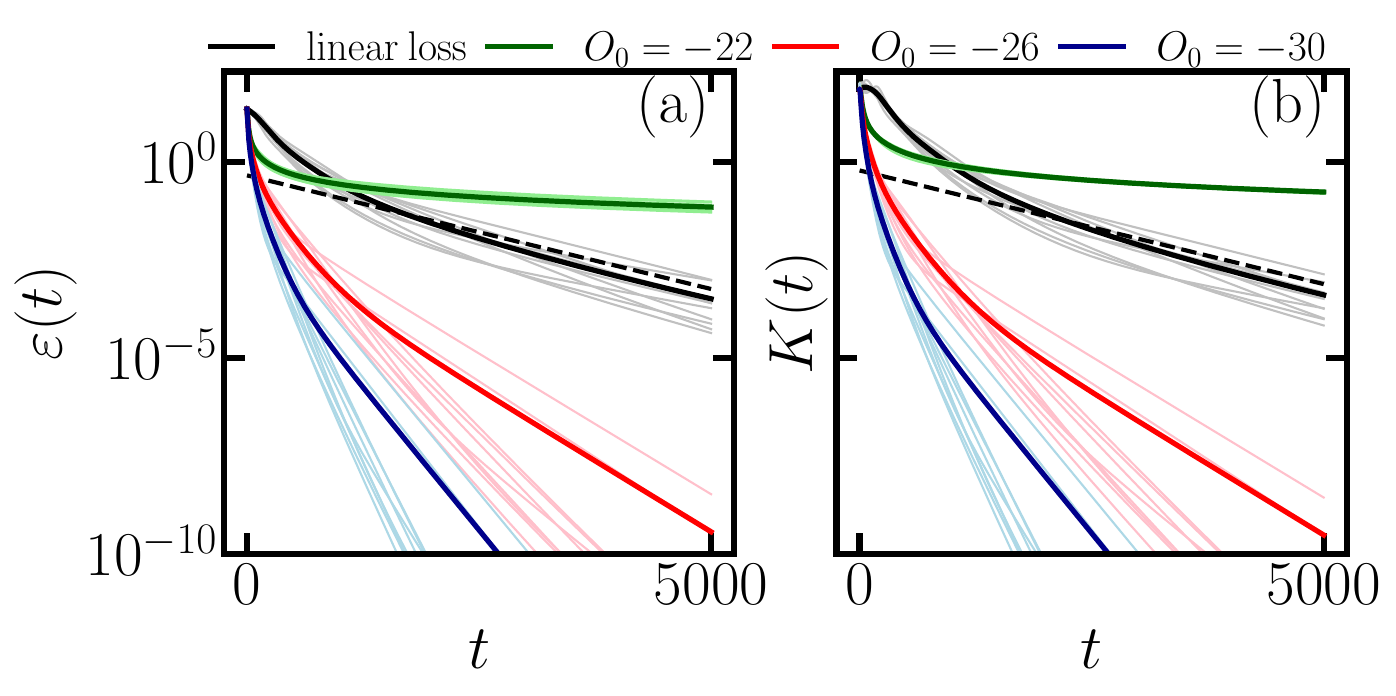}
    \caption{
    {\bf Dynamics in QNN in the example of XXZ model with different loss functions.} In (a) and (b), we show the dynamics of residual error $\varepsilon(t)$ (equals to total error $\epsilon(t)$) and QNTK $K(t)$ optimized with linear loss function (black solid) and quadratic loss functions with different $O_0$. $O_0 =-22$ (green) corresponds to $O_0 = O_{\rm min}$ at critical point and $O_0 = -26,-30$ (red and blue) correspond to $O_0 < O_{\rm min}$ in frozen error dynamics. Black dashed line indicates the exponential decay rate of the theoretical result in Eq.~\eqref{eq:sol_linear_main}. Thin lines with light colors represent dynamics with different initializations in each case, while the thick lines represent the ensemble average. Here random Pauli ansatz (RPA) consists of $L=192$ variational parameters ($D=L$ layers) on $n=6$ qubits, and the parameter of XXZ model is $J=2$.
    }
    \label{fig:linearloss_main}
\end{figure}

In the late-time limit, the results in `Unitary ensemble theory' section still applies to linear loss, and the relative dQNTK $\lambda(t)\equiv \mu(t)/K(t) = \lambda$ converges to a constant, leading to
\begin{align}
\label{eq:dy_eqs_linear_main}
\left\{ \begin{array}{ll}
         \partial_t \varepsilon(t) & = -\eta K(t),\\
        \partial_t K(t) &= -2\eta  \lambda K(t). \end{array} \right. 
\end{align}
Unlike the generalized LV model in Eqs.~\eqref{eq:LV-model-main} for the quadratic loss case, here the dynamics of $K(t)$ is self-determined, whereas the dynamics of $\varepsilon(t)=\epsilon(t)$ is fully determined by $K(t)$---the kernel-error duality is broken. Eqs.~\eqref{eq:dy_eqs_linear_main} can be directly solved as
\begin{align}
    2\lambda \epsilon(t) =2\lambda \varepsilon(t) = K(t) \propto e^{-2\eta \lambda t}.
    \label{eq:sol_linear_main}
\end{align}
Both $\epsilon(t)$ and $K(t)$ decay exponentially at a fixed rate $\propto \lambda$. In Fig.~\ref{fig:linearloss_main}, we present the numerical simulation results (black solid), and observe a good agreement with the theory (black dashed) from Eq.~\eqref{eq:sol_linear_main}.

With the linear-loss theory developed, we can now compare the convergence speed between the different choices of loss functions in solving the minimum value $O_{\rm min}$ and the corresponding ground state. As indicated in Eq.~\eqref{eq:sol_linear_main}, the linear loss function provides an exponential convergence with the exponent $2\eta \lambda$ being a constant (black). For quadratic loss functions, at the critical point setting $O_0 = O_{\rm min}$, the convergence is polynomial and exponent is zero (green lines in Fig.~\ref{fig:linearloss_main}), corresponding to a much slower convergence. However, recall that with a quadratic loss function, one can set $O_0< O_{\rm min}$ corresponding to the \textit{frozen error dynamics}, where the residual error $\varepsilon(t)$ decays exponentially with the exponent $2\eta \lambda R$ (see Eq.~\eqref{eq:K_nonfrozen_exp_main}). Here the residual $R$ is freely tunable by the target value $O_0$. Therefore, an appropriate choice of $O_0$ can provide a larger exponent and therefore faster convergence towards the solution, and we verify it in Fig.~\ref{fig:linearloss_main} through different values of $O_0$ (red and blue curves). Indeed, setting the target to be unachievable will still converge the output to the ground state, although the remaining error is frozen.

\section{Experimental results}
\begin{figure}
    \centering
    \includegraphics[width=0.45\textwidth]{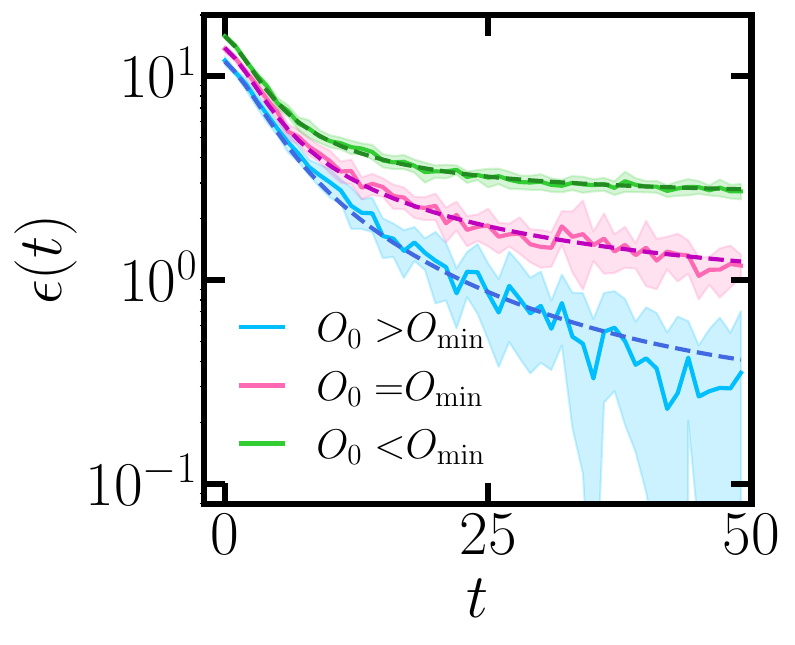}
    \caption{{\bf Dynamics of total error $\epsilon(t)$ on IBM quantum devices, Kolkata.} Solid and dashed curves represent experimental and theoretical results. An $n=2$ qubit $D=4$-layer hardware efficient ansatz (with $L=16$ parameters) is utilized to optimize with respect to XXZ model observable with $J=4$. The shaded areas represent the fluctuation (standard deviation) in the experimental data.}
    \label{fig:experiment}
\end{figure}

In this section, we consider the experimental-friendly hardware-efficient ansatz (HEA) to experimentally verify our results on real IBM quantum devices. Each layer of HEA consists of single qubit rotations along Y and Z directions, and followed by CNOT gates on nearest neighbors in a brickwall style~\cite{kandala2017hardware}.
Our experiments adopt the hardware \texttt{IBM Kolkata}, an IBM \texttt{Falcon r5.11} hardware with 27 qubits, via \texttt{IBM Qiskit}\cite{Qiskit}. The device has median $T_1\sim 
98.97 \text{us}$, median $T_2\sim 58.21 \text{us}$, median CNOT error $\sim 9.546 \times 10^{-3}$, median SX error $\sim 2.374\times 10^{-4}$, and median readout error $\sim 1.110\times 10^{-2}$. We randomly assign the initial variational angles, distributing them within the range of $[0,2\pi)$, and maintain consistency across all experiments. To suppress the impact of error, we average the results over 12 independent experiments conducted under the same setup for three distinct choices, $O_0={-10,-12,-14}$. 
In Fig.~\ref{fig:experiment}, the experimental data (solid) on \texttt{IBM Kolkata} agree well with the noisy theory model (dashed) and indicate the \textit{frozen-error dynamics} with constant error (green), the \textit{critical point} of polynomial decaying error (red) and the \textit{frozen-kernel dynamics} of exponential decaying error (blue). Individual training data and noisy theory model are presented in App.~\ref{app:hea}.

\section{Discussion}
Our results go beyond the early-time Haar random ensemble widely adopted in QNN study~\cite{mcclean2018barren,cerezo2021cost,liu2023analytic} and reveal rich physics in the dynamical transition controlled by the target loss function. The target-driven transcritical bifurcation transition in the dynamics of QNN suggests a different source to the transition without symmetry breaking. From the Schr\"odinger equation interpretation, there may exist other unexplored sources that can induce dynamical transition, especially when the QNN has limited depth and controllability. In practical applications, the dynamical transition guides us towards better design of loss functions to speedup the training convergence.

Another intriguing question pertains to the differences between classical and quantum machine learning within this formalism. In our examples, the target $O_0$ can be interpreted as a single piece of supervised data in a supervised machine learning task. Therefore, the dynamical transition we have discovered could be seen as a simplified version of a \emph{theory of data}. Classical machine learning also extensively explores dynamical transitions, whether in relation to learning rate dynamics \cite{lewkowycz2020large,2023arXiv230107737M} or the depth of classical neural networks \cite{lin2017does,roberts2021principles}. It is an open question whether some results similar to ours can be established for classical machine learning, especially in the context of the large-width regime of classical neural networks \cite{Batson:2021agz}. It is also an open problem how our results can generalize to the multiple data case.

Finally, we clarify the difference of our results to some related works. Firstly, while existing works~\cite{liu2022representation, liu2023analytic, liu2022laziness, wang2022symmetric, lee2017deep} on the quantum neural tangent kernel provide a perturbative explanation of gradient descent dynamics that fails to uncover the dynamical transition, our work uncovers the dynamical transition and formulates non-perturbative critical theories about the transition triggered by modifications in the quantum data. Secondly, we have developed a non-perturbative, phenomenological model using the generalized Lotka-Volterra equations to describe the dynamics as a transcritical bifurcation transition, providing a first-principle explanation using the restricted Haar ensemble. Thirdly, we provide an interpretation of the gradient descent dynamics using Schrödinger's equation in imaginary time, where the Hessian spectra can be mapped to the effective Hamiltonian using the language of physics, allowing us to study the effective spectral gap. Finally, using correlated dynamics of the Haar ensemble, we offer a more precise derivation of the statistics of the quantum neural tangent kernel, going beyond Ref.~\cite{liu2023analytic}.

\section{Methods}
\label{app:methods}

\subsection{QNN ansatz and details of the tasks}
The random Pauli ansatz (RPA) circuit is constructed as
\begin{align}
    \hat{U}(\bm \theta) = \prod_{\ell = 1}^{D} \hat{W}_\ell \hat{V}_{\ell}(\theta_\ell),
\end{align}
where $\bm \theta = (\theta_1,\dots,\theta_{L})$ are the variational parameters. For RPA, ${D}={L}$.
Here $\{\hat{W}_\ell\}_{\ell=1}^L \in \mathcal{U}_{\rm Haar}(d)$ is a set of fixed Haar random unitaries with dimension $d = 2^n$, and $\hat{V}_\ell$ is a $n$-qubit rotation gate defined to be
\begin{align}
    \hat{V}_\ell(\theta_\ell) = e^{-i\theta_\ell \hat{X}_\ell/2},
\end{align}
where $\hat{X}_\ell\in\{\hat{\sigma}^x,\hat{\sigma}^y,\hat{\sigma}^z\}^{\otimes n}$ is a random $n$-qubit Pauli operator nontrivially supported on every qubit. Once a circuit is constructed, $\{\hat{X}_\ell,\hat{W}_\ell\}_{\ell=1}^L$ are fixed through the optimization. Note that our results also hold for other typical universal ansatz of QNN, for instance, hardware efficient ansatz (see `Experimental results' and App.~\ref{app:hea}).

In the main text, some of our main results are derived for general observable $\hat{O}$. To simplify our expressions, we often consider $\hat{O}$ to be tracelss, for instance a spin Hamiltonian, which is not essential to our conclusions. A general traceless operator can be expressed as random mixture of Pauli strings (excluding identity)
\be 
\hat{O} = \sum_{i=1}^N c_i \hat{P}_i
\ee 
with real coefficients $c_i\in \mathbb{R}$ and nontrivial Pauli $\hat{P}_i\in \{\hat{\mathbb{I}},\hat{\sigma}^x,\hat{\sigma}^y,\hat{\sigma}^z\}^{\otimes n}/\{\hat{\mathbb{I}}^{\otimes n}\}$. 
To obtain explicit expressions, we also consider the XXZ model, described by
\begin{align}
    \hat{O}_{\rm XXZ} = -\sum_{i=1}^n \left[\hat\sigma^x_i \hat\sigma^x_{i+1} + \hat\sigma^y_i \hat\sigma^y_{i+1} + J \left(\hat\sigma^z_i \hat\sigma^z_{i+1} + \hat\sigma^z_i\right)\right].
\end{align}

To help understanding the non-frozen QNTK phenomena, we also consider a state preparation case with the observable $\hat O = \state{\Phi}$, where $\ket{\Phi}$ is the target state. 

\begin{figure*}
    \centering
    \includegraphics[width=0.8\textwidth]{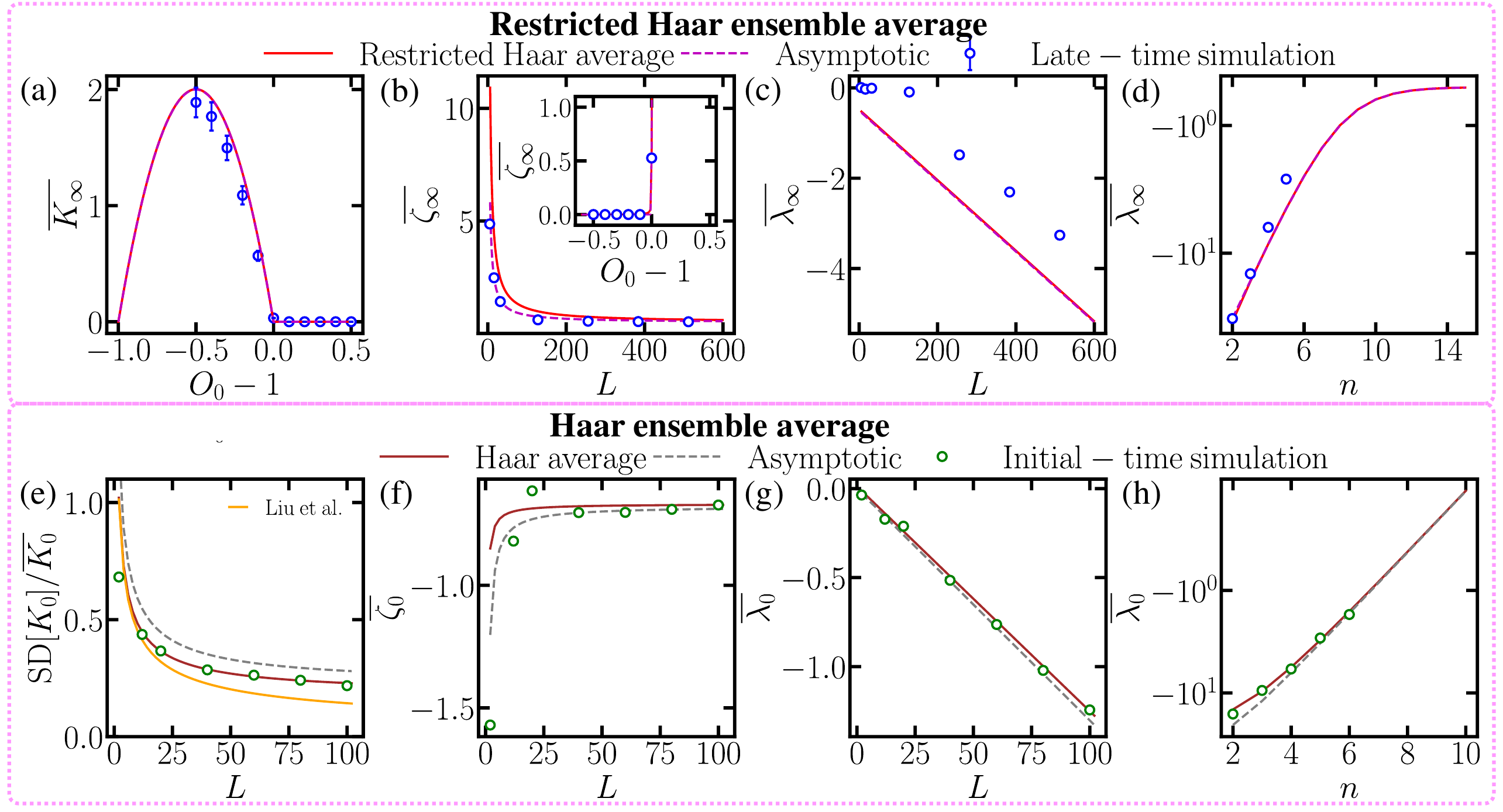}
    \caption{
    {\bf Ensemble average results under 
    restricted Haar ensemble (top) and Haar ensemble (bottom).} In top panel, we plot (a) $\overline{K_\infty}$ versus $O_0$ with $L=512$ fixed, (b) $\overline{\zeta_\infty}$ versus $L$,  $\overline{\lambda_\infty}$ versus (c) $L$ and (d) $n$ with $L=512$ at late time in state preparation. We set $O_0=1$ for (b) and (d), and $O_0 = 5$ for (c). Blue dots in top panel (a)-(c) represents numerical results from late-time optimization of $n=5$ qubit RPA. Red solid lines represent exact ensemble average with restricted Haar ensemble in  Eq.~\eqref{eq:K_rh},~\eqref{eq:zeta_rh},~\eqref{eq:lambda_rh} in App.~\ref{app:rh_result}. Magenta dashed lines represent asymptotic ensemble average with restricted Haar ensemble in Eq.~\eqref{eq:K_rh_main},~\eqref{eq:zeta_rh_main},~\eqref{eq:lambda_rh_main} which overlap with the exact results (red solid). The observable in all cases is $\ketbra{\Phi}{\Phi}$ with $\ket{\Phi}$ is a fixed Haar random state. In the inset of (b), we fix $L=512$.
    In bottom panel, we plot (e) fluctuation ${\rm SD}[K_0]/\overline{K_0}$ versus $L$, (f) $\overline{\zeta_0}$ versus $L$, $\overline{\lambda_0}$ versus (g) $L$ and (h) $n$ with $L=128$ under random initialization. Green dots in bottom panel from (e)-(g) represent numerical results from random initializations of $n=6$ qubit RPA. Brown solid lines represent exact ensemble average with Haar ensemble in  Eq.~\eqref{eq:Kfluc0},~\eqref{eq:zeta0},~\eqref{eq:lambda0} in App.~\ref{app:haar_result}. Gray dashed lines represent asymptotic ensemble average with restricted Haar ensemble in Eq.~\eqref{eq:relative_fluc_main},~\eqref{eq:zeta_haar_main},~\eqref{eq:lambda_haar_main}. The observable and target in (e)-(h) are XXZ model with $J=2$ and $O_0 = O_{\rm min}$. Orange solid line in (e) represents results from~\cite{liu2023analytic}.
    }
    \label{fig:rstHaar}
\end{figure*}

\subsection{Hamiltonian description and analytical solution of the LV dynamics}
From the conservation law in Eq.~\eqref{eq:LV_conserve}, we can introduce the canonical coordinates 
\begin{align}
P=\log(K),\;Q=\log(2\lambda\epsilon)
\end{align}
and the associated Hamiltonian
\begin{align}
H(Q,P)=
\eta( e^{Q}-e^{P}) \equiv \eta(2\lambda \epsilon-K),
\label{eq:LV_Hamiltonian}
\end{align}
from which the LV equations in \eqref{eq:LV-model-main} can be equivalently rewritten as the standard Hamiltonian equation generalizing Ref.~\citep{nutku1990hamiltonian},
\begin{align}
\label{eq:LV_Hamiltonian_eq}
\left\{ \begin{array}{ll}
         \frac{\textrm{d}Q}{\textrm{d}t} & 
         =\frac{\partial H}{\partial P}=\{Q,H\},
         \\
        \frac{\textrm{d}P}{\textrm{d}t} & 
        =-\frac{\partial H}{\partial Q}=\{P,H\},\end{array} \right. 
\end{align}
where $\{A,B\}=\frac{\partial A}{\partial Q}\frac{\partial B}{\partial P}-\frac{\partial A}{\partial P}\frac{\partial B}{\partial Q}$
denotes the Poisson bracket. From the position-momentum duality in Hamiltonian formulation, we identify an {\it error-kernel} duality between $e^{Q}\sim \epsilon$ and its gradient $e^{P}=|\partial\epsilon/\partial\boldsymbol{\theta}|^2$.

We can obtain analytical solution of Eqs.~\eqref{eq:LV-model-main} directly. When $C\neq 0$, we have
\begin{align}
\left\{ \begin{array}{ll}
\lambda\epsilon(t)&=C/\left[-2+B_1 e^{\eta C t}\right],
\\
K(t)&=C/\left[1-2B_1^{-1}e^{-\eta C t}\right],
\end{array} \right.
\label{eq:LV_Cnot0}
\end{align}
where $B_1$ is a constant fitting parameter as at early time we do not expect Eqs.~\eqref{eq:LV-model-main} to hold. 
When $C=0$, Eqs.~\eqref{eq:LV-model-main} leads to polynomial decay of both quantities
\begin{align}
K(t)=2\lambda \epsilon(t)&=2/\left(2\eta  t +B_2^{-1}\right),
\label{eq:LV_C0}
\end{align}
where $B_2$ is again a fitting parameter as at early time we do not expect Eqs.~\eqref{eq:LV-model-main} to hold. Indeed, we observe the bifurcation, and the convergence towards the fixed points are exponential for $C \lessgtr 0$ and polynomial for $C=0$.


\subsection{Details of restricted Haar ensemble}
Here we evaluate the average QNTK, relative dQNTK and dynamical index for the restricted Haar ensemble proposed in Eq.~\eqref{eq:restrict_U}. We focus on state preparation task to enable analytical calculation. As we aim to capture the late time dynamics with the state preparation task, we will be interested in the dynamics when the output state $\ket{\psi_0}$ has fidelity $\braket{\hat{O}} = |\braket{\psi_0|\Phi}|^2 = O_0 - R - \kappa$, with $\kappa \sim o(1)$ indicating late-time where observable is already close to its reachable target. Here the constant remaining term $R = O_0 - 1$ when $O_0 > 1$ and zero otherwise. Note that identity is the maximum reachable target value in state preparation. Under this ensemble, we have the following result (see details in App.~\ref{app:rh_result})
\begin{theorem}
\label{theorem:RH}
    For state projector observable $\hat{O} = \ketbra{\Phi}{\Phi}$, when the circuit satisfies restricted Haar ensemble, the ensemble average of QNTK, relative dQNTK and dynamical index
    \begin{align}
         \overline{K_\infty} &
         = \frac{{L}}{2d}\left(O_0 + R\right) \left( 1-O_0 - R\right), \label{eq:K_rh_main} \\
         \overline{\zeta_\infty} &= \frac{R}{O_0+R-1 }\left(1- \frac{1}{2(O_0 + R)} + \frac{d}{{L}}\right), \label{eq:zeta_rh_main}\\
         \overline{\lambda_\infty} &= \frac{{L}}{4d}\left(1-2O_0 - 2R\right) - \frac{O_0 + R}{2}, \label{eq:lambda_rh_main}
    \end{align}
    at the $L\gg1, d\gg1$ limit,
    where the target loss function value $O_0 \ge 0$, remaining constant $R = \min\{1-O_0, 0\}$.
\end{theorem}
Our results are verified numerically in Fig.~\ref{fig:rstHaar} in state preparation task, where we plot the above asymptotic equations as magenta dashed lines and the full formula in SI as solid red lines. Note that Lemma~\ref{lemma:RH_ratio} does not require $d\gg1$, but merely $D\gg1$. Indeed, full expressions in Theorem~\ref{theorem:RH} can also be derived for any finite $d$, just much more lengthy.

Subplot (a) plots $\overline{K_\infty}$ versus $O_0$. 
At late time, if the target $O_0 \ge 1$, from $O_0 = 1+ R$ we directly have $\overline{K_\infty} = 0$; if $O_0<1$, we have $R=0$ and $\overline{K_\infty} \propto O_0 (1-O_0)$ being a constant.

Subplot (b) shows the agreement of $\overline{\zeta_\infty}$ versus $L$, when we fix $O_0=1$. As predicted by the theory of Eq.~\eqref{eq:zeta_rh_main}, as $R=0$ in this case, $\overline{\zeta_\infty}= 1/2$ when $L\gg1$. Indeed, we see convergence towards $1/2$ as the depth increases. We also verify the $\overline{\zeta_\infty}$ versus $O_0$ relation in the inset, where $\overline{\zeta_\infty}= 0$ for $O_0< 1$, $1/2$ for $O_0=1$ and diverges for $O_0>1$. Note that for a circuit with medium depth $L \sim {\rm poly}(n)$, $\overline{\zeta_\infty} = 1/2+d/L$ would slightly deviate from $1/2$ for $O_0=1$ (Fig.~\ref{fig:rstHaar}(b)). This is indicating a `finite-size' effect affecting the dynamical transition, which we defer to future work.

Subplot (c) shows the agreement of $\overline{\lambda_\infty}$ versus $L$, where the linear relation is verified. As predicted by Eq.~\eqref{eq:lambda_rh_main}, this is the case regardless of $O_0$ value. We also verify the dependence of $\overline{\lambda_\infty}$ on $n$ (thus $d=2^n$) with a fixed $L$ in subplot (d), where we see as $n$ increases, $\overline{\lambda_\infty}$ converges to a constant only relying on $O_0$.


\subsection{Haar ensemble results}
We also evaluate the Haar ensemble expectation values for reference, which captures the early time QNN dynamics. 
Under the Haar random assumption, we find the following lemma
\begin{lemma}
\label{lemma:Haar}
For traceless operator $\hat{O}$, when the initial circuit satisfies Haar random ($4$-design) and circuit $L\gg 1$ and $d\gg 1$,
the ensemble averages of QNTK, relative dQNTK and dynamical index have leading order
\begin{align}
    \overline{K_0}=&L\frac{d\tr({\hat O}^2)}{2 (d-1) (d+1)^2} 
    \label{eq:K_haar_main}\\
    \overline{\zeta_0} = & -\frac{1}{L}\left[1 + \frac{\tr(\hat{O}^4)}{\tr(\hat{O}^2)^2}\right]
    \nonumber
    \\
    &\quad + \frac{1}{2}\left[\frac{\tr(\hat{O}^4)}{\tr(\hat{O}^2)^2}-d O_0 \frac{\tr(\hat{O}^3)}{\tr(\hat{O}^2)^2}-\frac{3}{d}\right], 
    \label{eq:zeta_haar_main}\\
    \overline{\lambda_0} = &\frac{L\tr(\hat{O}^3)}{4 d \tr(\hat{O}^2)}. \label{eq:lambda_haar_main}
\end{align}

\end{lemma}
Note that for observables with non-zero trace, evaluation is also possible, we present those lengthy formulae and the proofs in App.~\ref{app:haar_result}. Note that similar to Theorem~\ref{theorem:RH}, here the requirement of $d\gg1$ is for simplification of formula only and the full formula in SI apply to any finite $d$. Meanwhile, it is important to notice the dimension dependence of the trace terms.

Specifically, for the XXZ model we considered, when $d\gg1$, the above Lemma~\ref{lemma:Haar} leads to
\begin{align}
    \overline{K_0}_{\rm XXZ} &\simeq (1+J^2)\frac{Ln}{d}\\
    \overline{\zeta_0}_{\rm XXZ} &\simeq  -\frac{1}{L}\left(1 + \frac{3}{d}\right)-O_0 \frac{3J(1-J^2) }{4(1+J^2)^2 n},
    \label{zeta_XXZ}\\
    \overline{\lambda_0}_{\rm XXZ} &\simeq \frac{3J(1-J^2) L }{4 (1 + J^2) d}.
    \label{eq:lambda_XXZ}
\end{align}

We verified the Haar prediction on $\overline{\zeta_0}$ and $\overline{\lambda_0}$ with random initialized circuits in Fig.~\ref{fig:rstHaar}(f)-(h). Note that when $L$ is large enough, $\overline{\zeta_0}_{\rm XXZ}$ scales linearly with $O_0$, while $\overline{\lambda_0}_{\rm XXZ}$ converges to zero exponentially with $n$.

In the Haar case, we can also obtain the fluctuation properties.
\begin{theorem}
\label{theorem:K_concentration}
In the asymptotic limit of wide and deep QNN $d, L\gg 1$, we have the ensemble average of QNTK standard deviation (4-design)
\begin{align}
    {\rm SD}[K_0] 
    &= \left( \frac{3L}{4d^6}\left[ d^2\tr(\hat{O}^2)^2-2d \tr(\hat{O}^2) \tr(\hat{O})^2+ \tr(\hat{O})^4\right]\right.\nonumber\\
    &\ \ \left.+ \frac{L^2}{4d^5}\left[d\tr(\hat{O}^4) - 4\tr(\hat{O}^3)\tr(\hat{O})\right]\right)^{1/2}.
    \label{eq:Kstd_asym_main}
\end{align}
\end{theorem}
Note that similar to Theorem~\ref{theorem:RH} and Lemma~\ref{lemma:Haar}, here the requirement of $d\gg1$ is for simplification of formula only and the full formula in SI apply to any finite $d$.

For traceless operators, Eq.~\eqref{eq:Kstd_asym_main} can be further simplified and the relative sample fluctuation of QNTK is
\begin{align}
    \frac{{\rm SD}[K_0] }{\overline{K_0}} = \frac{1}{\sqrt{L}}\left(L\frac{\tr\left(\hat{O}^4\right)}{\tr(\hat{O}^2)^2}
    + 3\right)^{1/2}.
    \label{eq:relative_fluc_main}
\end{align}
This result refines Ref.~\cite{liu2023analytic} with more accurate ensemble averaging technique and provides an additional term $\sim \tr\left(\hat{O}^4\right)/\tr(\hat{O}^2)^2$. Therefore, the sample fluctuation also depends on the observable being optimized.
Specifically, for the XXZ model we considered, Eq.~\eqref{eq:relative_fluc_main} becomes 
\begin{align}
    \frac{{\rm SD}[K_0]}{\overline{K_0}} \simeq \sqrt{\frac{3}{L}\left(\frac{L}{d}
    + 1\right)}.
    \label{flutuation_relative_asympt}
\end{align}
When $L\gg d$, the relative fluctuation ${\rm SD}[K_0]/\overline{K_0}\sim 1/\sqrt{d}$ is constant. However, as $d=2^n$ is exponential while realistic number of layers $L$ is polynomial in $n$, therefore $d\gg L$ is more common, where the relative fluctuation ${\rm SD}[K_0]/\overline{K_0}\sim \sqrt{1/L}$ decays with the depth, consistent with Ref.~\cite{liu2023analytic}. We numerically evaluate the ensemble average in Fig.~\ref{fig:rstHaar}(e) and find a good agreement between our full analytical formula (red solid, Eq.~\eqref{eq:Kfluc0} in SI) and the numerical results (blue circle). The asymptotic result (magenta dashed, Eq.~\eqref{flutuation_relative_asympt}) also captures the scaling correctly. The results refine the calculation of Ref.~\cite{liu2023analytic}, which has a substantial deviation when $L$ and $d$ are comparable.


%

\begin{acknowledgements}
We thank David Simmons-Duffin for very helpful discussions. BZ and QZ acknowledges NSF CAREER Award CCF-2240641, ONR Grant No. N00014-23-1-2296, National Science Foundation OMA-2326746, Cisco Systems, Inc. and Halliburton Company. JL is supported in part by International Business Machines (IBM) Quantum through the Chicago Quantum Exchange, and the Pritzker School of Molecular Engineering at the University of Chicago through AFOSR MURI (FA9550-21-1-0209). LJ acknowledges support from the ARO (W911NF-23-1-0077), ARO MURI (W911NF-21-1-0325), AFOSR MURI (FA9550-19-1-0399, FA9550-21-1-0209, FA9550-23-1-0338), NSF (OMA-1936118, ERC-1941583, OMA-2137642, OSI-2326767, CCF-2312755), NTT Research, Packard Foundation (2020-71479), and the Marshall and Arlene Bennett Family Research Program. This material is based upon work supported by the U.S. Department of Energy, Office of Science, National Quantum Information Science Research Centers. WX acknowledges support from the Simons Collaboration on Ultra-Quantum Matter, which is a grant from the Simons Foundation (651440), and the Simons Investigator award (990660). 
\end{acknowledgements}


\begin{widetext}

\appendix

\tableofcontents

From now on, we will omit `$\hat{\ }$' on operators, as long as it is not confusing.

\section{Derivation of QNTK dynamics}
\label{app:diff_eq}

In this section, we provide details on the derivation of Eq.~\eqref{eq:K_de1} in the main text. 
The time difference of QNTK is
\begin{align}
    \delta K{(t)} &\equiv K(t+1)-K(t)\\
    &= \sum_\ell \delta \left(\frac{\partial\epsilon}{\partial \theta_\ell} \frac{\partial\epsilon}{\partial \theta_\ell}\right)\\
    &= \sum_\ell {\left.\left(\frac{\partial\epsilon (\bm \theta)}{\partial \theta_\ell}\frac{\partial\epsilon (\bm \theta)}{\partial \theta_\ell}\right)\right\rvert_{\bm \theta=\bm \theta(t+1)}} - {\left.\left(\frac{\partial\epsilon (\bm \theta) }{\partial \theta_\ell}\frac{\partial\epsilon (\bm \theta)}{\partial \theta_\ell}\right)\right\rvert_{\bm \theta=\bm \theta(t)}}\\
    &= \sum_\ell \left[{\left.\left(\frac{\partial\epsilon (\bm \theta)}{\partial \theta_\ell}\frac{\partial\epsilon (\bm \theta)}{\partial \theta_\ell}\right)\right\rvert_{\bm \theta=\bm \theta(t+1)}} - \left({\left.\frac{\partial\epsilon (\bm \theta)}{\partial \theta_\ell}\right\rvert_{\bm \theta=\bm \theta(t+1)} \left.\frac{\partial\epsilon(\bm \theta)}{\partial \theta_\ell}\right\rvert_{\bm \theta=\bm \theta(t)}}\right) + \left({\left.\frac{\partial\epsilon (\bm \theta)}{\partial \theta_\ell}\right\rvert_{\bm \theta=\bm \theta(t+1)} \left.\frac{\partial\epsilon(\bm \theta)}{\partial \theta_\ell}\right\rvert_{\bm \theta=\bm \theta(t)}}\right) \right.\nonumber\\
    &\quad \quad \quad - \left.{\left.\left(\frac{\partial\epsilon (\bm \theta)}{\partial \theta_\ell}\frac{\partial\epsilon (\bm \theta)}{\partial \theta_\ell}\right)\right\rvert_{\bm \theta=\bm \theta(t)}}\right] \\
    &= \sum_\ell {\left.\frac{\partial\epsilon (\bm \theta)}{\partial \theta_\ell}\right\rvert_{\bm \theta=\bm \theta(t+1)}} \delta\left({\left.\frac{\partial\epsilon (\bm \theta)}{\partial \theta_\ell}\right\rvert_{\bm \theta=\bm \theta(t)}}\right) + \delta\left({\left.\frac{\partial\epsilon (\bm \theta)}{\partial \theta_\ell}\right\rvert_{\bm \theta=\bm \theta(t)}} \right) {\left.\frac{\partial\epsilon (\bm \theta)}{\partial \theta_\ell}\right\rvert_{\bm \theta=\bm \theta(t)}} \\
    &= \sum_\ell 2\delta\left({\left.\frac{\partial\epsilon (\bm \theta)}{\partial \theta_\ell}\right\rvert_{\bm \theta=\bm \theta(t)}}\right){\left.\frac{\partial\epsilon (\bm \theta)}{\partial \theta_\ell}\right\rvert_{\bm \theta=\bm \theta(t)}} + {\left.\frac{\partial\epsilon (\bm \theta)}{\partial \theta_\ell}\right\rvert_{\bm \theta=\bm \theta(t+1)}} \delta\left({\left.\frac{\partial\epsilon (\bm \theta)}{\partial \theta_\ell}\right\rvert_{\bm \theta=\bm \theta(t)}}\right) - \delta\left({\left.\frac{\partial\epsilon (\bm \theta)}{\partial \theta_\ell}\right\rvert_{\bm \theta=\bm \theta(t)}}\right){\left.\frac{\partial\epsilon (\bm \theta)}{\partial \theta_\ell}\right\rvert_{\bm \theta=\bm \theta(t)}}\\
    &= \sum_\ell 2\delta\left({\left.\frac{\partial\epsilon (\bm \theta)}{\partial \theta_\ell}\right\rvert_{\bm \theta=\bm \theta(t)}}\right){\left.\frac{\partial\epsilon (\bm \theta)}{\partial \theta_\ell}\right\rvert_{\bm \theta=\bm \theta(t)}} + \delta\left({\left.\frac{\partial\epsilon (\bm \theta)}{\partial \theta_\ell}\right\rvert_{\bm \theta=\bm \theta(t)}}\right) \delta\left({\left.\frac{\partial\epsilon (\bm \theta)}{\partial \theta_\ell}\right\rvert_{\bm \theta=\bm \theta(t)}}\right).
\end{align}
The second term in the last formula has two $\delta$, so it is in higher orders in $\eta$, and we only focus on the first term. We utilize the leading order Taylor expansion on $\delta \partial \epsilon (\bm \theta)/\partial \theta_\ell$ as
\begin{align}
\delta \left( {\left.\frac{\partial\epsilon (\bm \theta)}{\partial \theta_\ell}\right\rvert_{\bm \theta=\bm \theta(t)}}\right) =\underset{{{\ell }_{1}}}{\mathop \sum }\, \left.\frac{{{\partial }^{2}}\epsilon {(\bm \theta)}}{\partial {{\theta }_{\ell_1 }}\partial {{\theta }_{{{\ell }}}}}  \delta {{\theta }_{{{\ell }_{1}}}} \right\rvert_{\bm \theta=\bm \theta(t)}  + \mathcal{O}({{\eta }^{2}})
= -\eta \underset{{{\ell }_{1}}}{\mathop \sum }\, \left.\frac{{{\partial }^{2}}\epsilon {(\bm \theta)} }{\partial {{\theta }_{\ell_{1} }}\partial {{\theta }_{{{\ell }}}}} \frac{\partial \epsilon {(\bm \theta)} }{\partial {{\theta }_{{{\ell }_{1}}}}}\epsilon{(\bm \theta)}\right\rvert_{\bm \theta = \bm \theta(t)} +\mathcal{O}({{\eta }^{2}}).
\end{align}
So we have
\begin{align}
\sum_{\ell} \delta\left({\left.\frac{\partial\epsilon (\bm \theta)}{\partial \theta_\ell}\right\rvert_{\bm \theta=\bm \theta(t)}}\right){\left.\frac{\partial\epsilon (\bm \theta)}{\partial \theta_\ell}\right\rvert_{\bm \theta=\bm \theta(t)}} =-\eta \underset{\ell ,{{\ell }_{1}}}{\mathop \sum }\, \left.\frac{{{\partial }^{2}}\epsilon {(\bm \theta)} }{\partial {{\theta }_{\ell }}\partial {{\theta }_{{{\ell }_{1}}}}}\frac{\partial \epsilon {(\bm \theta)} }{\partial {{\theta }_{{{\ell }_{1}}}}}\frac{\partial \epsilon {(\bm \theta)} }{\partial {{\theta }_{\ell }}}\epsilon{(\bm \theta)}\right\rvert_{\bm \theta = \bm \theta(t)} +\mathcal{O}({{\eta }^{2}})=-\eta \epsilon{(\bm \theta)} \mu{(\bm \theta)}|_{\bm \theta = \bm \theta(t)} +\mathcal{O}({{\eta }^{2}}),
\end{align}
which leads to the gradient descent dynamical equation of $K(t)$ as
\begin{align}
    \delta K{(t)} = -2\eta \epsilon{(t)} \mu{(t)} + \mathcal{O}(\eta^2),
    \label{eq:K_diff_eq}
\end{align}
where we omit the explicit dependence on $\bm \theta$ and only present the $t$-dependence. It recovers Eq.~\eqref{eq:K_de1} of the main text.

\section{Schr\"odinger equation interpretation}
\label{app:Schrondinger}

\QZ{In this section, we provide more details on applying statistical physics tools to study the properties associated with the dynamical transition.}
If we consider an unnormalized `differential state' as a superposition of two output states of the QNN, 
\begin{align}
|\Psi(\boldsymbol{\theta})\rangle=|\psi(\boldsymbol{\theta})\rangle-|\psi(\boldsymbol{\theta}^{*})\rangle=N (\bm \theta) \left(\bm \theta - \bm \theta^*\right),
\end{align}
where $N (\bm \theta) =\partial|\psi(\boldsymbol{\theta})\rangle/\partial\boldsymbol{\theta}$ at $\boldsymbol{\theta}=\boldsymbol{\theta}^{*}$ is a $d\times {L}$ dimensional matrix. Formally, the pseudoinverse of the linear relation is denoted by $N (\bm \theta)^{-1}$. From Eq.~\eqref{delta_theta_M} \QZ{of the main text}, the effective Hamiltonian $H_\infty =NMN^{-1}$ can be obtained from 
\begin{align}    
\delta|\Psi(\boldsymbol{\theta})\rangle=N(\bm \theta) \cdot\delta\boldsymbol{\theta}=-\eta N(\bm \theta) M(\bm \theta) N(\bm \theta) ^{-1}|\Psi(\boldsymbol{\theta})\rangle.
\end{align}
which is the Schr\"odinger equation with imaginary time $i\eta$. Note that the quantum mechanical Hamiltonian is valid only under the late-time limit $t\to\infty$. 

\begin{figure}[t]
    \centering
    \includegraphics[width=0.5\textwidth]{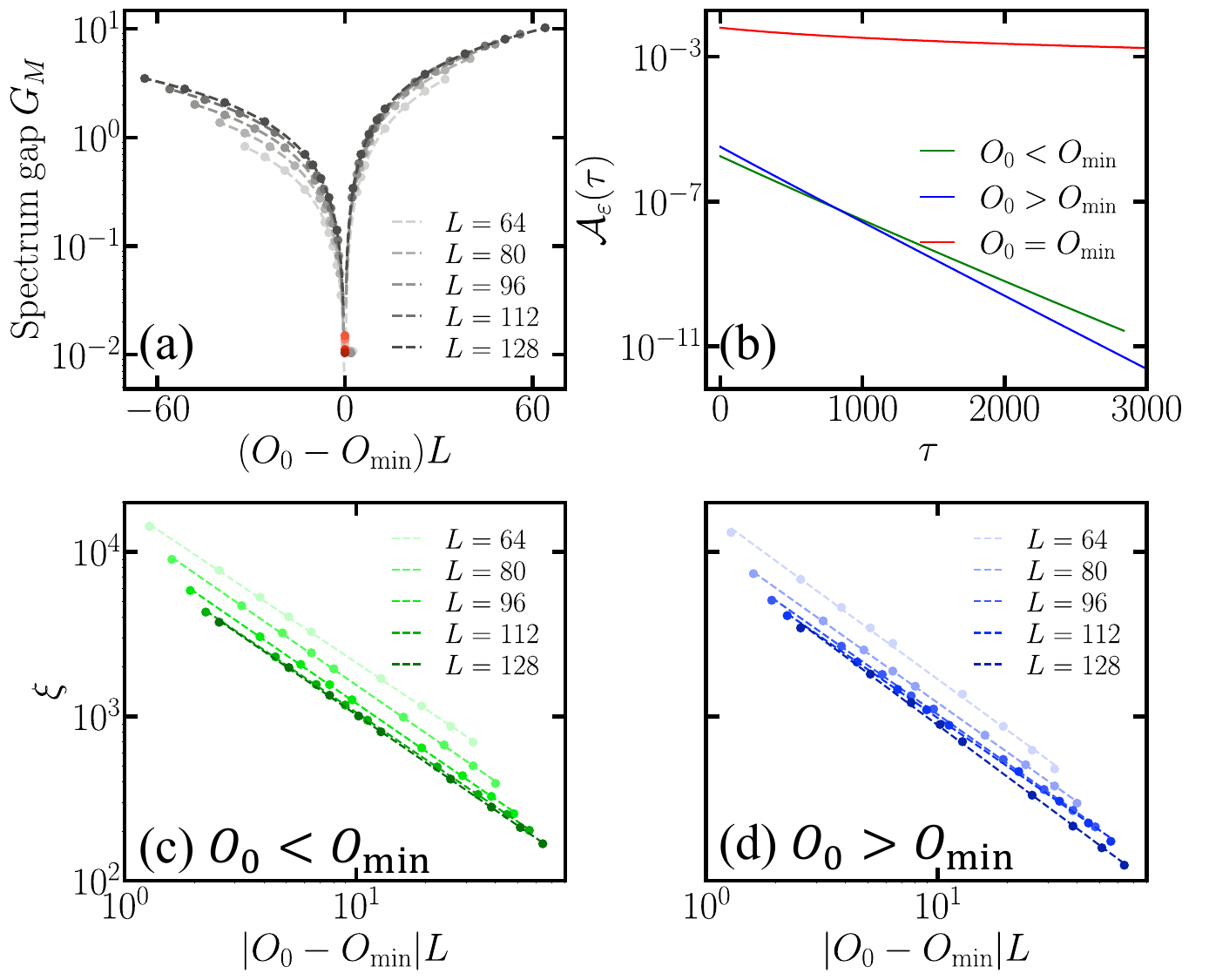}
    \caption{{\bf Spectrum gap and correlation functions in the Schr\"odinger equation interpretation.} 
    \QZ{(a) The spectrum gap versus the scaled target value $(O_0 - O_{\rm min})L$ with different number of parameters. We choose $O_0 - O_{\rm min} \in [-0.5, 0.5]$. Dots from light to dark represent gaps with increasing $L$. The dashed lines with corresponding colors show $G_M \sim (|O_0 - O_{\rm min}|L)^{\nu_1}$ with $\nu_1 \simeq 1$ from fitting. Red dots represent the critical point $O_0 = O_{\rm min}$. The nonvanishing gap at $O_0 = O_{\rm min}$ is due to finite training time. }
    In \QZ{(b)}, we plot the decay of autocorrelators $\calA_\epsilon(\tau)$ with different $O_0 \lesseqgtr O_{\rm min}$ (green for `$<$', red for `$=$' and blue for `$>$'). In \QZ{(c)} and \QZ{(d)}, we show the scaling of correlation length $\xi \sim (|O_0 - O_{\rm min}|L)^{-\nu_2}$ with $\nu_2 \simeq 1$ (dashed lines) by fitting for both $O_0 \lessgtr O_{\rm min}$. Dots from light to dark represent $\xi$ with increasing $L$ variational parameters, and the dashed lines with the corresponding color represent fitting result. Here we choose $O_0 - O_{\rm min} \in [-0.5, 0.5]$. In \QZ{(b)} the RPA consists of ${D}=96$ layers (equivalently ${L}=96$ parameters). In all cases, the RPA is applied on a system of $n=4$ qubits and the parameter in XXZ model is $J=2$.
    }
    \label{fig:scalings}
\end{figure}

\QZ{
To show that the large depth limit is well-defined, we evaluate the Hessian gap of QNNs at late time for various different circuit depth, or equivalently the number of parameters (i.e. the `parameter size'). As we see in Fig.~\ref{fig:scalings}(a), the curves of the spectrum gap versus a rescaled target value $(O_0-O_{\rm min})L$ collapse well as the parameter size $L$ increases, indicating a well-defined transition at the large-depth limit.
We notice a linear-closing gap around the critical point (red triangle in Fig.~\ref{fig:main_results_CMT} of the main text), and verify the scaling in Fig.~\ref{fig:scalings}(a) via fitted the gap $G_M$ to 
\begin{align}
    G_M \sim (|O_0 - O_{\rm min}|L)^{\nu_1},
\end{align}
resulting in $\nu_1 = 0.996\pm 0.004, 1.09 \pm 0.021$ for $O_0 \lessgtr O_{\rm min}$ (dashed lines). 
However, we also notice that the minimum gap in the numerical study has no significant dependence on the parameter size $L$---it is dominated by the finite training time in the numerical simulation which fails to achieve the infinite time limit. As at the critical point $O_0=O_{\rm min}$, the QNN training dynamics converges polynomially, which makes accessing the infinite-time limit numerically difficult. However, we do expect that the Hessian gap vanishes exactly at infinite time as both error and QNTK will vanish. Such an exact gap closing within finite size is in contrast to normal phase transitions in statistical physics and therefore we regard the transition not as a conventional phase transition in the statistical physics sense.
}


\QZ{
Despite not being a genuine phase transition, we can still adopt tools from statistical physics to provide more insight into the gap-closing transition.
Regarding the QNN as a Schr\"odinger system described by a Hamiltonian dependent on both $L$ and $t$, we then study the correlators.}
Both $K(t)$ and $\epsilon(t)$ has the same correlator behavior and we focus on $\epsilon$ here (see App.~\ref{app:correlators}) and define the autocorrelator
$
    \calA_\epsilon(\tau) \equiv 
        \mathbb{E}\left[\varepsilon(t)\varepsilon(t+\tau)\right],
$
where the average is over ensemble of trajectories and we will consider the $t\gg1$ region. Here $\varepsilon$ is adopted as it captures the residual error for the study of fluctuations.
Away from the critical point, from the mean-field dynamics at late time according to Eq.~\eqref{eq:LV_Cnot0} of main text, we expect the autocorrelator
\begin{align}
    \calA_\epsilon (\tau) \sim \exp\left[-|\tau|/\xi\right]
\end{align} 
to decay exponentially with a finite correlation length $\xi$, which is verified in Fig.~\ref{fig:scalings}(b). In Fig.~\ref{fig:scalings}(c)(d), the correlation length $\xi$ versus the rescaled target value $|O_0 - O_{\rm min}|L$ also collapses with the increasing number of parameters $L$, which aligns with the behavior of the spectrum gap in Fig.~\ref{fig:scalings}(a).
We further reveal the scaling of correlation length as 
\begin{align}
    \xi \sim 1/|C| \sim (|O_0 - O_{\rm min}|L)^{-\nu_2},
\end{align}
where $\nu_2$ is found to be $0.961\pm 0.012, 1.01\pm 0.025$ for $O_0 \lessgtr O_{\rm min}$, shown in Fig.~\ref{fig:scalings}(c) and (d). The numerical values of $\nu_1$ (Fig.~\ref{fig:scalings}(a)) and $\nu_2$ are indeed identical up to numerical precision, as expected.

At the critical point $O_0 = O_{\rm min}$, any physical quantity $F$ is expected to exhibit power-law correlation $\calA_F (\tau)\sim 1/|\tau|^{2\Delta[F]}
$ for $|\tau|\gg t$, defining the scaling dimension $\Delta[F]$. Based on the definitions Eqs.~\eqref{eq:K_def},~\eqref{eq:mu_def} and~\eqref{eq:lambda_def} of the main text, one can establish the following scaling relations
\begin{align}
    &\Delta[\epsilon] = 2\Delta[K] - \Delta[\mu] \nonumber\\
    & \Delta[\lambda] = \Delta[\mu] - \Delta[K]. 
    \label{eq:scaling_laws}
\end{align}
As shown in Fig.~\ref{fig:scalings}(c), our numerical result suggests $\Delta[\epsilon]=1/2$. This seems consistent with the solution Eq.~\eqref{eq:LV_C0} of main text, assuming the correlation is factorizable. In App.~\ref{app:correlators}, we also find $\Delta[K] = \Delta[\mu] = 1/2$, while $\Delta[\lambda] = 0$ as $\lambda$ is a constant. We see that the scaling relations in Eqs.~\eqref{eq:scaling_laws} are indeed fulfilled.

In summary, we have taken tools from statistical physics to study the dynamical transition, which is shown to mimic certain properties of a continuous gap-closing transition in a quantum mechanical system described by the time-independent effective Hamiltonian $H_{\infty}$. 
We also want to mention that reaching the truly infinite-time limit poses challenges both numerically and experimentally. In our estimation of the correlation function in App.~\ref{app:correlators}, we rely on taking the subtle limit $\tau\gg t\gg1$ and the fluctuations being small.

\section{Details of autocorrelators}
\label{app:correlators}

In this section, we provide a mean-field approach to provide an insight to the scaling of autocorrelators. For any time-dependent quantity $F(t)$ that has ensemble fluctuations, we define the late-time autocorrelator as
\begin{align}
    \calA_F(\tau) \equiv 
        \mathbb{E}\left[\left(F(t)-F(\infty)\right) \left(F(t+\tau)-F(\infty)\right)\right],
\label{eq:autocorrelator_def}
\end{align}
where the average is over the ensemble of trajectories and we will consider $t\gg1$ region. Here $F(\infty)=\lim_{T\to\infty}\int_T^{2T}{\rm d}tF(t)/T$ is the smoothed late-time value of the function. 
For $\epsilon(t)$, the definition in Eq.~\eqref{eq:autocorrelator_def} leads to $
    \calA_\epsilon(\tau) \equiv 
        \mathbb{E}\left[\varepsilon(t)\varepsilon(t+\tau)\right]
$, which is the one adopted in the main text.

When $O_0 > O_{\rm min}$ with $C > 0$, utilizing the solution Eq.~\eqref{eq:LV_Cnot0} of the main text, the mean-field approximated autocorrelator in Eq.~\eqref{eq:autocorrelator_def} becomes
\begin{align}
    \calA_\epsilon(\tau) &=  \int d\Gamma(\lambda) d\Gamma(B_1) \frac{C^2/\lambda^2}{\left(B_1 e^{\eta C t} - 2\right) \left(B_1 e^{\eta C (t+\tau)} - 2\right)}\\
    &\simeq \int d\Gamma(\lambda) d\Gamma(B_1) \frac{C^2\lambda^2}{B_1^2 e^{2\eta Ct} e^{\eta C \tau}}\\
    &\sim \frac{C^2/\overline{\lambda}^2}{\overline{B_1}^2 e^{2\eta Ct} e^{\eta C \tau}} \sim e^{-\eta C \tau},
\end{align}
where $\Gamma(\lambda), \Gamma(B_1)$ is the distribution of conserved quantity and fitting parameter in different initialization. Similarly, for $O_0 < O_{\rm min}$ with $C < 0$, we have
\begin{align}
    \calA_\epsilon(\tau) &=  \int d\Gamma(\lambda) d\Gamma(B_1) \left(\frac{C/\lambda}{B_1 e^{\eta C t} - 2} - R\right) \left(\frac{C/\lambda}{B_1 e^{\eta C (t+\tau)} - 2} - R\right)\\
    &= \int d\Gamma(\lambda) d\Gamma(B_1) \left(\frac{-2R}{B_1 e^{\eta C t} - 2} - R\right) \left(\frac{-2R}{B_1 e^{\eta C (t+\tau)} - 2} - R\right)\\
    &= R^2 \int d\Gamma(\lambda) d\Gamma(B_1) \frac{1}{1-2B_1^{-1}e^{-\eta C t}} \frac{1}{1-2B_1^{-1}e^{-\eta C (t+\tau)}}\\
    &\sim \frac{R^2 \overline{B_1}^2 e^{2\eta C t} e^{\eta C \tau}}{4} \sim e^{\eta C \tau}.
\end{align}
We numerically show the decay of autocorrelators with different $O_0$ in Fig.~\ref{fig:correlation_noncrit}(a), (d).
In both cases, we see the exponential decay of autocorrelators, and the correlation length defined by 
$\calA_F(\tau) \sim \exp\left(-\tau/\xi\right)$ is
\begin{align}
    \xi \sim 1/(\eta |C|) \sim |O_0 - O_{\rm min}|^{-\nu_2}.
    \label{eq:xi}
\end{align}
For $O_0 < O_{\rm min}$, we directly have $|C| \propto |O_0 - O_{\rm min}|$ which leads to $\nu_2 = 1$. This is verified in Fig.~\ref{fig:correlation_noncrit}(c), with the fitted exponent $\nu_2 = 1.006$. For $O_0 > O_{\rm min}$, we have $|C| = |K| = G_M$, with $G_M$ being the spectrum gap of Hessian at late time, which indicates that $\nu_2 = \nu_1$. This is again verified in Fig.~\ref{fig:correlation_noncrit}(f) with the fitted exponent $\nu_2 = 1.038$.

From the duality between $\epsilon$ and $K$, the autocorrelator of $K$ also decays exponentially, when the system is not at the critical point, shown in Fig.~\ref{fig:correlation_noncrit}(b), (e). The corresponding correlation length exponent can also be found as $\nu_2 = 1.006, 1.034$ for $O_0 \lessgtr O_{\rm min}$.
In general, we have 
\begin{align}
    \nu_2 = 1,
\end{align}
within our numerical precision.

\begin{figure}[t]
    \centering
    \includegraphics[width=0.8\textwidth]{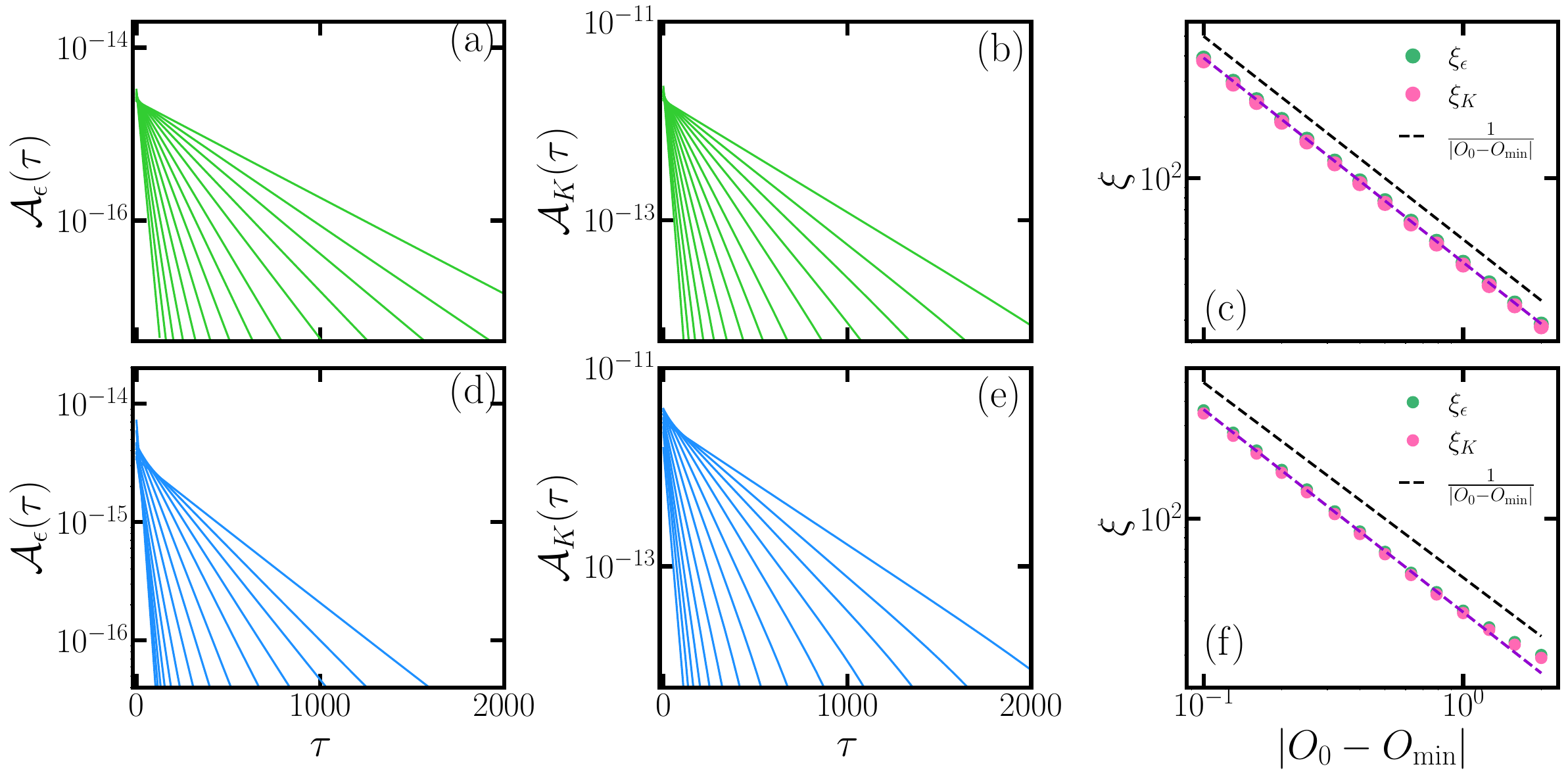}
    \caption{Decay of autocorrelators and corresponding correlation length with $O_0$ away from critical points as $O_0 < O_{\rm min}$ (top) and $O_0 > O_{\rm min}$ (bottom). The first two columns plot the autocorrelators. In (c), (f), the overlapping red and green dots represent correlation length fitted from Eq.~\eqref{eq:xi} and dashed lines with same color show the fitting results. Black dashed lines represent its scaling as $1/|O_0 - O_{\rm min}|$. The observable is the Hamiltonian of XXZ model with $J=2$, and circuit ansatz is $n=2$ qubit RPA with $L=64$ layers.}
    \label{fig:correlation_noncrit}
\end{figure}

On the other hand, at the critical point $O_0 = O_{\rm min}$ we have
\begin{align}
    \calA_\epsilon(\tau) &= \int d\Gamma(\lambda) d\Gamma(B_2) \frac{1/\lambda^2}{\left(B_2^{-1} +  2\eta t\right) \left(B_2^{-1} + 2 \eta (t+\tau)\right)}\\
    &\simeq \int d\Gamma(\lambda) d\Gamma(B_2) \frac{1/\lambda^2}{4\eta^2 t(t+\tau)}\\
    &\sim \frac{1/\overline{\lambda}^2}{4\eta^2 t(t+\tau)}\sim \frac{1}{\tau},
\end{align}
which decays polynomially with $\tau$ (see Fig.~\ref{fig:correlation_crit}). Note that the scaling of $\calA_\epsilon(\tau) \sim 1/\tau$ holds only if $\tau \gg t$, otherwise it is nearly a constant. As $\lambda = \mu/K$ approaches a constant, we have that $\mu \sim K \sim 1/t$, and thus $\calA_\mu(\tau)\sim 1/\tau$. 
From the definition of scaling dimension, $\calA_F(\tau) = 1/\tau^{2\Delta [F]}$, one can find that
\begin{align}
    \Delta[\epsilon] = \Delta[K] = \Delta[\mu] = 1/2,
\end{align}
which is verified in Fig.~\ref{fig:correlation_crit}, and $\Delta[\mu] = 0$.

\begin{figure}[t]
    \centering
    \includegraphics[width=0.5\textwidth]{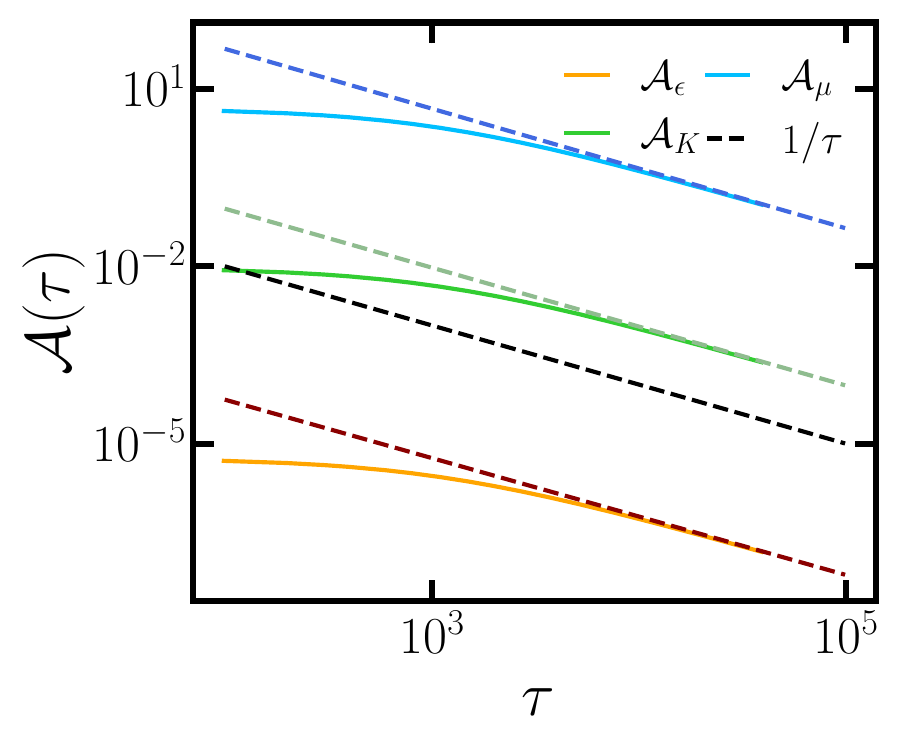}
    \caption{Decay of autocorrelators for $\epsilon(t), K(t), \mu(t)$ at critical $O_0 = -6$ (solid curves). Dashed lines with corresponding color show the fitting results with $\Delta[\epsilon] = 0.494$, $\Delta[K] = 0.499$ and $\Delta[\mu] = 0.506$. Black dashed line represent the scaling $1/\tau$.}
    \label{fig:correlation_crit}
\end{figure}

\section{Observable trace properties}
\label{app:trace_property}

In this work, we {mainly} focus on the traceless observables, where a typical example is the spin Hamiltonian {of many-body system}. In general, a $n$-qubit observable can always be written in the form of linear combinations of nontrivial Paulis $O=\sum_{i=1}^N c_i P_i$ with $P_i \in \{\mathbb{I},\sigma^x,\sigma^y,\sigma^z\}^{\otimes n}/\{\mathbb{I}^{\otimes n}\}$, where $1\le N \le 4^n - 1$ is the number of unqiue Paulis in the observable. We discuss the scaling of the trace of its powers up to four with respect to Hilbert space dimension $d$ and number of terms $N$. To begin with,
\begin{align}
    \tr(O) &= 0,\\
    \tr(O^2) &= \sum_{i_1,i_2=1}^N c_{i_1} c_{i_2} \tr(P_{i_1} P_{i_2}) = \sum_i c_{i}^2 \tr(P_{i}^2) + \sum_{i_1\neq i_2} c_{i_1} c_{i_2} \tr(P_{i_1} P_{i_2}) \sim Nd. \label{eq:trO2}
\end{align}
For higher orders, we focus on some typical cases of observables to provide an insight to its scaling.

\subsection{One-body observable}

For the simplest case, a linear combination of $1$-local Paulis $O = \sum_i c_i P_i$, where $P_i$ is nontrivially supported on only one qubit and $c_i \in \mathbb{R}$, the trace of its third and fourth power is
\begin{align}
    \tr(O_{1-{\rm local}}^3) &= \sum_{i_1,i_2,i_3} c_{i_1}c_{i_2}c_{i_3}\tr(P_{i_1}P_{i_2}P_{i_3}) = 0, \label{eq:trO3_1local}\\
    \tr(O_{1-{\rm local}}^4) &= \sum_{i_1,i_2,i_3, i_4} c_{i_1}c_{i_2}c_{i_3}c_{i_4} \tr(P_{i_1}P_{i_2}P_{i_3}P_{i_4}) \\
    &= \sum_{i_1,i_2}c_{i_1}^2 c_{i_2}^2 \tr(P_{i_1}^2 P_{i_2}^2) + 2\sum_{i_1,i_2\neq i_3} c_{i_1}^2 c_{i_2} c_{i_3} \tr(P_{i_1}^2 P_{i_2} P_{i_3})+ \sum_{\substack{i_1\neq i_2\\ i_3\neq i_4}} c_{i_1} c_{i_2} c_{i_3} c_{i_4} \tr(P_{i_1}P_{i_2} P_{i_3} P_{i_4})\\
    &= \sum_{i_1,i_2} c_{i_1}^2 c_{i_2}^2 \tr(\bI) + 0 + 2\sum_{\substack{i_1\neq i_2}} c_{i_1}^2 c_{i_2}^2 \tr(P_{i_1}^2 P_{i_2}^2)\\
    &\sim N^2 d + 2 N(N-1) d \sim 3N^2 d ,\label{eq:trO4_1local}
\end{align}
where the contribution from Paulis nontrivially supported on the same qubit is overestimated. In a special case where $O$ incorporates all possible $1$-local Pauli with equal weights, the rigorous result is $\tr(O^4) = (3N^2-6N)d\sim 3N^2 d$, leading to a sub-order correction to the estimation in Eq.~\eqref{eq:trO4_1local}.

\subsection{Two-body observable: XXZ model}

In the $2$-local Pauli case, we consider the Hamiltonian consists of Paulis at most non-trivially supported on two qubit, and specifically, the two qubit are nearest neighbors. Here we take the Heisenberg model as an example, 
\begin{align}
    O_{\rm HM} &= -\sum_{i=1}^{n-1} \left(J_x \sigma^x_i \sigma^x_{i+1} + J_y \sigma^y_i \sigma^y_{i+1} + J_z \sigma^z_i \sigma^z_{i+1}\right) - h\sum_{i=1}^n \sigma^z_i . 
\end{align}
Specifically, when $J_x=J_y$ and $J_z=h$ but $J_z\neq J_x$, the general Heisenberg model is reduced to the XXZ model as
\begin{align}
    O_{\rm XXZ} &= -\sum_{i=1}^{n-1} \left(\sigma^x_i \sigma^x_{i+1} + \sigma^y_i y_{i+1} + J \sigma^z_i \sigma^z_{i+1}\right) - J\sum_{i=1}^n \sigma^z_i .
\end{align}
which is studied in the main text.

The trace of its power from second to fourth can be exactly solved as
\begin{align}
    \tr(O^2_{\rm XXZ}) &= \sum_{i,j=1}^{n-1} \tr(\sigma^x_i \sigma^x_{i+1} \sigma^x_j \sigma^x_{j+1}) + \tr(\sigma^y_i \sigma^y_{i+1} \sigma^y_j \sigma^y_{j+1}) + J^2 \tr(\sigma^z_i  \sigma^z_{i+1} \sigma^z_j \sigma^z_{j+1}) + \sum_{i,j=1}^n J^2 \tr(\sigma^z_i  \sigma^z_j)\\
    &= \sum_{i=1}^{n-1} \left(2\tr(\bI) + J^2 \tr(\bI)\right) + \sum_{i=1}^n J^2 \tr(\bI)\\
    &= \left[(J^2+2)(n-1) + J^2 n\right]d
    \label{eq:trO2_xxz}\\
    &\simeq 2(J^2+1) n d,
\end{align}
where in the second line we only keep the nonzero terms and omit the zero contributions. 

With one more step, we can find the trace of its third power as
\begin{align}
    \tr(O_{\rm XXZ}^3) &= -6J\sum_{i,j,k=1}^{n-1} \tr(\sigma^x_i \sigma^x_{i+1} \sigma^y_j \sigma^y_{j+1} \sigma^z_k \sigma^z_{k+1})- 3J^3\sum_{i,j,k=1}^{n-1}\tr(\sigma^z_i \sigma^z_{i+1} \sigma^z_j \sigma^z_k)\\
    &= 6(n-1)J d - 6J^3 (n-1) d\\
    &= 6J(1-J^2) (n-1)d,
    \label{eq:trO3_xxz}\\
    &\simeq 6J(1-J^2) n d,
\end{align}
where again the first equation is an effective equation for all non-zero contributions.
When $J \lessgtr 1$, we have $\tr(O_{\rm XXZ}^3) \sim \mp Nd$, and at the critical $J = 1$, we have $\tr(O_{\rm XXZ}^3) = 0$. 

The trace of fourth power is
\small
\begin{align}
    &\tr(O_{\rm XXZ}^4)
    \nonumber
    \\
    &= \sum_{\substack{i,j,\\k,l=1}}^{n-1} \left[ 2\tr(\sigma^x_i \sigma^x_{i+1} \sigma^x_j \sigma^x_{j+1} \sigma^x_k \sigma^x_{k+1} \sigma^x_l \sigma^x_{l+1}) + 4\tr(\sigma^x_i \sigma^x_{i+1} \sigma^x_j \sigma^x_{j+1} \sigma^y_k \sigma^y_{k+1} \sigma^y_l \sigma^y_{l+1}) + 8J^2 \tr(\sigma^x_i \sigma^x_{i+1} \sigma^x_j \sigma^x_{j+1} \sigma^z_k \sigma^z_{k+1} \sigma^z_l \sigma^z_{l+1}) \right.\nonumber\\
    &\left. \quad  + 2\tr(\sigma^x_i \sigma^x_{i+1} \sigma^y_j \sigma^y_{j+1} \sigma^x_k \sigma^x_{k+1} \sigma^y_l \sigma^y_{l+1}) + 4J^2 \tr(\sigma^x_i \sigma^x_{i+1} \sigma^z_j \sigma^z_{j+1} \sigma^z_k \sigma^z_{k+1} \sigma^z_l \sigma^z_{l+1}) + J^4 \tr(\sigma^z_i \sigma^z_{i+1} \sigma^z_j \sigma^z_{j+1} \sigma^z_k \sigma^z_{k+1} \sigma^z_l \sigma^z_{l+1}) \right]\nonumber\\
    & \quad +  \sum_{i,j=1}^{n-1}\sum_{k,l=1}^n \left[ 8J^2\tr(\sigma^x_i \sigma^x_{i+1} \sigma^x_j \sigma^x_{j+1} \sigma^z_k \sigma^z_l) + 8J^2\tr(\sigma^x_i \sigma^x_{i+1} \sigma^y_j \sigma^y_{j+1} \sigma^z_k \sigma^z_l) + 6J^4\tr(\sigma^z_i \sigma^z_{i+1} \sigma^z_j \sigma^z_{j+1} \sigma^z_k \sigma^z_l)\right]\nonumber\\
    &\quad + \sum_{i,k=1}^{n-1}\sum_{j,l=1}^n \left[ 4 J^2\tr(\sigma^x_i \sigma^x_{i+1} \sigma^z_j \sigma^x_k \sigma^x_{k+1} \sigma^z_l) + 4J^2 \tr(\sigma^x_i \sigma^x_{i+1} \sigma^z_j \sigma^y_k \sigma^y_{k+1} \sigma^z_l)\right] + \sum_{i,j,k,l=1}^n J^4\tr(\sigma^z_i \sigma^z_j \sigma^z_k \sigma^z_l)\\
    &= 2(n-1)(3n-5)d + 4(n-1)^2 d + 8J^2 (n-1)^2 d + 2(n-3)^2 d + 4J^2 (n-3)^2 d + J^4 (n-1)(3n-5)d\nonumber\\
    &\quad  + 8J^2 n (n-1)d + 8J^2 \left(-2(n-1)d\right) + 6J^4 (n^2 + 3n- 8)d + 4J^2 (n-1)(n-4)d + 4J^2\left(2 (n-1)d\right) + J^4 n(3n-2)d\\
    &= \left[12 \left(J^2+1\right)^2 n^2+4 \left(2 J^4-19 J^2-9\right) n-43 J^4+68 J^2+32\right]d
    \label{eq:trO4_xxz}\\
    &\simeq 12(J^2 + 1)^2 n^2 d \sim N^2 d,
\end{align}
\normalsize
where in the first equation we only show the nonzero unique contributions and the coefficient ahead of each term counts its repetitions. 

We leave observables with more body interaction for future work though it does not change the major conclusion/scaling of this work.

\section{Method in ensemble average calculation}

To assist the following discussion, we present the expression of first order gradient of residual error by commutators as
\begin{align}
    \frac{\partial\epsilon}{\partial \theta_\ell} = \partial_{\theta_\ell}\braket{\psi_0|U^\dagger(\bm \theta) O \hat{U}(\bm \theta)|\psi_0}= \frac{i}{2}\braket{\psi_0|U^\dagger _{\ell^-}\left[X_\ell, U_{\ell^+}^\dagger OU_{\ell^+}\right]U_{\ell^-}|\psi_0} =  \frac{i}{2}\braket{\psi_0|U^\dagger _{\ell^-}\left[X_\ell, O_{\ell^+} \right]U_{\ell^-}|\psi_0},
\end{align}
where $\ket{\psi_0}$ is the initial pure state of system.
Here we define the unitary notations $U_{\ell^-}$ as
\begin{align}
    U_{\ell^-} = \prod_{k=1}^{\ell-1} W_k V_k (\theta_k), U_{\ell^+} = \prod_{k=\ell}^L W_k V_k (\theta_k),
    \label{eq:U_ell_def}
\end{align}
and $O_{\ell^+} = U_{\ell^+}^\dagger O U_{{\ell^+}}$ for simplicity.

The second order gradient assuming $\ell_1<\ell_2$ and $\ell_1 = \ell_2 = \ell$ can be written in a similar way as
\begin{align}
    \frac{\partial^2 \epsilon}{\partial \theta_{\ell_1}\partial\theta_{\ell_2}}  &= -\frac{1}{4}\braket{\psi_0|U_{\ell_1^-}^\dagger[X_{\ell_1},U_{\ell_1 \shortto \ell_2}^\dagger[X_{\ell_2}, U_{\ell_2^+}^\dagger O U_{\ell_2^+}]U_{\ell_1\shortto \ell_2}]U_{\ell_1^-}|\psi_0} = -\frac{1}{4}\braket{\psi_0|U_{\ell_1^-}^\dagger[X_{\ell_1},U_{\ell_1 \shortto \ell_2}^\dagger[X_{\ell_2}, O_{\ell_2^+}] U_{\ell_1\shortto \ell_2}]U_{\ell_1^-}|\psi_0}\\
    \frac{\partial^2 \epsilon}{\partial\theta_\ell^2} &= -\frac{1}{4}\braket{\psi_0|U_{\ell^-}^\dagger [X_\ell, [X_\ell, O_{\ell^+}]]U_{\ell^-}|\psi_0},
\end{align}
where
\begin{align}
    U_{\ell_1\shortto \ell_2} = \prod_{k = \ell_1}^{\ell_2 - 1} W_k V_k (\theta_k).
    \label{eq:U_1to2_def}
\end{align}


\section{Definition of restricted Haar random unitary ensemble}
\label{app:rst_def}

{In this section, we provide the definition of restricted Haar (abbreviated as ``RH'') random ensemble, which will be used in App.~\ref{app:k_design} and App.~\ref{app:rh_result}. For the optimization problem considered in the main text, we always search for the optimal output state satisfying the constraint from minimum loss function. Equivalently, we can regard it as a state preparation problem with observable $O = \ketbra{\Phi}{\Phi}$. Therefore, the optimal unitary has to be able to map the trivial input state towards the target state while its mapping on the rest states orthogonal to input state can be arbitrary. We can then define the unitary to implement this operation as 
\begin{align}
    U_{\rm RH} = 
    \begin{pmatrix}
    1 & {\bm 0} \\
    {\bm 0} & V
\end{pmatrix},
    \label{eq:toyU_app}
\end{align}
where $V$ is a unitary of dimension $d-1$ and we assume $V \sim \calU_{\rm Haar}(d-1)$. Here the columns correspond to orthonormal basis including $\ket{\psi_0}$ while rows correspond to the basis including $\ket{\Phi}$. 
}

\section{Frame potential with restricted Haar ensemble}
\label{app:k_design}

\subsection{Frame potential applied to QNN}

To quantify the randomness of an ensemble of unitaries, we evaluate the $k$th frame potential $\calF^{(k)}$. For an arbitrary unitary ensemble $\calE$,
\begin{align}
    \calF_\mathcal{E}^{(k)} = \frac{1}{|\mathcal{E}|^2}\sum_{U,U^\prime \in \mathcal{E}}|\tr(U^\dagger U^\prime)|^{2k}\ge \calF_{\rm Haar}^{(k)} = k!,
\end{align}
where the minimum is achieved by Haar ensemble~\cite{roberts2017chaos}. 

To provide insight into the ensemble in the case of $O_0\le O_{\rm min}$, we evaluate the frame potential of the restricted Haar ensemble
\begin{align}
     \calF^{(k)}_{\rm RH}
     &= \sum_{\substack{k_1,k_2=0\\ k_1+2k_2 \le k}}^k \frac{k!}{k_1! (k_2!)^2 (k-k_1-2k_2)!} \calF_{\rm Haar}^{(k_1 + k_2)} \label{eq:F_k_main}\\
     &\ge \calF_{\rm Haar}^{(k+1)}=(k+1)!.
\end{align}
We verify the above Eq.~\eqref{eq:F_k_main} and lower bound $(k+1)!$ in Fig.~\ref{fig:framepot}(a).

To verify that the ensemble distribution of unitaries of the QNN satisfy the restricted Haar ensemble in Eq.~\eqref{eq:restrict_U}, ideally we want to consider the different unitaries from random initialization that lead to the same converged state, so that the ensemble averaged values can provide insight into a specific training dynamics where a specific converged state is observed. However, this is in general challenging as random initialization in general will lead to convergence to different local optimums, unless in the case of $O_0 \le O_{\rm min}$, where the converged state will be the ground state, up to a finite degeneracy. In this case, we can directly evaluate the frame potential over late time unitaries from different initializations.
We numerically evaluate the dynamics of $2$-order frame potential $\calF^{(2)}$ in Fig.~\ref{fig:framepot}(b) where different $O_0$ is considered. For $k=2$, the frame potential $\calF^{(k)}$ over the restricted Haar unitary ensemble is $\calF^{(2)} = 7$ according to Eq.~\eqref{eq:F_k_main} and $\calF^{(2)}=2$ for Haar random ensemble. Indeed, we see that for $O_0 \le O_{\rm min}$, the ensemble frame potential approaches to the prediction of restricted Haar ensemble. When $O_0 > O_{\rm min}$, $\calF^{(2)}$ is far away from it, this is because the converged state is not unique for $O_0 > O_{\rm min}$ and different random initializations fail to provide the ensemble of unitaries with a fixed converged state: When one consider different random initialization, each training trajectory converges to a different state and the entire unitary ensemble under random initialization does not capture the restrictions and in fact approach Haar random. While in each single trajectory, the convergence still places a restriction on the typical unitary that maps the initial state to the final state.

\begin{figure}[t]
    \centering
    \includegraphics[width=0.7\textwidth]{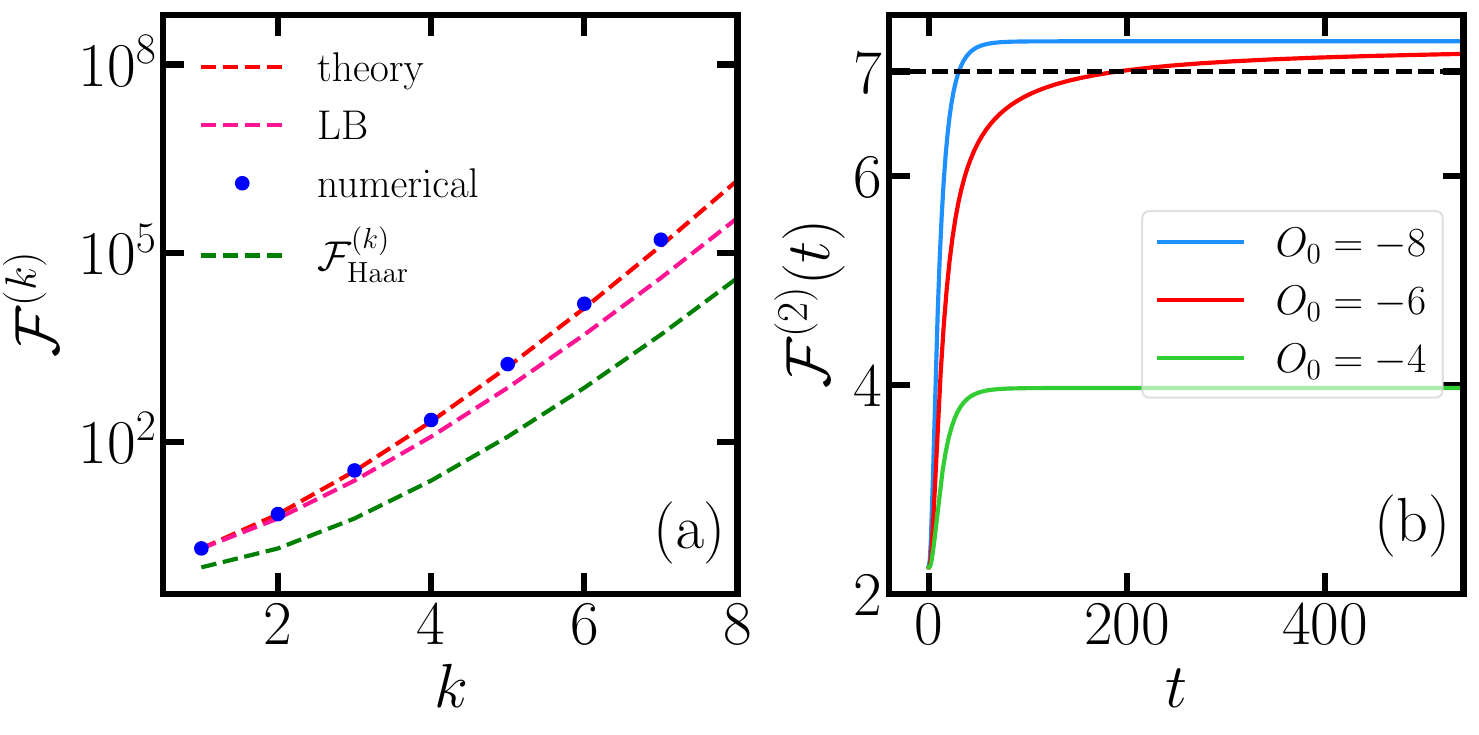}
    \caption{
    (a) Frame potential $F^{(k)}$ for the restricted Haar ensemble with dimension $2^n = 4$. Red dashed line is the exact theory prediction in Eq.~\eqref{eq:F_k} and magenta dashed line is its lower bound $(k+1)!$.
    (b) The evolution of $2$-order frame potential $\calF^{(2)}$ for ensemble of RPA circuit unitary with different target $O_0$. The observable is $O_{\rm XXZ}$ with $J=2$, with $O_{\rm min}=-6$. Black dashed line is exact value of $\calF^{(k)}$ in Eq.~\ref{eq:F_k_main}.
    }
    \label{fig:framepot}
\end{figure}

\subsection{Details of formula}

For simplicity, we assume $V$ is a Haar random unitary in Eq.~\eqref{eq:toyU_app}. The $k$th frame potential of the unitary ensemble is thus
\begin{align}
    \calF^{(k)}_{\rm RH} &= \frac{1}{|\calE|^2} \sum_{U,U^\prime \in \calE} |\tr(U^\dagger U^\prime)|^{2k}\\
    &= \frac{1}{|\calE|^2} \sum_{U,U^\prime \in \calE} |1 + \tr(V^\dagger V^\prime)|^{2k}\\
    &= \int_{\rm Haar} dV dV^\prime \left(1 + \tr(V^\dagger V^\prime) + \tr(V^\dagger V^\prime)^* + |\tr(V^\dagger V^\prime)|^2\right)^k.
\end{align}
{where the second line comes from the definition in Eq.~\eqref{eq:toyU_app}.}

For simplicity, we denote $\tr(V^\dagger V^\prime) \equiv z$ and then have
\begin{align}
     \calF^{(k)}_{\rm RH} &=\int_{\rm Haar} dV dV^\prime \left(1 + z + z^* + |z|^2\right)^k\\
     &= \sum_{\substack{k_1,k_2,k_3=0\\ k_1+k_2+k_3 \le k}}^k \binom{k}{k_1, k_2, k_3} \int_{\rm Haar} dV dV^\prime  |z|^{2k_1} z^{* k_2} z^{k_3}\\
     &= \sum_{\substack{k_1,k_2=0\\ k_1+2k_2 \le k}}^k \frac{k!}{k_1! (k_2!)^2 (k-k_1-2k_2)!} \int_{\rm Haar} dV dV^\prime  |z|^{2k_1} |z|^{2k_2} \label{eq:Frh_eq1}\\
     &= \sum_{\substack{k_1,k_2=0\\ k_1+2k_2 \le k}}^k \frac{k!}{k_1! (k_2!)^2 (k-k_1-2k_2)!} \calF_{\rm Haar}^{(k_1 + k_2)} \label{eq:F_k}\\
     &\ge \calF^{(k)}_{\rm Haar} + k^2 \calF^{(k-1)}_{\rm Haar} = (k+1)\calF_{\rm Haar}^{(k)} = \calF_{\rm Haar}^{(k+1)}.
\end{align}
{where in Eq.~\eqref{eq:Frh_eq1} we only keep $k_2 = k_3$ from the above line to keep frame potential to be real.}

\section{Details on dynamics with linear loss function}
\label{app:linear_loss_detail}

{In this section, we provide details on derive analytical theories for dynamics with linear loss functions. Following the same formalism, we can still parameterize the quantum neural network via a parameterized quantum circuit, for instance, random Haar ansatz we considered in the main text. The loss function to be considered is 
\begin{align}
    \calL(\bm \theta) = \braket{\psi_0|U^\dagger(\bm \theta) O U(\bm \theta)|\psi_0},
\end{align}
where $\psi_0$ is a trivial input state and $U(\bm \theta)$ is the parameterized quantum circuit as usual. With sufficiently deep circuit, the loss function will be minimized to the ground state energy $O_{\rm min}$ of the observable, and thus we describe its converge via the residual error $\varepsilon(t) = \braket{O} - O_{\rm min}$.}

{Via gradient descent, each parameter is shifted as
\begin{align}
    \delta \theta_\ell(t) = -\eta \frac{\partial \calL (\bm \theta)}{\partial \theta_\ell} = -\eta \frac{\partial \varepsilon (\bm \theta) }{\partial \theta_\ell}.
\end{align}
Thus under small learning rate $\eta \ll 1$, the total error is updated as
\begin{align}
    \delta \varepsilon(t) &\simeq \sum_{\ell}\frac{\partial\epsilon (\bm \theta)}{\partial\theta_\ell} \delta \theta_\ell + \frac{1}{2}\sum_{\ell_1, \ell_2}\frac{\partial^2\epsilon (\bm \theta)}{
    \partial\theta_{\ell_1}\partial\theta_{\ell_2}
    }\delta \theta_{\ell_1}\delta \theta_{\ell_2}\\
    &= -\eta K(\bm \theta)|_{\bm \theta = \bm \theta (t)} + \frac{1}{2}\eta^2 \mu(\bm \theta)|_{\bm \theta = \bm \theta (t)},
\end{align}
where the QNTK $K(\bm \theta)$ and dQNTK $\mu(\bm \theta)$ is defined the same as in the main text. The dynamical equation in the first order of $\eta$ is
\begin{align}
    \delta \varepsilon(t) = -\eta K(t) + \mathcal{O}(\eta^2).
    \label{eq:epseq_linear}
\end{align}
}

{
We can also derive the dynamical equation of $K(t)$ as following. Recall from App.~\ref{app:diff_eq},
\begin{align}
    \delta K(t) = \sum_\ell 2\delta\left(\left.\frac{\partial\varepsilon (\bm \theta)}{\partial \theta_\ell}\right\rvert_{\bm \theta=\bm \theta(t)}\right)\left.\frac{\partial\varepsilon (\bm \theta)}{\partial \theta_\ell}\right\rvert_{\bm \theta=\bm \theta(t)} + \delta\left(\left.\frac{\partial\varepsilon (\bm \theta)}{\partial \theta_\ell}\right\rvert_{\bm \theta=\bm \theta(t)}\right) \delta\left(\left.\frac{\partial\varepsilon (\bm \theta)}{\partial \theta_\ell}\right\rvert_{\bm \theta=\bm \theta(t)}\right),
\end{align}
where the second term is second order of $\eta$, and thus can be omitted in our calculation. The first term becomes
\begin{align}
\delta \left( \left.\frac{\partial \varepsilon (\bm \theta)}{\partial {{\theta }_{\ell }}}\right\rvert_{\bm \theta = \bm \theta(t)} \right) =\underset{{{\ell }_{1}}}{\mathop \sum }\, \left.\frac{{{\partial }^{2}}\varepsilon (\bm \theta)}{\partial {{\theta }_{\ell_1 }}\partial {{\theta }_{{{\ell }}}}}\delta {{\theta }_{{{\ell }_{1}}}} \right\rvert_{\bm \theta = \bm \theta(t)} +\mathcal{O}({{\eta }^{2}})
= -\eta \underset{{{\ell }_{1}}}{\mathop \sum }\, \left.\frac{{{\partial }^{2}}\varepsilon (\bm \theta) }{\partial {{\theta }_{\ell_{1} }}\partial {{\theta }_{{{\ell }}}}}\frac{\partial \varepsilon (\bm \theta)}{\partial {{\theta }_{{{\ell }_{1}}}}} \right\rvert_{\bm \theta = \bm \theta(t)} +\mathcal{O}({{\eta }^{2}}).
\end{align}
So we have
\begin{align}
\sum_{\ell} \delta\left(\left. \frac{\partial\varepsilon(\bm \theta)}{\partial \theta_\ell} \right\rvert_{\bm \theta = \bm \theta(t)}\right) \left.\frac{\partial \varepsilon (\bm \theta)}{\partial \theta_\ell}\right\rvert_{\bm \theta = \bm \theta(t)} =-\eta \underset{\ell ,{{\ell }_{1}}}{\mathop \sum }\, \left.\frac{{{\partial }^{2}}\varepsilon (\bm \theta) }{\partial {{\theta }_{\ell }}\partial {{\theta }_{{{\ell }_{1}}}}}\frac{\partial \varepsilon (\bm \theta) }{\partial {{\theta }_{{{\ell }_{1}}}}}\frac{\partial \varepsilon (\bm \theta) }{\partial {{\theta }_{\ell }}} \right\rvert_{\bm \theta = \bm \theta(t)} +\mathcal{O}({{\eta }^{2}})=-\eta \mu(\bm \theta)|_{\bm \theta = \bm \theta(t)}  +\mathcal{O}({{\eta }^{2}}),
\end{align}
which leads to the gradient descent dynamical equation of $K(t)$ as
\begin{align}
    \delta K(t) = -2\eta \mu(t) + \mathcal{O}(\eta^2).
    \label{eq:Keq_linear}
\end{align}
}

{
Recall that the relative dQNTK is defined as $\lambda(t) = \mu(t)/K(t)$, and again we assume it converges to a constant $\lambda$ in late time (see Fig.~\ref{fig:linearloss_detail}(c)), the above dynamical equations in Eqs.~\eqref{eq:epseq_linear},~\eqref{eq:Keq_linear} can be reduced to 
\begin{align}
\left\{ \begin{array}{ll}
        \partial_t K(t) &= -2\eta  \lambda K(t)\\
        \partial_t \varepsilon(t) & = -\eta K(t)
\end{array} \right. 
\end{align}
Notice that the dynamics of $K(t)$ is self-consistent and $\varepsilon(t)$ is fully controlled by $K(t)$, we can then directly solve the dynamics of $\varepsilon(t)$ and $K(t)$ as
\begin{align}
2\lambda \varepsilon(t) = K(t) = Ae^{-2\eta \lambda t},
\label{eq:sol_linear}
\end{align}
where is exactly the solution we propose in Eq.~\eqref{eq:sol_linear_main} in the main text. Here $A$ is a free parameter to be fitted. Both residual error and QNTK exponentially decays to zero in the late time at a fixed rate, and the dynamical transition does not persist.
} 

\begin{figure}
    \centering
    \includegraphics[width=0.8\textwidth]{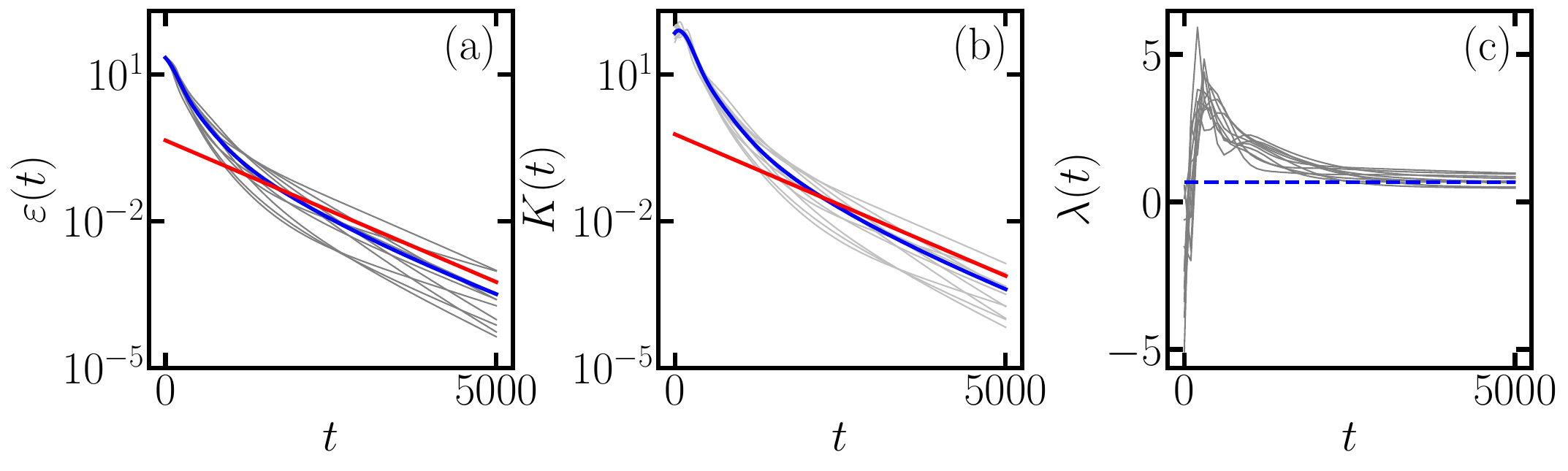}
    \caption{
    {
    Dynamics in QNN in the example of XXZ model with linear loss function. Dynamics of residual error (a) $\varepsilon(t)$ and (b) QNTK $K(t)$ are plotted. Grey lines represent simulation with random initializations and blue lines represent the ensemble average. Red dashed lines show the theory model from Eq.~\eqref{eq:sol_linear}. In (c), we show the convergence of dQNTK $\lambda(t)$. Blue dashed line represent the average as $\overline{\lambda} = \overline{\mu}/\overline{K}$. Here the random Pauli ansatz (RPA) consists of $L=192$ parameters on $n=6$ qubits, and the observable is XXZ model with $J=2$.
    }
    }
    \label{fig:linearloss_detail}
\end{figure}

\section{More details on numerical results}
\label{app:numeric_detail}

{
We first present the training dynamics of another well-established controllable QNN ansatz, hardware efficient ansatz (HEA)~\cite{kandala2017hardware}, and optimize the corresponding quadratic loss function. Suppose the HEA consists of $D$ layers on $n$ qubits, in each layer RY and RZ single-qubit gates are applied on every qubit, where each of them includes a trainable parameters, and single-qubit gates are followed by CNOT gate arranged in brickwall style on nearest neighbors. Therefore, for a $D$-layer HEA, the total number of trainable parameters is $L = 2nD$. 
In the optimization problem, we choose the observable to be the Hamiltonian of transverse-field Ising model (TFIM) as
\begin{align}
    O_{\rm TFIM} &= -\sum_{i} \sigma^z_i \sigma^z_{i+1} - h\sum_i \sigma^x_i,
    \label{eq:tfim}
\end{align}
where $h$ describes the strength of transverse field. In Fig.~\ref{fig:hea_tfim}, we show the training dynamics of error and kernel with $O_0 \gtreqless O_{\rm min}$, where the two branches of dynamics and the critical point can also be identified. We still see alignment between numerical simulation (blue) and our theories (red), indicating the applicable of our theory to characterize the dynamics for general deep controllable QNNs.
}

\begin{figure}
    \centering
    \includegraphics[width=0.8\textwidth]{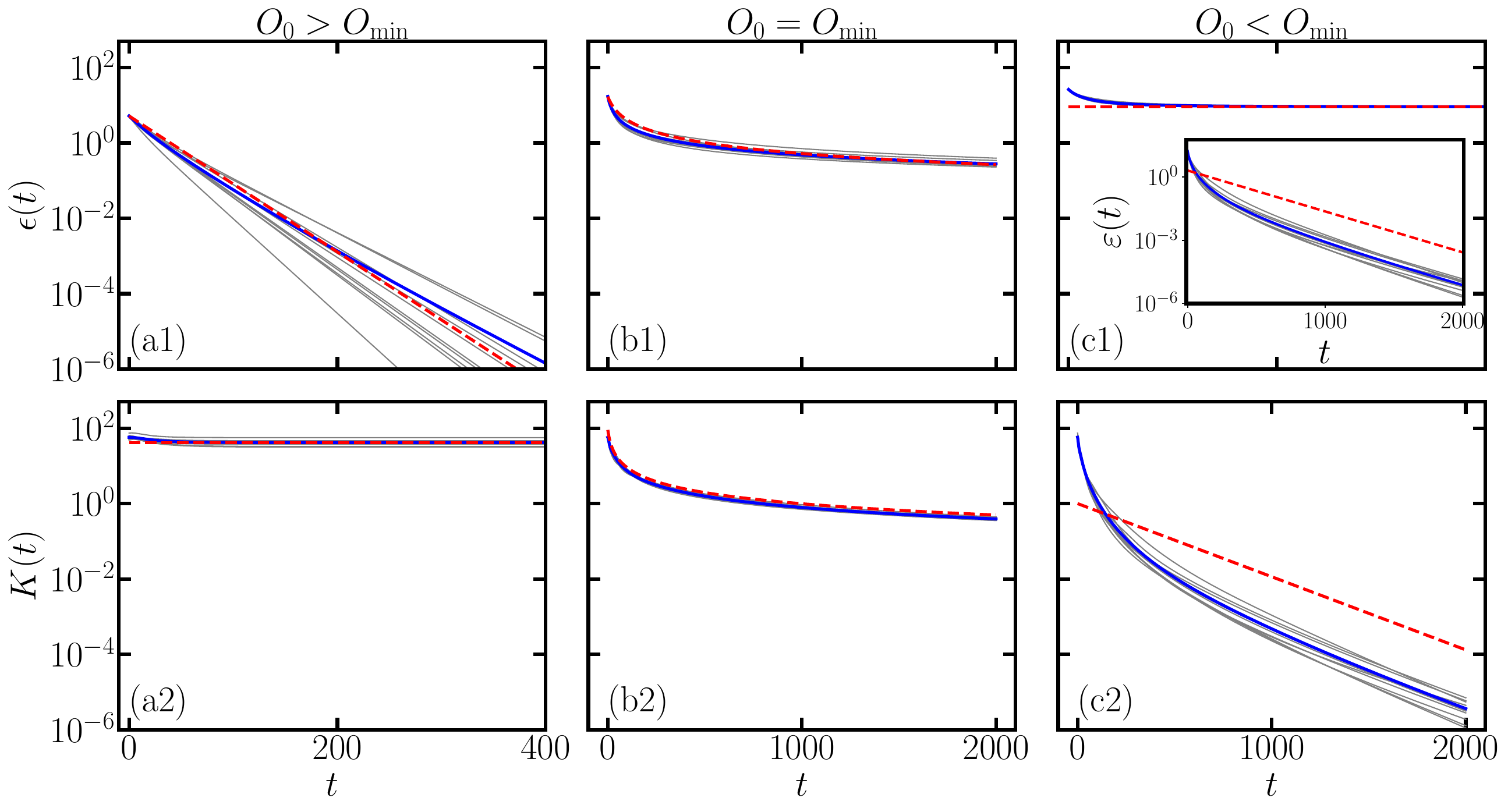}
    \caption{{Dynamics in QNN in the example of TFIM model.  
    The top and bottom panel shows the dynamics of total error ${\epsilon}(t)$ and QNTK ${K}(t)$ with respect to the three cases $O_0 \gtreqless O_{\rm min}$. Blue solid curves represent numerical ensemble average result. Red dashed curves in panels represents theoretical predictions on the dynamics of total error in Eq.~\eqref{eq:eps_t_frozen_exp},~\eqref{eq:K_poly_asym_main},~\eqref{eq:K_nonfrozen_exp_main} (from left to right). Grey solid lines show the dynamics for each random sample. The inset in (c1) shows the exponentially decay of residual error $\varepsilon(t)$. Here hardware efficient ansatz (HEA) consists of $D=48$ layers, equivalently $L=768$ variational parameters, on $n=8$ qubits, and the parameter in TFIM model is $h=2$.}}
    \label{fig:hea_tfim}
\end{figure}

{
Next, we provide numerical evidence to support the assumption in the main text that at late time, relative dQNTK $\lambda(t)$ converges to a constant in a relatively shallow QNN. In Fig.~\ref{fig:lambda_shallow}, we show the dynamics of $\lambda(t)$ of random samples in Fig.~\ref{fig:main_results_finite} in the main text, and see that $\lambda(t)$ converges to a constant no matter $O_{\rm min}\gtreqless O_{\rm min}(L)$ though the fluctuation among different samples is comparably large in shallow QNNs.
}

\begin{figure}
    \centering
    \includegraphics[width=0.8\textwidth]{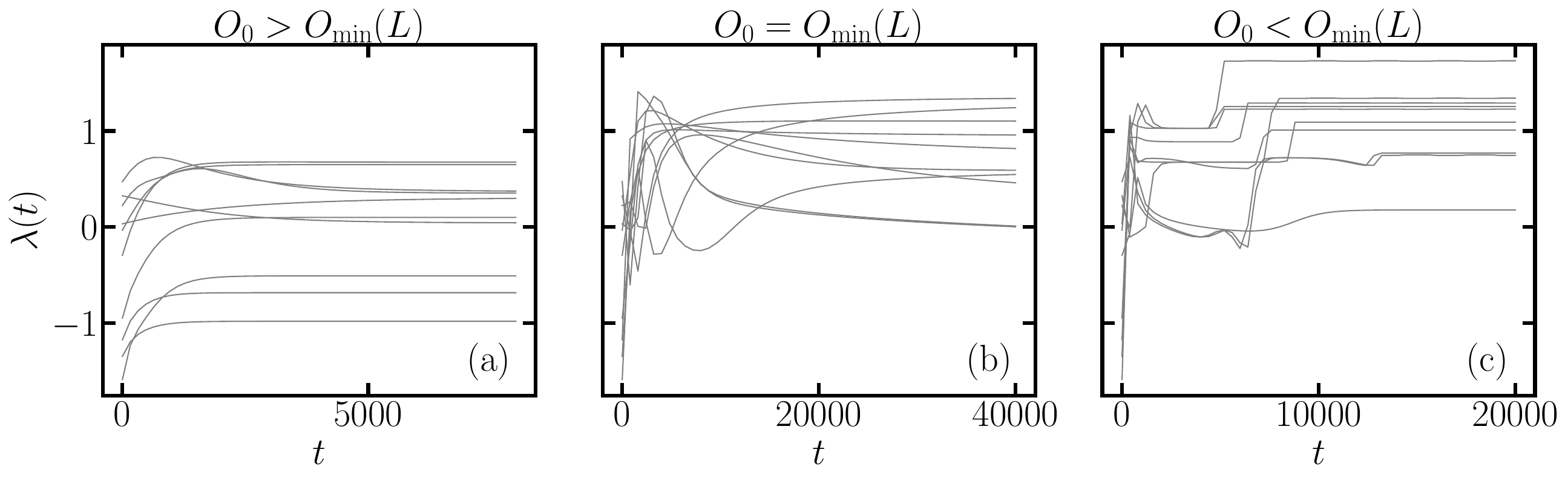}
    \caption{
    {
    Dynamics of relative dQNTK $\lambda(t)$ in shallow QNN in the example of XXZ model with $O_0 \gtreqless O_{\rm min}(L)$ (from left to right). The critical point $O_{\rm min}(L)$ for shallow QNN depends on $L$. Here random Pulai ansatz (RPA) consists of $L=6$ variational parameters ($D=L$ for RPA) on $n=6$ qubits, and the parameter in XXZ model is $J=2$.}
    }
    \label{fig:lambda_shallow}
\end{figure}

{
In the following, we show more numerical results associated with the restricted Haar ensemble.  Recall the definition of $\lambda_\infty \equiv \mu_\infty/K_\infty$, we can have the average of $\lambda_\infty$ defined via the following two as 
$
\mathbb{E}[\lambda_\infty] \equiv \overline{\mu_\infty/K_\infty}
$, and 
$
\overline{\lambda_\infty} \equiv \overline{\mu_\infty}/\overline{K_\infty}
$.
In Fig.~\ref{fig:mean_compare}(a1)(b1), we see that with increasing of $L$, the discrepancy between $\mathbb{E}[\lambda_\infty]$ and $\overline{\lambda_\infty}$ vanishes for different kind of observables like state projector and XXZ model, which suggests it is free to exchange the definition of average as we have done in definition of relative sample fluctuation of $\lambda_\infty$ in the main text. Similarly, for $\zeta_\infty \equiv \epsilon_\infty \mu_\infty/K_\infty^2$, we see similar results for $\mathbb{E}[\zeta_\infty] \equiv \overline{\epsilon_\infty \mu_\infty/K_\infty^2}$ and $\overline{\zeta_\infty}\equiv \overline{\epsilon_\infty \mu_\infty} / \overline{K_\infty}^2$.
In Fig.~\ref{fig:flucs_xxz}, we show the relative sample fluctuation of $\lambda_\infty$ and $\zeta_\infty$ versus parameters $L$ (thus depth $D$ in RPA) with observable to be the XXZ model. The findings discussed in the main text for state projector observable also hold here.}

\begin{figure}[t]
    \centering
    \includegraphics[width=0.7\textwidth]{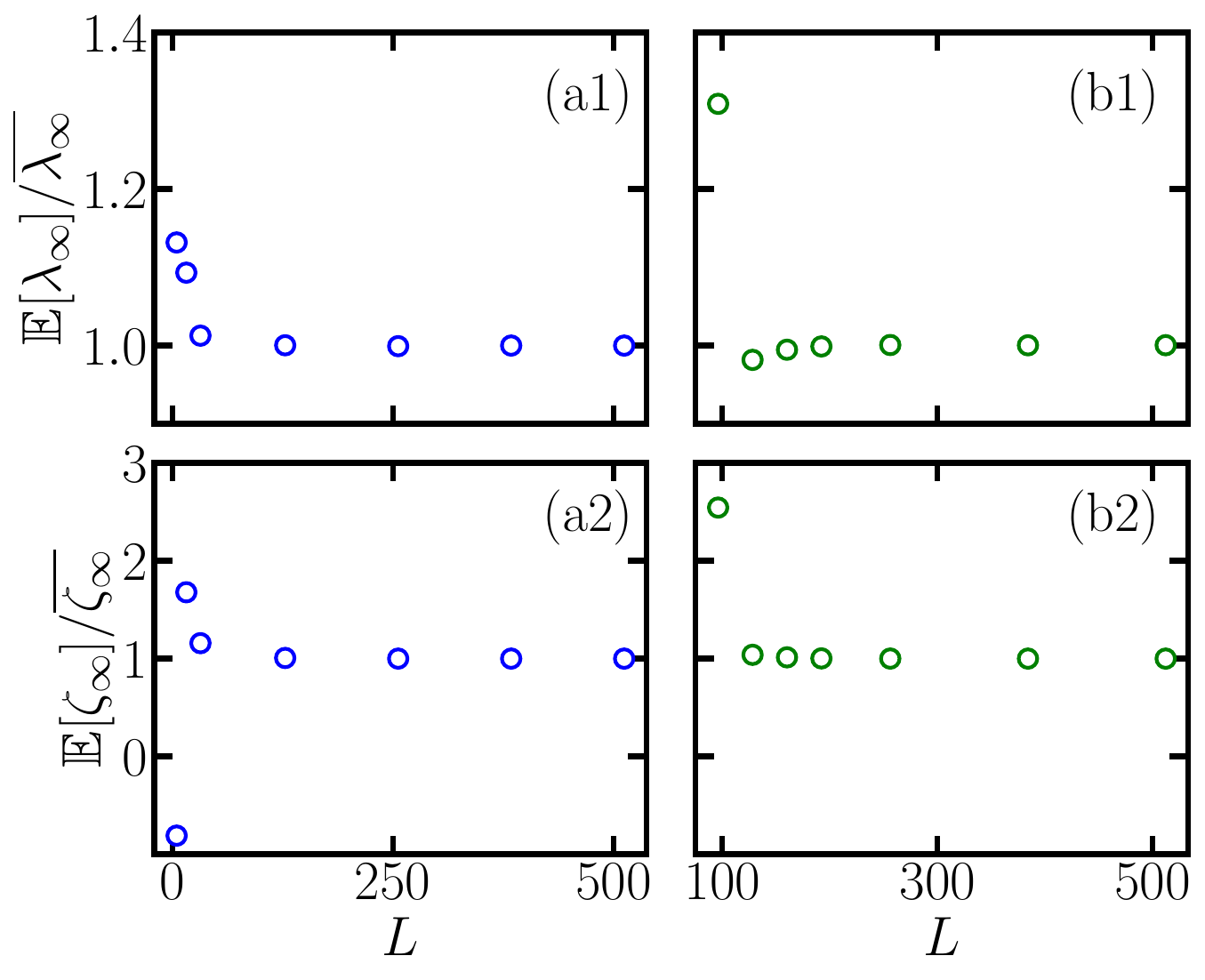}
    \caption{{
    Comparison between $\mathbb{E}[\lambda_\infty]$ and $\overline{\lambda_\infty}$ (top) as well as $\zeta_\infty$ (bottom) versus $L$. Blue dots left panel represent results for state projector observable in $n=5$ qubit system with $O_0 = 1$. Green dots in right panel show results for XXZ model with $J=2$ in $n=6$ qubit system with $O_0 = O_{\rm min}$.}
    }
    \label{fig:mean_compare}
\end{figure}

\begin{figure}[t]
    \centering
    \includegraphics[width=0.7\textwidth]{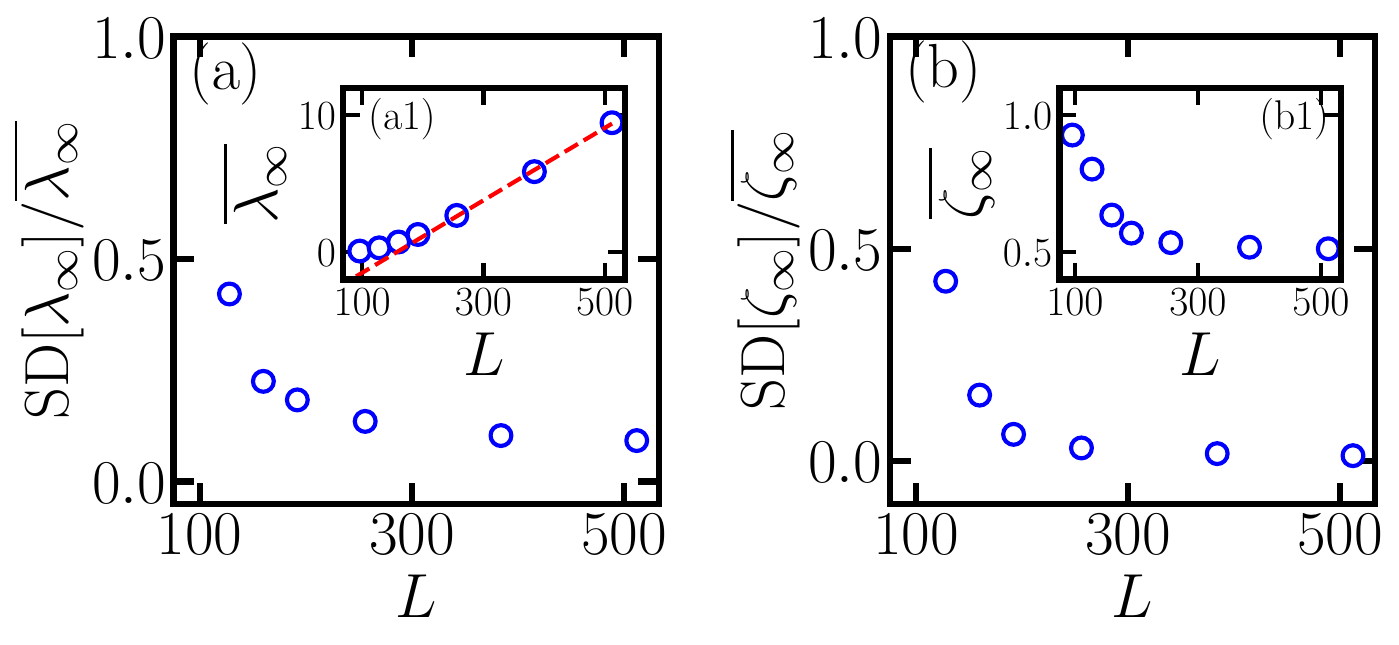}
    \caption{{
    Late-time scaling of relative dQNTK $\lambda_\infty$ and dynamical index $\zeta_\infty$ on number of parameters $D$. Relative sample fluctuation of (a) $\lambda_\infty$ and (b) $\zeta_\infty$ versus $L$ at late time. Inset (a1) and (b1) shows the average $\overline{\lambda_\infty} \propto L$ and $\overline{\zeta_\infty} \to 1/2$ separately. Red dashed line in (a1) is the linear fitting of $\overline{\lambda_\infty}$ over $L$. Here the RPA is applied on $n=6$ qubits with different $L$ parameters (equivalently $D$ layers). The observable and target is XXZ model with $J=2$ and the target is $O_0 = O_{\rm min}=-22$.}
    }
    \label{fig:flucs_xxz}
\end{figure}

{
At the end, we provide more details on the training dynamics with linear loss function. In Fig.~\ref{fig:linearloss_detail}(a)(b), we see good agreement between our theory (red dashed lines) in Eq.~\eqref{eq:sol_linear} and numerical simulation results (blue lines). Note that the exponential decay of total error is also found in a recent work~\cite{you2022convergence} with similar interest via the Riemann gradient flow formalism.
}

{
All our numerical simulation is performed with \texttt{TensorCircuit}~\cite{zhang2023tensorcircuit} on \texttt{Jax} backend.
}
\begin{figure}[b]
    \centering
    \includegraphics[width=0.3\textwidth]{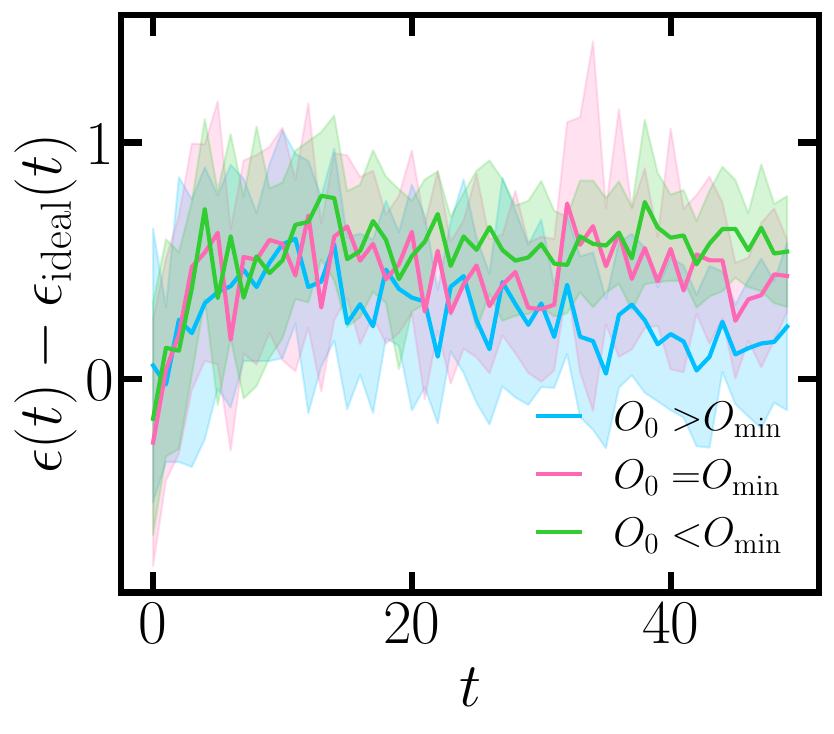}
    \caption{
    {Difference between experimental result $\epsilon(t)$ and ideal noiseless simulation result $\epsilon_{\rm ideal}(t)$ for different $O_0$. Shaded areas of different colors represent the fluctuation with corresponding $O_0$. Here the hardware efficient ansatz (HEA) consists of $D=4$ layers on $n=2$ qubits, and the observable is XXZ model with $J=2$.}
        \label{fig:noise_deviation}
        }
    \centering
    \includegraphics[width=0.8\textwidth]{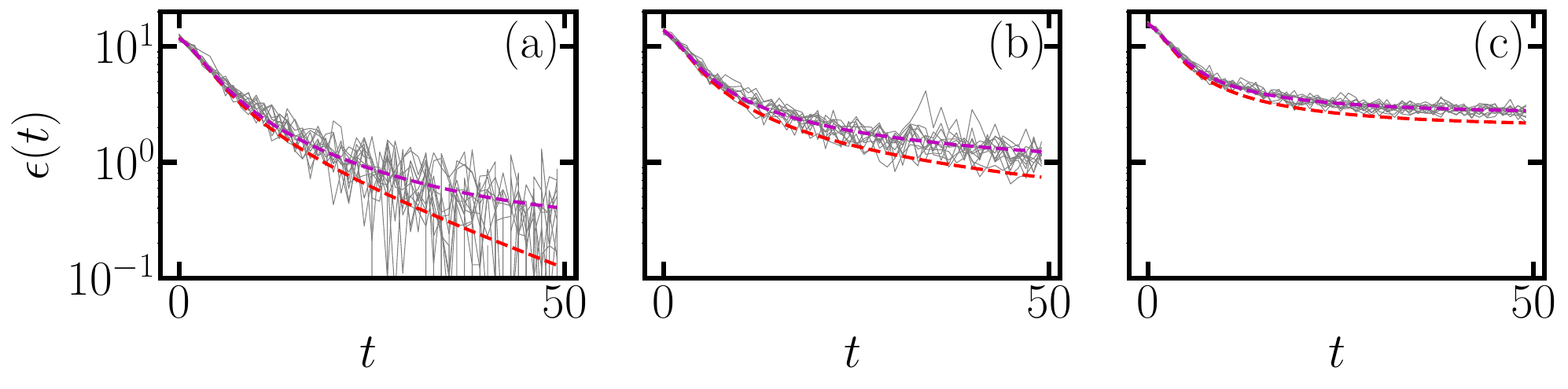}
    \caption{
    {Dynamical trajectories (grey) of each measurement trial of experiment on \texttt{IBM Kolkata} in Fig.~\ref{fig:experiment}. From (a)-(c) we present $O_0 \gtreqless O_{\rm min}$. Red dashed lines represent the ideal simulation for reference. Magenta dashed lines represent the noisy model prediction in Eq.~\eqref{eq:noise_model} with $p=0.028, 0.044, 0.051$ (from left to right).
    }\label{fig:experiment_detail}
    }
    \includegraphics[width=0.3\textwidth]{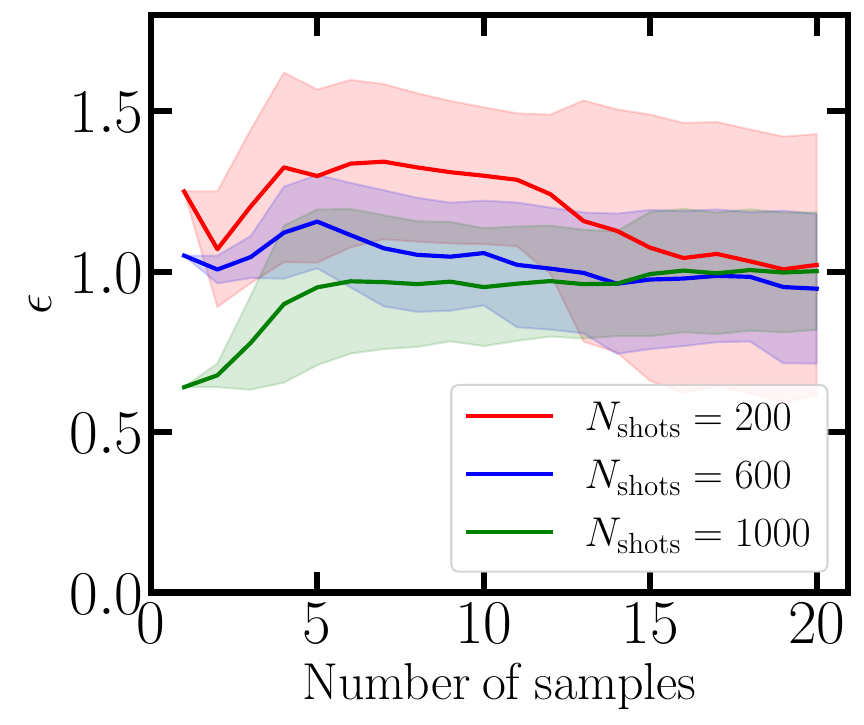}
    \caption{Experiment result of the error $\epsilon(t)$ for different number of shots in the measurement repetitions for estimating the observable in an individual sample. Multiple samples of training are performed to show the sample fluctuation. As the number of samples increase, the average $\epsilon$ becomes almost independent of the number of the shots. The shaded region indicate the sample fluctuation in standard deviation. $O_0=10$ and the results are at late time of $t= 40$.
    \label{fig:experiment_shots}
    }
\end{figure}

\section{Details of experiments}
\label{app:hea}

{
In this section, we provide our noisy model to characterize the dynamics on noisy devices. Given a noiseless idea prediction $\epsilon_{\rm ideal}(t) = \tr(U \rho_0 U^\dagger O) - O_0$, we consider the depolarizing noise model $\calN_p (\rho) = (1-p) \rho + p \bI/d$ where $\bI$ is the identity matrix of dimension $d$. The depolarizing noisy total error $\epsilon_{\rm dp}(t)$ then becomes
\begin{align}
    \epsilon_{\rm dp}(t) &= \tr(\calN_p(U\rho_0 U^\dagger) O) - O_0\\
    &= (1-p)\left[\tr(U \rho_0 U^\dagger O) - O_0\right] + p\left[\tr(O)/d - O_0\right]\\
    &= (1-p)\epsilon_{\rm ideal}(t) - p O_0,
    \label{eq:noise_model}
\end{align}
where in the last line we assume $O$ is a traceless observable. Therefore, compared to ideal case, the total error under depolarizing noise model is simply shifted by a constant depending on target. The residual error can also be studied similarly with $\varepsilon_{\rm dp}(t) = \epsilon_{\rm dp}(t) - R_{\rm dp}$, where $R_{\rm dp} = \lim_{t \to \infty} \epsilon_{\rm dp}(t)$ could be different from ideal case due to noise. For $O_0 \ge O_{\rm min}$, we have $R_{\rm dp} = -pO_0$ while for $O_0 < O_{\rm min}$, we have $R_{\rm dp} = (1-p)O_{\rm min} - O_0$.
In Fig.~\ref{fig:noise_deviation}, we provide intuitions via the difference $\epsilon(t) - \epsilon_{\rm ideal}(t)$, and for different $O_0$, we see that the differences quickly approach to a constant. We can further estimate the depolarizing probability $p$ via least error as follows. For a given $O_0$, we have multiple data points from ideal simulation $\{\epsilon_{\rm ideal}(t_i)\}_{i=1}^M$, and true experimental data $\{\epsilon(t_i)\}_{i=1}^M$. From Eq.~\eqref{eq:noise_model}, the SSR (sum squared error) is
\begin{align}
    {\rm SSR} &= \sum_{i=1}^M \left[\epsilon(t_i) - (1-p)\epsilon_{\rm ideal}(t_i) + p O_0\right]^2,
\end{align}
and the minimum with respect to $p$ is
\begin{align}
    0&=\frac{\partial \rm SSR}{\partial p} = 2\sum_i \left[\epsilon(t_i) - (1-p)\epsilon_{\rm ideal}(t_i) + p O_0\right]\left[\epsilon_{\rm ideal}(t_i) + O_0\right]\\
    p &= \frac{\sum_i \left[\epsilon_{\rm ideal}(t_i) - \epsilon(t_i)\right]\left[\epsilon_{\rm ideal}(t_i) + O_0\right]}{\sum_i \left[\epsilon_{\rm ideal}(t_i) + O_0\right]^2}.
\end{align}
For the three different $O_0\gtreqless O_{\rm min}$, we find $p = 0.028\pm 0.007, 0.044 \pm 0.011, 0.051\pm 0.005$ which are close and the fluctuation could origin from other types of error other than depolarizing noise and the drift of device.
We then verify our noisy model via different experimental trajectories in Fig.~\ref{fig:experiment_detail}
}. 
In contrast to the ideal simulation (red dashed line), the noisy model (magenta) shows a better agreement.
We see that each trial also follows the scaling though larger fluctuation exists due to the {other} circuit and measurement noise.

To rule out other noise sources causing the deviation, we also consider changing the number of shots of measurement in the estimation of the observable. As we see in Fig.~\ref{fig:experiment_shots}, the average error all converge to around the same value, regardless of the number of shots in the estimation. For $20$ rounds of experiment, we find the error as $\epsilon = 1.02\pm 0.41, 0.95\pm 0.23, 1.00\pm 0.18$ for number of measurement shots $N_{\rm shots}=200, 600, 1000$. The difference is much smaller than the sample fluctuation between multiple runs of the experiments.


\section{Results with Haar random ensemble}
\label{app:haar_result}

In this section, we present results evaluated from Haar random unitary ensemble, which provides characterizations of QNN dynamics at early time. The rest of the contents are regarding ensemble averaging over Haar (App.~\ref{app:haar_result}) or restricted Haar ensemble (App.~\ref{app:rh_result}), where we have utilized symbolic tools {\texttt{RTNI}} from Ref.~\cite{fukuda2019rtni}. 

\subsection{Average QNTK under Haar random ensemble}
\label{app:K0}

With random initialization, the circuit forms a Haar random ensemble, and the ensemble average of QNTK is $\overline{K_0} = \sum_\ell \mathbb{E}_{{\calU_{\rm Haar}}}\left[\left(\frac{\partial\epsilon}{\partial \theta_{\ell}}\right)^2\right]$, where the ensemble average inside the summation is 
\begin{align}
    \mathbb{E}_{{\calU_{\rm Haar}}} \left[\left(\frac{\partial\epsilon}{\partial \theta_\ell}\right)^2\right] &= -\frac{1}{4}\int_{{\calU_{\rm Haar}}} dU_{\ell^-}dU_{\ell^+} \braket{\psi_0|U_{\ell^-}^\dagger\left[X_\ell, O_{\ell^+}\right]U_{\ell^-}|\psi_0}\braket{\psi_0|U^\dagger_{\ell^-}\left[X_\ell, O_{\ell^+}\right]U_{\ell^-}|\psi_0}\\
    &= -\frac{1}{4}\int_{{\calU_{\rm Haar}}} dU_{\ell^-}dU_{\ell^+} \tr\left(\rho_0 U_{\ell^-}^\dagger\left[X_\ell, O_{\ell^+}\right]U_{\ell^-}\rho_0 U_{\ell^-}^\dagger\left[X_\ell, O_{\ell^+}\right]U_{\ell^-}\right)\\
    &= -\int_{{\calU_{\rm Haar}}} dU_{\ell^+}\frac{\tr\left(\left[X_\ell, O_{\ell^+}\right]^2\right)}{4(d^2+d)},
\end{align}
where $\rho_0 = \ketbra{\psi_0}{\psi_0}$. Using the trace identity $\tr([A,B]^2) = 2\tr(ABAB)-2\tr(ABBA)$, we then have
\begin{align}
    \mathbb{E}_{{\calU_{\rm Haar}}} \left[\left(\frac{\partial\epsilon}{\partial \theta_{\ell}}\right)^2\right] &= -\int_{{\calU_{\rm Haar}}} dU_{\ell^+}\frac{\tr\left(\left[X_\ell, O_{\ell^+}\right]^2\right)}{4(d^2+d)} = -\frac{2}{4(d^2+d)}\left[\int_{{\calU_{\rm Haar}}} dU_{\ell^+} \tr\left(X_\ell O_{\ell^+}X_\ell O_{\ell^+}\right) - \tr(O^2)\right]\\
    &= \frac{d\tr(O^2)-\tr(O)^2}{2 (d-1) (d+1)^2}.
\end{align}
Note that here we assume both $U_{\ell^-}$ and $U_{\ell^+}$ form a Haar random ensemble ($2$-design) separately.
Therefore, the ensemble average of QNTK is
\begin{align}
    \overline{K_0} = \sum_\ell \mathbb{E}_{{\calU_{\rm Haar}}}\left[\left(\frac{\partial\epsilon}{\partial \theta_\ell}\right)^2\right] = L\frac{d\tr(O^2)-\tr(O)^2}{2 (d-1) (d+1)^2}\simeq \frac{L}{2d^3}\left(d\tr(O^2) - \tr(O)^2\right),
    \label{eq:K0}
\end{align}
where we approximate it by $d\gg 1$ in the last equation {to simplify expression}. Specifically, for the traceless operator we considered in the main text, we have
\begin{align}
    \overline{K_0} \simeq \frac{L}{2d^2} \tr(O^2).
    \label{eq:K0_traceless}
\end{align}


\subsection{Average relative dQNTK under Haar random ensemble}

We define the average $\overline{\lambda}$ as
\begin{align}
    \overline{\lambda} = \overline{\mu}/\overline{K},
\end{align}
where dQNTK $\mu$ is defined as
\begin{align}
    \mu = \sum_{\ell_1,\ell_2} \frac{\partial^2 \epsilon}{\partial \theta_{\ell_1}\partial\theta_{\ell_2}}\frac{\partial \epsilon}{\partial \theta_{\ell_1}} \frac{\partial \epsilon}{\partial \theta_{\ell_2}}.
\end{align}
In the following, we calculate the Haar ensemble average of dQNTK as
\begin{align}
    \overline{\mu_0} &= 2\sum_{\ell_1<\ell_2}\mathbb{E}_{{\calU_{\rm Haar}}}\left[\frac{\partial^2 \epsilon}{\partial \theta_{\ell_1}\partial\theta_{\ell_2}}\frac{\partial \epsilon}{\partial \theta_{\ell_1}} \frac{\partial \epsilon}{\partial \theta_{\ell_2}}\right] + \sum_\ell \mathbb{E}_{{\calU_{\rm Haar}}}\left[\frac{\partial^2 \epsilon}{\partial\theta_\ell^2}\left(\frac{\partial \epsilon}{\partial \theta_\ell}\right)^2 \right]\\
    &= L(L-1)\mathbb{E}_{{\calU_{\rm Haar}}}\left[\frac{\partial^2 \epsilon}{\partial \theta_{\ell_1}\partial\theta_{\ell_2}} \frac{\partial \epsilon}{\partial \theta_{\ell_1}} \frac{\partial \epsilon}{\partial \theta_{\ell_2}}\right] + L\mathbb{E}_{{\calU_{\rm Haar}}}\left[\frac{\partial^2 \epsilon}{\partial\theta_\ell^2}\left(\frac{\partial \epsilon}{\partial \theta_\ell}\right)^2 \right].
\end{align}

\paragraph{Calculation of $\frac{\partial^2 \epsilon}{\partial\theta_\ell^2}\left(\frac{\partial \epsilon}{\partial \theta_\ell}\right)^2$  with Haar random ensemble\\}

We first evaluate the summation of $\frac{\partial^2 \epsilon}{\partial\theta_\ell^2}\left(\frac{\partial \epsilon}{\partial \theta_\ell}\right)^2$
\begin{align}
    \mathbb{E}_{{\calU_{\rm Haar}}}\left[\frac{\partial^2 \epsilon}{\partial\theta_\ell^2}\left(\frac{\partial \epsilon}{\partial \theta_\ell}\right)^2 \right] &= \frac{1}{16}\int_{{\calU_{\rm Haar}}} dU_{\ell^-}dU_{\ell^+}\braket{\psi_0|U_{\ell^-}^\dagger [X_\ell, [X_\ell, O_{\ell^+}]]U_{\ell^-}|\psi_0}\braket{\psi_0|U_{\ell^-}^\dagger\left[X_\ell, O_{\ell^+}\right]U_{\ell^-}|\psi_0}^2\\
    &= \frac{1}{16}\int_{{\calU_{\rm Haar}}} dU_{\ell^-}dU_{\ell^+} \tr\left(\rho_0 U_{\ell^-}^\dagger [X_\ell, [X_\ell, O_{\ell^+}]] U_{\ell^-}\rho_0 U_{\ell^-}^\dagger\left[X_\ell, O_{\ell^+}\right]U_{\ell^-}\rho_0 U_{\ell^-}^\dagger\left[X_\ell, O_{\ell^+}\right]U_{\ell^-}\right)\\
    &= \int_{{\calU_{\rm Haar}}} dU_{\ell^+} \frac{\tr\left(\left[X_\ell, [X_\ell, O_{\ell^+}]\right]\left[X_\ell, O_{\ell^+}\right]^2\right)}{8d(2+3d+d^2)}\\
    &= 0,
    \label{eq:gllglgl}
\end{align}
where the last equation is obtained from the trace cyclic identity.

\paragraph{Calculation of $\frac{\partial^2 \epsilon}{\partial \theta_{\ell_1}\partial\theta_{\ell_2}} \frac{\partial \epsilon}{\partial \theta_{\ell_1}} \frac{\partial \epsilon}{\partial \theta_{\ell_2}}$ with Haar random ensemble\\}

We next consider the term $\frac{\partial^2 \epsilon}{\partial \theta_{\ell_1}\partial\theta_{\ell_2}} \frac{\partial \epsilon}{\partial \theta_{\ell_1}} \frac{\partial \epsilon}{\partial \theta_{\ell_2}}$ assuming $\ell_1 < \ell_2$ as
\begin{align}
    &\mathbb{E}_{{\calU_{\rm Haar}}}\left[\frac{\partial^2 \epsilon}{\partial \theta_{\ell_1}\partial\theta_{\ell_2}} \frac{\partial \epsilon}{\partial \theta_{\ell_1}} \frac{\partial \epsilon}{\partial \theta_{\ell_2}}\right]\nonumber\\
    &= \frac{1}{16}\int_{{\calU_{\rm Haar}}} dU_{\ell_1^-}dU_{\ell_1\shortto\ell_2}dU_{\ell_2^+}\braket{\psi_0|U_{\ell_1^-}^\dagger[X_{\ell_1},U_{\ell_1\shortto\ell_2}^\dagger[X_{\ell_2}, O_{\ell_2^+}]U_{\ell_1\shortto\ell_2}]U_{\ell_1^-}|\psi_0} \braket{\psi_0|U_{\ell_1^-}^\dagger[X_{\ell_1},O_{\ell_1^+}]U_{\ell_1^-}|\psi_0}\nonumber\\
    & \quad \quad \quad \quad \quad \quad \quad \quad \quad \quad \quad \quad  \times\braket{\psi_0|U_{\ell_1^-}^\dagger U_{\ell_1\shortto\ell_2}^\dagger[X_{\ell_2}, O_{\ell_2^+}]U_{\ell_1\shortto\ell_2}U_{\ell_1^-}|\psi_0}\\
    &= \frac{1}{16}\int_{{\calU_{\rm Haar}}} dU_{\ell_1^-}dU_{\ell_1\shortto\ell_2}dU_{\ell_2^+}
    \\
    &\quad \quad 
    \tr\left(\rho_0 U_{\ell_1^-}^\dagger [X_{\ell_1}, U_{\ell_1\shortto\ell_2}^\dagger \left[X_{\ell_2}, O_{\ell_2^+}\right]U_{\ell_1\shortto\ell_2}] U_{\ell_1^-}\rho_0 U_{\ell_1^-}^\dagger \left[X_{\ell_1}, O_{\ell_1^+}\right] U_{\ell_1^-}\rho_0 U_{\ell_1^-}^\dagger U_{\ell_1\shortto\ell_2}^\dagger \left[X_{\ell_2}, O_{\ell_2^+}\right] U_{\ell_1\shortto\ell_2} U_{\ell_1^-} \right)\\
    &= \frac{1}{16(d^3+3d^2+2d)} \int_{{\calU_{\rm Haar}}} dU_{\ell_1\shortto\ell_2}dU_{\ell_2^+}\left[\tr\left(\left[X_{\ell_1}, U_{\ell_1\shortto\ell_2}^\dagger \left[X_{\ell_2}, O_{\ell_2^+}\right]U_{\ell_1\shortto\ell_2}\right]\left[X_{\ell_1}, O_{\ell_1^+}\right] U_{\ell_1\shortto\ell_2}^\dagger \left[X_{\ell_2}, O_{\ell_2^+}\right] U_{\ell_1\shortto\ell_2}\right) \right.\nonumber\\
    &\left. \quad \quad \quad \quad \quad \quad \quad \quad \quad \quad \quad \quad  \quad \quad \quad \quad +\tr\left(\left[X_{\ell_1}, O_{\ell_1^+}\right][X_{\ell_1}, U_{\ell_1\shortto\ell_2}^\dagger \left[X_{\ell_2}, O_{\ell_2^+}\right]U_{\ell_1\shortto\ell_2}] U_{\ell_1\shortto\ell_2}^\dagger \left[X_{\ell_2}, O_{\ell_2^+}\right] U_{\ell_1\shortto\ell_2}\right)\right],
\end{align}
where in the last equation we do the Haar integration ($4$-design) over $U_{\ell_1^-}$. Next we evaluation the Haar integral over $U_{\ell_1\shortto \ell_2}, U_{\ell_2^+}$ separately.

The first term $\tr\left(\left[X_{\ell_1}, U_{\ell_1\shortto\ell_2}^\dagger \left[X_{\ell_2}, O_{\ell_2^+}\right]U_{\ell_1\shortto\ell_2}\right]\left[X_{\ell_1}, O_{\ell_1^+}\right] U_{\ell_1\shortto\ell_2}^\dagger \left[X_{\ell_2}, O_{\ell_2^+}\right] U_{\ell_1\shortto\ell_2}\right)$ is further integrated as
\begin{align}
    &\int_{{\calU_{\rm Haar}}} dU_{\ell_1\shortto\ell_2}dU_{\ell_2^+}\tr\left(\left[X_{\ell_1}, U_{\ell_1\shortto\ell_2}^\dagger \left[X_{\ell_2}, O_{\ell_2^+}\right] U_{\ell_1\shortto\ell_2}\right]\left[X_{\ell_1}, O_{\ell_1^+}\right] U_{\ell_1\shortto\ell_2}^\dagger \left[X_{\ell_2}, O_{\ell_2^+}\right] U_{\ell_1\shortto\ell_2}\right)\nonumber\\
    &= \int_{{\calU_{\rm Haar}}} dU_{\ell_1\shortto\ell_2}dU_{\ell_2^+}\left[\tr\left(X_{\ell_1}U_{\ell_1\shortto\ell_2}^\dagger \left[X_{\ell_2}, O_{\ell_2^+}\right]U_{\ell_1\shortto\ell_2}X_{\ell_1}U_{\ell_1\shortto\ell_2}^\dagger O_{\ell_2^+}\left[X_{\ell_2}, O_{\ell_2^+}\right]U_{\ell_1\shortto\ell_2}\right) \right.\nonumber\\
    &\left. \quad \quad \quad \quad \quad \quad \quad \quad \quad \quad 
    +\tr\left(\left[X_{\ell_2}, O_{\ell_2^+}\right]^2 U_{\ell_1\shortto\ell_2} X_{\ell_1}U_{\ell_1\shortto\ell_2}^\dagger O_{\ell_2^+}U_{\ell_1\shortto\ell_2}X_{\ell_1}U_{\ell_1\shortto\ell_2}^\dagger\right)  \right.\nonumber\\
    &\left. \quad \quad \quad \quad \quad \quad \quad \quad \quad \quad  - \tr\left(X_{\ell_1}U_{\ell_1\shortto\ell_2}^\dagger \left[X_{\ell_2}, O_{\ell_2^+}\right]O_{\ell_2^+}U_{\ell_1\shortto\ell_2}X_{\ell_1}U_{\ell_1\shortto\ell_2}^\dagger \left[X_{\ell_2}, O_{\ell_2^+}\right]U_{\ell_1\shortto\ell_2}\right) -\tr\left(\left[X_{\ell_2}, O_{\ell_2^+}\right]^2 O_{\ell_2^+}\right)\right]\\
    &= \int_{{\calU_{\rm Haar}}}  dU_{\ell_2^+}\left[\frac{\tr\left(O_{\ell_2^+}\left[X_{\ell_2}, O_{\ell_2^+}\right]^2\right)}{1-d^2} + \frac{d \tr(\left[X_{\ell_2}, O_{\ell_2^+}\right]^2) \tr(O)-\tr(O_{\ell_2^+}\left[X_{\ell_2}, O_{\ell_2^+}\right]^2)}{d^2-1} - \frac{\tr\left(O_{\ell_2^+}\left[X_{\ell_2}, O_{\ell_2^+}\right]^2\right)}{1-d^2}\right.\nonumber\\
    &\left. \quad \quad \quad \quad \quad \quad \quad \quad -\tr(O_{\ell_2^+}\left[X_{\ell_2}, O_{\ell_2^+}\right]^2)\right] \label{eq:eq1} \\
    &= \int_{{\calU_{\rm Haar}}}  dU_{\ell_2^+}\frac{d \left[\tr(\left[X_{\ell_2}, O_{\ell_2^+}\right]^2) \tr(O)-d \tr(O_{\ell_2^+}\left[X_{\ell_2}, O_{\ell_2^+}\right]^2)\right]}{d^2-1}\\
    &= \frac{d}{d^2-1}\left(\tr(O)\frac{2 d \left[\tr(O)^2-d \tr(O^2)\right]}{d^2-1} - d\frac{d \left[\tr(O) \tr(O^2)-d \tr(O^3)\right]}{d^2-1} \right) \label{eq:eq2}\\
    &= \frac{d^2 \left[d^2 \tr(O^3)-3 d \tr(O^2) \tr(O)+2 \tr(O)^3\right]}{\left(d^2-1\right)^2},
    \label{eq:g12g1g2_1}
\end{align}
{where we perform Haar unitary integral over $U_{\ell_1\shortto \ell_2}$ and $U_{\ell_2^+}$ to obtain Eqs.~\eqref{eq:eq1},~\eqref{eq:eq2} correspondingly.}

The second term $\tr\left(\left[X_{\ell_1}, O_{\ell_1^+}\right]\left[X_{\ell_1}, U_{\ell_1\shortto\ell_2}^\dagger \left[X_{\ell_2}, O_{\ell_2^+}\right]U_{\ell_1\shortto\ell_2}\right] U_{\ell_1\shortto\ell_2}^\dagger \left[X_{\ell_2}, O_{\ell_2^+}\right] U_{\ell_1\shortto\ell_2}\right)$ becomes
\begin{align}
    &\int_{{\calU_{\rm Haar}}} dU_{\ell_1\shortto\ell_2} dU_{\ell_2^+}\tr\left(\left[X_{\ell_1}, O_{\ell_1^+}\right]\left[X_{\ell_1}, U_{\ell_1\shortto\ell_2}^\dagger \left[X_{\ell_2}, O_{\ell_2^+}\right]U_{\ell_1\shortto\ell_2}\right] U_{\ell_1\shortto\ell_2}^\dagger \left[X_{\ell_2}, O_{\ell_2^+}\right] U_{\ell_1\shortto\ell_2}\right)\nonumber\\
    &= \int_{{\calU_{\rm Haar}}} dU_{\ell_1\shortto\ell_2} dU_{\ell_2^+}\left[\tr\left(\left[X_{\ell_2}, O_{\ell_2^+}\right]^2 U_{\ell_1\shortto\ell_2} X_{\ell_1}U_{\ell_1\shortto\ell_2}^\dagger O_{\ell_2^+}U_{\ell_1\shortto\ell_2}X_{\ell_1}U_{\ell_1\shortto\ell_2}^\dagger\right)\right.\nonumber\\
    &\left. \quad \quad \quad \quad \quad \quad \quad \quad \quad \quad+ \tr\left(X_{\ell_1}U_{\ell_1\shortto\ell_2}^\dagger \left[X_{\ell_2}, O_{\ell_2^+}\right]U_{\ell_1\shortto\ell_2}X_{\ell_1}U_{\ell_1\shortto\ell_2}^\dagger \left[X_{\ell_2}, O_{\ell_2^+}\right]O_{\ell_2^+} U_{\ell_1\shortto\ell_2}\right)\right.\nonumber\\
    &\left. \quad \quad \quad \quad \quad \quad \quad \quad \quad \quad - \tr\left(X_{\ell_1}U_{\ell_1\shortto\ell_2}^\dagger O_{\ell_2^+}\left[X_{\ell_2}, O_{\ell_2^+}\right]U_{\ell_1\shortto\ell_2}X_{\ell_1}U_{\ell_1\shortto\ell_2}^\dagger \left[X_{\ell_2}, O_{\ell_2^+}\right]U_{\ell_1\shortto\ell_2}\right) - \tr\left(\left[X_{\ell_2}, O_{\ell_2^+}\right]^2 O_{\ell_2^+}\right)\right]\\
    &= \int_{{\calU_{\rm Haar}}} dU_{\ell_2^+}\left[\frac{d \tr(\left[X_{\ell_2}, O_{\ell_2^+}\right]^2) \tr(O)-\tr(O_{\ell_2^+}\left[X_{\ell_2}, O_{\ell_2^+}\right]^2)}{d^2-1} + \frac{\tr(O_{\ell_2^+}\left[X_{\ell_2}, O_{\ell_2^+}\right]^2)}{1-d^2} - \frac{\tr(O_{\ell_2^+}\left[X_{\ell_2}, O_{\ell_2^+}\right]^2)}{1-d^2}\right.\nonumber\\
    &\left. \quad \quad \quad \quad \quad \quad \quad \quad - \tr(O_{\ell_2^+}\left[X_{\ell_2}, O_{\ell_2^+}\right]^2)\right]\\
    &= \int_{{\calU_{\rm Haar}}} dU_{\ell_2^+}\frac{d \left[\tr(\left[X_{\ell_2}, O_{\ell_2^+}\right]^2) \tr(O)-d \tr(O_{\ell_2^+}\left[X_{\ell_2}, O_{\ell_2^+}\right]^2)\right]}{d^2-1}\\
    &= \frac{d^2 \left[d^2 \tr(O^3)-3 d \tr(O^2) \tr(O)+2 \tr(O)^3\right]}{\left(d^2-1\right)^2}.
    \label{eq:g12g1g2_2}
\end{align}

Combine Eq.~\eqref{eq:g12g1g2_1} and~\eqref{eq:g12g1g2_2}, we then have the average of $\frac{\partial^2 \epsilon}{\partial \theta_{\ell_1}\partial\theta_{\ell_2}} \frac{\partial \epsilon}{\partial \theta_{\ell_1}} \frac{\partial \epsilon}{\partial \theta_{\ell_2}}$ as
\begin{align}
    \mathbb{E}_{{\calU_{\rm Haar}}}\left[\frac{\partial^2 \epsilon}{\partial \theta_{\ell_1}\partial\theta_{\ell_2}} \frac{\partial \epsilon}{\partial \theta_{\ell_1}} \frac{\partial \epsilon}{\partial \theta_{\ell_2}}\right] &= \frac{1}{16(d^3+3d^2+2d)}\left[\tr\left(\left[X_{\ell_1}, U_{\ell_1\shortto\ell_2}^\dagger \left[X_{\ell_2}, O_{\ell_2^+}\right]U_{\ell_1\shortto\ell_2}\right]\left[X_{\ell_1}, O_{\ell_1^+}\right] U_{\ell_1\shortto\ell_2}^\dagger \left[X_{\ell_2}, O_{\ell_2^+}\right] U_{\ell_1\shortto\ell_2}\right) \right.\nonumber\\
    &\left.  \quad \quad \quad \quad \quad \quad \quad \quad +\tr\left(\left[X_{\ell_1}, O_{\ell_1^+}\right]\left[X_{\ell_1}, U_{\ell_1\shortto\ell_2}^\dagger \left[X_{\ell_2}, O_{\ell_2^+}\right]U_{\ell_1\shortto\ell_2}\right] U_{\ell_1\shortto\ell_2}^\dagger \left[X_{\ell_2}, O_{\ell_2^+}\right] U_{\ell_1\shortto\ell_2}\right)\right]\\
    &=\frac{d \left[d^2 \tr(O^3)-3 d \tr(O^2) \tr(O)+2 \tr(O)^3\right]}{8 (d-1)^2 (d+1)^3 (d+2)}.
    \label{eq:g12g1g2}
\end{align}

\paragraph{Summary of average relative dQNTK $\overline{\lambda_0}$ under Haar random ensemble\\}

Summarizing from Eq.~\eqref{eq:gllglgl} and~\eqref{eq:g12g1g2}, the mean of dQNTK $\overline{\mu_0}$ is
\begin{align}
    \overline{\mu_0} &= L(L-1)\mathbb{E}_{{\calU_{\rm Haar}}}\left[\frac{\partial^2 \epsilon}{\partial \theta_{\ell_1}\partial\theta_{\ell_2}} \frac{\partial \epsilon}{\partial \theta_{\ell_1}} \frac{\partial \epsilon}{\partial \theta_{\ell_2}}\right] + L\mathbb{E}_{{\calU_{\rm Haar}}}\left[\frac{\partial^2 \epsilon}{\partial\theta_\ell^2}\left(\frac{\partial \epsilon}{\partial \theta_\ell}\right)^2 \right] \\
    &= L(L-1)\frac{d \left[d^2 \tr(O^3)-3 d \tr(O^2) \tr(O)+2 \tr(O)^3\right]}{8 (d-1)^2 (d+1)^3 (d+2)} \label{eq:mu0}\\
    &\simeq \frac{L^2}{8d^5}\left[d^2\tr(O^3) - 3d\tr(O^2) \tr(O) + 2\tr(O)^3\right],
    \label{eq:mu0_asym}
\end{align}
where the last line is approximated in the wide QNN limit with $d\gg 1$. Since $\tr(O^3)$ can be nonzero depending on specific choice of $O$, our result characterizes more of $\overline{\mu_0}$'s scaling compared to the result of $\overline{\mu_0}=0$ from Ref.~\cite{liu2023analytic} where $U_{\ell_1^-},U_{\ell_1^+}, U_{\ell_2^-},U_{\ell_2^+}$ is considered to be independent Haar random unitaries, {instead, only $U_{\ell_1^-}, U_{\ell_1\shortto \ell_2}, U_{\ell_2^+}$ should be considered as independent Haar random unitaries.}

According to the definition of $\overline{\lambda_0} = \overline{\mu_0}/\overline{K_0}$, we have
\begin{align}
    \overline{\lambda_0} &= \overline{\mu_0}/\overline{K_0} \nonumber\\
    &= (L-1)\frac{d \left[d^2 \tr(O^3)-3 d \tr(O^2) \tr(O)+2 \tr(O)^3\right]}{4 (d-1) (d+1) (d+2) \left[d\tr(O^2)-\tr(O)^2\right]} \label{eq:lambda0} \\
    &\simeq \frac{L}{4d^2}\frac{d^2 \tr(O^3)-3 d \tr(O^2) \tr(O)+2 \tr(O)^3 }{d\tr(O^2)-\tr(O)^2}.
    \label{eq:lambda0_asym}
\end{align}
Specifically for traceless observable, it is further reduced to
\begin{align}
    \overline{\lambda_0} \simeq \frac{L}{4d}\frac{\tr(O^3)}{\tr(O^2)}.
\end{align}

\subsection{Average dynamical index with Haar random ensemble}
\label{app:epsmu0}

We define the average $\overline{\zeta}$ as
\begin{align}
    \overline{\zeta} = \overline{\epsilon\mu}/\overline{K}^2.
\end{align}
The Haar ensemble average of {$\epsilon \mu$} becomes
\begin{align}
    \overline{\epsilon_0 \mu_0} &= \sum_{\ell_1,\ell_2} \mathbb{E}_{{\calU_{\rm Haar}}} \left[\epsilon \frac{\partial^2 \epsilon}{\partial \theta_{\ell_1}\partial\theta_{\ell_2}}\frac{\partial \epsilon}{\partial \theta_{\ell_1}} \frac{\partial \epsilon}{\partial \theta_{\ell_2}}\right] \\
    &= \sum_{\ell_1,\ell_2} \mathbb{E}_{{\calU_{\rm Haar}}}\left[ {\tr(\rho_0U^\dagger O U)} \frac{\partial^2 \epsilon}{\partial \theta_{\ell_1}\partial\theta_{\ell_2}}\frac{\partial \epsilon}{\partial \theta_{\ell_1}} \frac{\partial \epsilon}{\partial \theta_{\ell_2}}\right] - O_0 \overline{\mu_0} \label{eq:eq3}\\
    &= 2\sum_{\ell_1<\ell_2} \mathbb{E}_{{\calU_{\rm Haar}}} \left[{\tr(\rho_0U^\dagger O U)} \frac{\partial^2 \epsilon}{\partial \theta_{\ell_1}\partial\theta_{\ell_2}}\frac{\partial \epsilon}{\partial \theta_{\ell_1}} \frac{\partial \epsilon}{\partial \theta_{\ell_2}}\right] + \sum_\ell \mathbb{E}_{{\calU_{\rm Haar}}} \left[{\tr(\rho_0U^\dagger O U)} \frac{\partial^2 \epsilon}{\partial\theta_\ell^2}\left(\frac{\partial \epsilon}{\partial \theta_\ell}\right)^2 \right] - O_0 \overline{\mu_0} \\
    &= L(L-1) \mathbb{E}_{{\calU_{\rm Haar}}} \left[{\tr(\rho_0U^\dagger O U)} \frac{\partial^2 \epsilon}{\partial \theta_{\ell_1}\partial\theta_{\ell_2}}\frac{\partial \epsilon}{\partial \theta_{\ell_1}} \frac{\partial \epsilon}{\partial \theta_{\ell_2}}\right] + L\mathbb{E}_{{\calU_{\rm Haar}}} \left[{\tr(\rho_0U^\dagger O U)} \frac{\partial^2 \epsilon}{\partial\theta_\ell^2}\left(\frac{\partial \epsilon}{\partial \theta_\ell}\right)^2 \right] - O_0 \overline{\mu_0}.
\end{align}
{where in Eq.~\eqref{eq:eq3} we expand $\epsilon = \tr(\rho_0U^\dagger O U) - O_0$ following its definition.}
As $\overline{\mu_0}$ is already solved above, we only need to evaluate the first two parts.

\paragraph{Calculation of ${\tr(\rho_0U^\dagger O U)}\frac{\partial^2 \epsilon}{\partial\theta_\ell^2}\left(\frac{\partial \epsilon}{\partial \theta_\ell}\right)^2 $ with Haar random ensemble\\}

The ensemble average of ${\tr(\rho_0U^\dagger O U)}\frac{\partial^2 \epsilon}{\partial\theta_\ell^2}\left(\frac{\partial \epsilon}{\partial \theta_\ell}\right)^2 $ is
\begin{align}
    &\mathbb{E}_{{\calU_{\rm Haar}}} \left[{\tr(\rho_0U^\dagger O U)} \frac{\partial^2 \epsilon}{\partial\theta_\ell^2}\left(\frac{\partial \epsilon}{\partial \theta_\ell}\right)^2 \right]\nonumber\\
    &= \frac{1}{16}\int_{{\calU_{\rm Haar}}} dU_{\ell^-}dU_{\ell^+} \braket{\psi_0|U_{\ell^-}^\dagger O_{\ell^+} U_{\ell^-}|\psi_0} \braket{\psi_0|U_{\ell^-}^\dagger [X_\ell, [X_\ell, O_{\ell^+}]]U_{\ell^-}|\psi_0}\braket{\psi_0|U_{\ell^-}^\dagger\left[X_\ell, O_{\ell^+}\right]U_{\ell^-}|\psi_0}^2\\
    &= \frac{1}{16}\int_{{\calU_{\rm Haar}}} dU_{\ell^-}dU_{\ell^+} \tr\left(\rho_0 U_{\ell^-}^\dagger O_{\ell^+} U_{\ell^-}\rho_0 U_{\ell^-}^\dagger [X_\ell,[X_\ell, O_{\ell^+}]]U_{\ell^-}\rho_0 U_{\ell^-}^\dagger[X_\ell, O_{\ell^+}]U_{\ell^-}\rho_0 U_{\ell^-}^\dagger[X_\ell, O_{\ell^+}] U_{\ell^-}\right)\\
    &= \int_{{\calU_{\rm Haar}}} dU_{\ell^+} \frac{1}{d(d+1)(d+2)(d+3)}\left[\tr([X_\ell, O_{\ell^+}]^2) \tr([X_\ell,[X_\ell, O_{\ell^+}]]O_{\ell^+}) +2 \tr([X_\ell, O_{\ell^+}]^2[X_\ell,[X_\ell, O_{\ell^+}]]O_{\ell^+}) \right.\nonumber\\
    & \left. \quad \quad  \quad \quad  \quad \quad  \quad \quad  \quad \quad  \quad \quad  \quad \quad  \quad \quad  +2 \tr([X_\ell, O_{\ell^+}][X_\ell,[X_\ell, O_{\ell^+}]][X_\ell, O_{\ell^+}] O_{\ell^+}) +2 \tr([X_\ell,[X_\ell, O_{\ell^+}]][X_\ell, O_{\ell^+}]^2 O_{\ell^+})\right],
\end{align}
where we do the Haar integral ($4$-design) over $U_{\ell^-}$ {to obtain the last equation}.

The integration over $U_{\ell^+}$ of the first term $\tr([X_\ell, O_{\ell^+}]^2) \tr([X_\ell,[X_\ell, O_{\ell^+}]]O_{\ell^+})$ is
\begin{align}
    &\int_{{\calU_{\rm Haar}}}  dU_{\ell^+} \tr([X_\ell, O_{\ell^+}]^2) \tr([X_\ell,[X_\ell, O_{\ell^+}]]O_{\ell^+})\nonumber\\
    &= \int_{{\calU_{\rm Haar}}}  dU_{\ell^+} \left[8\tr(X_\ell O_{\ell^+} X_\ell O_{\ell^+})\tr(O^2)-4\tr(X_\ell O_{\ell^+} X_\ell O_{\ell^+})^2-4\tr(O^2)^2\right]\nonumber\\
    &= \frac{8 \tr(O^2) \left[d \tr(O)^2-\tr(O^2)\right]}{d^2-1} - 4\tr(O^2)^2 \nonumber\\
    &\quad -\frac{4}{\left(d^2-1\right)\left(d^2-4\right) \left(d^2-9\right)}\left[\left(d^4-d^3-9 d^2+4 d+20\right) \tr(O)^4 +2 d \left(-3 d^2+5 d+7\right) \tr(O^2) \tr(O)^2  \right.\nonumber\\
    &\quad \quad \quad \quad \quad \quad \quad \quad \quad \quad \quad \quad \quad \quad  \left.+\left(d^5+d^4-12 d^3-5 d^2+12 d+24\right) \tr(O^2)^2 -2 d \left(d^3+2 d^2-14 d+2\right) \tr(O^4)\right]\\
    &= \frac{4}{\left(d^2-9\right) \left(d^2-4\right) \left(d^2-1\right)}\left[-\left(d^4-d^3-9 d^2+4 d+20\right) \tr(O)^4+2 d \left(d^4-10 d^2-5 d+29\right) \tr(O^2) \tr(O)^2\right.\nonumber\\
    &\quad \quad \quad \quad \quad \quad \quad \quad \quad \quad \quad \quad \quad  -8 \left(3 d^2-5 d-7\right) \tr(O^3) \tr(O)-\left(d^6+d^5-11 d^4-12 d^3+18 d^2+12 d+60\right) \tr(O^2)^2 \nonumber\\
    & \quad \quad \quad \quad \quad \quad \quad \quad \quad \quad \quad \quad \quad \left. +2 d \left(d^3+2 d^2-14 d+2\right) \tr(O^4)\right].
\end{align}
{where in last line we do simple algebra to reduce all terms to same denominator.}

The integration of the second term $\tr([X_\ell, O_{\ell^+}]^2[X_\ell,[X_\ell, O_{\ell^+}]]O_{\ell^+})$ becomes
\begin{align}
    &\int_{{\calU_{\rm Haar}}}  dU_{\ell^+} \tr([X_\ell, O_{\ell^+}]^2[X_\ell,[X_\ell, O_{\ell^+}]]O_{\ell^+})\nonumber\\
    &= \int_{{\calU_{\rm Haar}}}  dU_{\ell^+}\left[8\tr(X_\ell O_{\ell^+} X_\ell O_{\ell^+}^3)-4\tr(X_\ell O_{\ell^+}^2 X_\ell O_{\ell^+}^2)-2\tr(X_\ell O_{\ell^+} X_\ell O_{\ell^+} X_\ell O_{\ell^+} X_\ell O_{\ell^+})-2\tr(O^4)\right]\nonumber\\
    &=\frac{2 d \left[4 \left(d^2-7\right) \tr(O^3) \tr(O)+\left(21-2 d^2\right) \tr(O^2)^2-d \left(d^2-7\right) \tr(O^4)-2 d \tr(O^2) \tr(O)^2+\tr(O)^4\right]}{(d^2-1)(d^2-9)}.
\end{align}

The third term $\tr([X_\ell, O_{\ell^+}][X_\ell,[X_\ell, O_{\ell^+}]][X_\ell, O_{\ell^+}] O_{\ell^+})$ becomes
\begin{align}
    &\int_{{\calU_{\rm Haar}}}  dU_{\ell^+} \tr([X_\ell, O_{\ell^+}] [X_\ell,[X_\ell, O_{\ell^+}]][X_\ell, O_{\ell^+}] O_{\ell^+})\nonumber\\
    &= \int_{{\calU_{\rm Haar}}}  dU_{\ell^+} \left[4\tr(X_\ell O_{\ell^+}^2 X_\ell O_{\ell^+}^2)+2\tr(X_\ell O_{\ell^+} X_\ell O_{\ell^+} X_\ell O_{\ell^+} X_\ell O_{\ell^+})+2\tr(O^4)-8\tr(X_\ell O_{\ell^+} X_\ell O_{\ell^+}^3)\right]\\
    &= -\frac{2 d \left[4 \left(d^2-7\right) \tr(O^3) \tr(O)+\left(21-2 d^2\right) \tr(O^2)^2-d \left(d^2-7\right) \tr(O^4)-2 d \tr(O^2) \tr(O)^2+\tr(O)^4\right]}{(d^2-1)(d^2-9)}.
\end{align}

The last term $\tr([X_\ell,[X_\ell, O_{\ell^+}]][X_\ell, O_{\ell^+}]^2 O_{\ell^+})$ becomes
\begin{align}
    &\int_{{\calU_{\rm Haar}}}  dU_{\ell^+} \tr([X_\ell,[X_\ell, O_{\ell^+}]][X_\ell, O_{\ell^+}]^2 O_{\ell^+})\nonumber\\
    &= \int_{{\calU_{\rm Haar}}}  dU_{\ell^+} \left[8\tr(X_\ell O_{\ell^+} X_\ell O_{\ell^+}^3)-4\tr(X_\ell O_{\ell^+}^2 X_\ell O_{\ell^+}^2)-2\tr(X_\ell O_{\ell^+} X_\ell O_{\ell^+} X_\ell O_{\ell^+} X_\ell O_{\ell^+})-2\tr(O^4)\right]\\
    &=\frac{2 d \left[4 \left(d^2-7\right) \tr(O^3) \tr(O)+\left(21-2 d^2\right) \tr(O^2)^2-d \left(d^2-7\right) \tr(O^4)-2 d \tr(O^2) \tr(O)^2+\tr(O)^4\right]}{(d^2-1)(d^2-9)}.
\end{align}

Therefore, we have the {Haar} ensemble average over ${\tr(\rho_0U^\dagger O U)} \frac{\partial^2 \epsilon}{\partial\theta_\ell^2}\left(\frac{\partial \epsilon}{\partial \theta_\ell}\right)^2$ as
\small
\begin{align}
    &\mathbb{E}_{{\calU_{\rm Haar}}} \left[{\tr(\rho_0U^\dagger O U)} \frac{\partial^2 \epsilon}{\partial\theta_\ell^2}\left(\frac{\partial \epsilon}{\partial \theta_\ell}\right)^2 \right] = -\frac{1}{4d(d-1)(d+1)^2 (d-2)  (d+2)^2 (d-3) (d+3)^2}\left[\left(d^4-2 d^3-9 d^2+8 d+20\right) \tr(O)^4\right.\nonumber\\
    &+2 d \left(-d^4+d^3+10 d^2+d-29\right) \tr(O^2) \tr(O)^2-4 \left(d^5-11 d^3-6 d^2+38 d+14\right) \tr(O^3)\tr(O)\nonumber\\
    &\left.+\left(d^6+3 d^5-11 d^4-41 d^3+18 d^2+96 d+60\right) \tr(O^2)^2+d \left(d^5-13 d^3-4 d^2+56 d-4\right) \tr(O^4)\right].
    \label{eq:Ogllglgl}
\end{align}
\normalsize

\paragraph{Calculation ${\tr(\rho_0U^\dagger O U)} \frac{\partial^2 \epsilon}{\partial \theta_{\ell_1}\partial\theta_{\ell_2}}\frac{\partial \epsilon}{\partial \theta_{\ell_1}} \frac{\partial \epsilon}{\partial \theta_{\ell_2}} $ with Haar random ensemble \\}

The {Haar} ensemble average of ${\tr(\rho_0U^\dagger O U)} \frac{\partial^2 \epsilon}{\partial \theta_{\ell_1}\partial\theta_{\ell_2}}\frac{\partial \epsilon}{\partial \theta_{\ell_1}} \frac{\partial \epsilon}{\partial \theta_{\ell_2}}$ assuming $\ell_1 < \ell_2$ can be written as
\begin{align}
    &\mathbb{E}_{{\calU_{\rm Haar}}} \left[{\tr(\rho_0U^\dagger O U)} \frac{\partial^2 \epsilon}{\partial \theta_{\ell_1}\partial\theta_{\ell_2}}\frac{\partial \epsilon}{\partial \theta_{\ell_1}} \frac{\partial \epsilon}{\partial \theta_{\ell_2}}\right]\nonumber\\
    &= \frac{1}{16}\int_{{\calU_{\rm Haar}}} dU_{\ell_1^-}dU_{\ell_1\shortto\ell_2}dU_{\ell_2^+} \left[\braket{\psi_0|U_{\ell_1^-}^\dagger U_{\ell_1\shortto\ell_2}^\dagger O_{\ell_2^+}U_{\ell_1\shortto\ell_2}U_{\ell_1^-}|\psi_0}\braket{\psi_0|U_{\ell_1^-}^\dagger \left[X_{\ell_1},U_{\ell_1\shortto\ell_2}^\dagger \left[X_{\ell_2}, O_{\ell_2^+}\right] U_{\ell_1\shortto\ell_2}\right] U_{\ell_1^-}|\psi_0} \right. \nonumber\\
    &\left. \quad \quad \quad \quad \quad \quad \quad \quad \quad \quad \quad  \quad \quad \times \braket{\psi_0|U_{\ell_1^-}^\dagger \left[X_{\ell_1}, O_{\ell_1^+}\right] U_{\ell_1^-}|\psi_0} \braket{\psi_0|U_{\ell_1^-}^\dagger U_{\ell_1\shortto\ell_2}^\dagger \left[X_{\ell_2}, O_{\ell_2^+}\right] U_{\ell_1\shortto\ell_2}U_{\ell_1^-}|\psi_0}\right]\\
    &= \frac{1}{16}\int_{{\calU_{\rm Haar}}} dU_{\ell_1^-}dU_{\ell_1\shortto\ell_2}dU_{\ell_2^+}\tr\left(\rho_0 U_{\ell_1^-}^\dagger U_{\ell_1\shortto\ell_2}^\dagger O_{\ell_2^+}U_{\ell_1\shortto\ell_2}U_{\ell_1^-}\rho_0 U_{\ell_1^-}^\dagger \left[X_{\ell_1}, U_{\ell_1\shortto\ell_2}^\dagger \left[X_{\ell_2}, O_{\ell_2^+}\right]U_{\ell_1\shortto\ell_2}\right] U_{\ell_1^-}\rho_0\right.\nonumber\\
    & \quad \quad \quad \quad \quad \quad \quad \quad \quad \quad \quad  \quad \quad \quad \left.\cdot U_{\ell_1^-}^\dagger \left[X_{\ell_1}, O_{\ell_1^+}\right] U_{\ell_1^-}\rho_0 U_{\ell_1^-}^\dagger U_{\ell_1\shortto\ell_2}^\dagger \left[X_{\ell_2}, O_{\ell_2^+}\right] U_{\ell_1\shortto\ell_2} U_{\ell_1^-}\right).
\end{align}
The integration of $U_{\ell_1^-}$ over {Haar unitary ensemble} becomes
\begin{align}
    &\int_{{\calU_{\rm Haar}}} dU_{\ell_1^-}dU_{\ell_1\shortto\ell_2}dU_{\ell_2^+}\tr\left(\rho_0 U_{\ell_1^-}^\dagger U_{\ell_1\shortto\ell_2}^\dagger O_{\ell_2^+}U_{\ell_1\shortto\ell_2}U_{\ell_1^-}\rho_0 U_{\ell_1^-}^\dagger \left[X_{\ell_1}, U_{\ell_1\shortto\ell_2}^\dagger \left[X_{\ell_2}, O_{\ell_2^+}\right] U_{\ell_1\shortto\ell_2}\right] U_{\ell_1^-}\rho_0\right.\nonumber\\
    & \quad \quad \quad \quad \quad \quad \quad \quad \quad \quad \quad \quad \left. \cdot  U_{\ell_1^-}^\dagger \left[X_{\ell_1}, O_{\ell_1^+}\right] U_{\ell_1^-}\rho_0 U_{\ell_1^-}^\dagger U_{\ell_1\shortto\ell_2}^\dagger \left[X_{\ell_2}, O_{\ell_2^+}\right] U_{\ell_1\shortto\ell_2} U_{\ell_1^-}\right)\nonumber\\
    &= \int_{{\calU_{\rm Haar}}} dU_{\ell_1\shortto\ell_2}dU_{\ell_2^+}\frac{1}{d^4+6 d^3+11 d^2+6 d}\left[\tr(O) \tr\left(\left[X_{\ell_1}, O_{\ell_1^+}\right]U_{\ell_1\shortto\ell_2}^\dagger \left[X_{\ell_2}, O_{\ell_2^+}\right]U_{\ell_1\shortto\ell_2}\left[X_{\ell_1},U_{\ell_1\shortto\ell_2}^\dagger \left[X_{\ell_2}, O_{\ell_2^+}\right]U_{\ell_1\shortto\ell_2}\right]\right)\right.\nonumber\\
    &+\tr(O) \tr\left(U_{\ell_1\shortto\ell_2}^\dagger \left[X_{\ell_2}, O_{\ell_2^+}\right]U_{\ell_1\shortto\ell_2}\left[X_{\ell_1}, O_{\ell_1^+}\right]\left[X_{\ell_1},U_{\ell_1\shortto\ell_2}^\dagger \left[X_{\ell_2}, O_{\ell_2^+}\right]U_{\ell_1\shortto\ell_2}\right]\right)\nonumber\\
    &+\tr\left(U_{\ell_1\shortto\ell_2}^\dagger \left[X_{\ell_2}, O_{\ell_2^+}\right]U_{\ell_1\shortto\ell_2}\left[X_{\ell_1}, O_{\ell_1^+}\right]\right) \tr\left(\left[X_{\ell_1},U_{\ell_1\shortto\ell_2}^\dagger \left[X_{\ell_2}, O_{\ell_2^+}\right]U_{\ell_1\shortto\ell_2}\right]U_{\ell_1\shortto\ell_2}^\dagger O_{\ell_2^+}U_{\ell_1\shortto\ell_2}\right) \nonumber\\
    &+\tr\left(\left[X_{\ell_1}, O_{\ell_1^+}\right]U_{\ell_1\shortto\ell_2}^\dagger O_{\ell_2^+}U_{\ell_1\shortto\ell_2}\right) \tr\left(U_{\ell_1\shortto\ell_2}^\dagger \left[X_{\ell_2}, O_{\ell_2^+}\right]U_{\ell_1\shortto\ell_2}\left[X_{\ell_1},U_{\ell_1\shortto\ell_2}^\dagger \left[X_{\ell_2}, O_{\ell_2^+}\right]U_{\ell_1\shortto\ell_2}\right]\right)\nonumber\\
    &+\tr\left(\left[X_{\ell_1}, O_{\ell_1^+}\right]\left[X_{\ell_1},U_{\ell_1\shortto\ell_2}^\dagger \left[X_{\ell_2}, O_{\ell_2^+}\right]U_{\ell_1\shortto\ell_2}\right]U_{\ell_1\shortto\ell_2}^\dagger \left[X_{\ell_2}, O_{\ell_2^+}\right]O_{\ell_2^+}U_{\ell_1\shortto\ell_2}\right)\nonumber\\
    &+\tr\left(\left[X_{\ell_1}, O_{\ell_1^+}\right]\left[X_{\ell_1},U_{\ell_1\shortto\ell_2}^\dagger \left[X_{\ell_2}, O_{\ell_2^+}\right]U_{\ell_1\shortto\ell_2}\right]U_{\ell_1\shortto\ell_2}^\dagger O_{\ell_2^+}\left[X_{\ell_2}, O_{\ell_2^+}\right]U_{\ell_1\shortto\ell_2}\right)\nonumber\\
    &+\tr\left(\left[X_{\ell_1},U_{\ell_1\shortto\ell_2}^\dagger \left[X_{\ell_2}, O_{\ell_2^+}\right]U_{\ell_1\shortto\ell_2}\right]\left[X_{\ell_1}, O_{\ell_1^+}\right]U_{\ell_1\shortto\ell_2}^\dagger \left[X_{\ell_2}, O_{\ell_2^+}\right]O_{\ell_2^+}U_{\ell_1\shortto\ell_2}\right) \nonumber\\
    &+\tr\left(\left[X_{\ell_1},U_{\ell_1\shortto\ell_2}^\dagger \left[X_{\ell_2}, O_{\ell_2^+}\right]U_{\ell_1\shortto\ell_2}\right]\left[X_{\ell_1}, O_{\ell_1^+}\right]U_{\ell_1\shortto\ell_2}^\dagger O_{\ell_2^+}\left[X_{\ell_2}, O_{\ell_2^+}\right]U_{\ell_1\shortto\ell_2}\right)\nonumber\\
    &+\tr\left(\left[X_{\ell_1}, O_{\ell_1^+}\right]U_{\ell_1\shortto\ell_2}^\dagger \left[X_{\ell_2}, O_{\ell_2^+}\right]U_{\ell_1\shortto\ell_2}\left[X_{\ell_1},U_{\ell_1\shortto\ell_2}^\dagger \left[X_{\ell_2}, O_{\ell_2^+}\right]U_{\ell_1\shortto\ell_2}\right]U_{\ell_1\shortto\ell_2}^\dagger O_{\ell_2^+}U_{\ell_1\shortto\ell_2}\right)\nonumber\\
    &+\tr\left(\left[X_{\ell_1},U_{\ell_1\shortto\ell_2}^\dagger \left[X_{\ell_2}, O_{\ell_2^+}\right]U_{\ell_1\shortto\ell_2}\right]U_{\ell_1\shortto\ell_2}^\dagger \left[X_{\ell_2}, O_{\ell_2^+}\right]U_{\ell_1\shortto\ell_2}\left[X_{\ell_1}, O_{\ell_1^+}\right]U_{\ell_1\shortto\ell_2}^\dagger O_{\ell_2^+}U_{\ell_1\shortto\ell_2}\right)\nonumber\\ &\left.+\tr\left(\left[X_{\ell_1}, O_{\ell_1^+}\right]\left[X_{\ell_1},U_{\ell_1\shortto\ell_2}^\dagger \left[X_{\ell_2}, O_{\ell_2^+}\right]U_{\ell_1\shortto\ell_2}\right]\right) \tr(\left[X_{\ell_2}, O_{\ell_2^+}\right]O_{\ell_2^+})\right].
\end{align}
We will evaluate the integration over $U_{\ell_1\shortto \ell_2}, U_{\ell_2^+}$ on each item in the following. 

The first two are already solved before (see Eqs.~\eqref{eq:g12g1g2_1} and~\eqref{eq:g12g1g2_2}) {up to a cyclic transformation in trace}
\begin{align}
    &\int_{{\calU_{\rm Haar}}} dU_{\ell_1\shortto\ell_2}dU_{\ell_2^+}\tr(O) \tr\left(\left[X_{\ell_1}, O_{\ell_1^+}\right]U_{\ell_1\shortto\ell_2}^\dagger \left[X_{\ell_2}, O_{\ell_2^+}\right]U_{\ell_1\shortto\ell_2}\left[X_{\ell_1},U_{\ell_1\shortto\ell_2}^\dagger \left[X_{\ell_2}, O_{\ell_2^+}\right]U_{\ell_1\shortto\ell_2}\right]\right)\nonumber\\
    &= \tr(O)  \frac{d^2 \left[d^2 \tr(O^3)-3 d \tr(O^2) \tr(O)+2 \tr(O)^3\right]}{\left(d^2-1\right)^2}\\
    &\int_{{\calU_{\rm Haar}}} dU_{\ell_1\shortto\ell_2}dU_{\ell_2^+} \tr(O) \tr\left(U_{\ell_1\shortto\ell_2}^\dagger \left[X_{\ell_2}, O_{\ell_2^+}\right]U_{\ell_1\shortto\ell_2}\left[X_{\ell_1}, O_{\ell_1^+}\right]\left[X_{\ell_1},U_{\ell_1\shortto\ell_2}^\dagger \left[X_{\ell_2}, O_{\ell_2^+}\right]U_{\ell_1\shortto\ell_2}\right]\right) \nonumber\\
    &= \tr(O)\frac{d^2 \left[d^2 \tr(O^3)-3 d \tr(O^2) \tr(O)+2 \tr(O)^3\right]}{\left(d^2-1\right)^2}.
\end{align}

The integral of the third one becomes
\begin{align}
    &\int_{{\calU_{\rm Haar}}}  dU_{\ell_1\shortto\ell_2}dU_{\ell_2^+}\tr\left(U_{\ell_1\shortto\ell_2}^\dagger \left[X_{\ell_2}, O_{\ell_2^+}\right]U_{\ell_1\shortto\ell_2}\left[X_{\ell_1}, O_{\ell_1^+}\right]\right) \tr\left(\left[X_{\ell_1},U_{\ell_1\shortto\ell_2}^\dagger \left[X_{\ell_2}, O_{\ell_2^+}\right]U_{\ell_1\shortto\ell_2}\right] U_{\ell_1\shortto\ell_2}^\dagger O_{\ell_2^+}U_{\ell_1\shortto\ell_2}\right)\nonumber\\
    &= \sum_{i_1,i_2}\int_{{\calU_{\rm Haar}}}  dU_{\ell_1\shortto\ell_2}dU_{\ell_2^+}\left[2\tr\left(P_{i_2,i_1} X_{\ell_1} U_{\ell_1\shortto\ell_2}^\dagger O_{\ell_2^+} \left[X_{\ell_2}, O_{\ell_2^+}\right] U_{\ell_1\shortto\ell_2} P_{i_1,i_2} X_{\ell_1}U_{\ell_1\shortto\ell_2}^\dagger \left[X_{\ell_2}, O_{\ell_2^+}\right] O_{\ell_2^+}  U_{\ell_1\shortto\ell_2}\right)\right.\nonumber\\
    & \quad \quad \quad \quad \quad \quad \quad \quad \quad \quad  \quad - \tr\left(P_{i_2,i_1} X_{\ell_1} U_{\ell_1\shortto\ell_2}^\dagger O_{\ell_2^+} \left[X_{\ell_2}, O_{\ell_2^+}\right]   U_{\ell_1\shortto\ell_2} P_{i_1,i_2} X_{\ell_1}U_{\ell_1\shortto\ell_2}^\dagger O_{\ell_2^+} \left[X_{\ell_2}, O_{\ell_2^+}\right]U_{\ell_1\shortto\ell_2}\right)\nonumber\\
    & \quad \quad \quad \quad \quad \quad \quad \quad \quad \quad \quad \left.- \tr\left(P_{i_2,i_1} X_{\ell_1} U_{\ell_1\shortto\ell_2}^\dagger  \left[X_{\ell_2}, O_{\ell_2^+}\right]  O_{\ell_2^+}  U_{\ell_1\shortto\ell_2} P_{i_1,i_2} X_{\ell_1}U_{\ell_1\shortto\ell_2}^\dagger \left[X_{\ell_2}, O_{\ell_2^+}\right]  O_{\ell_2^+}U_{\ell_1\shortto\ell_2}\right)\right] \label{eq:eq4}\\
    &= \int_{{\calU_{\rm Haar}}}  dU_{\ell_2^+} 2\left[\frac{d \tr(O_{\ell_2^+}^2 \left[X_{\ell_2}, O_{\ell_2^+}\right]^2)-\tr(O_{\ell_2^+}\left[X_{\ell_2}, O_{\ell_2^+}\right])^2}{d^2-1}
    \right.
    \nonumber
    \\
    &\left.\quad \quad  \quad \quad \quad \quad \quad \quad  +\frac{\tr(O_{\ell_2^+}\left[X_{\ell_2}, O_{\ell_2^+}\right])^2-d \tr(O_{\ell_2^+} \left[X_{\ell_2}, O_{\ell_2^+}\right] O_{\ell_2^+} \left[X_{\ell_2}, O_{\ell_2^+}\right])}{d^2-1}\right]\\
    &= \int_{{\calU_{\rm Haar}}} dU_{\ell_2^+}\frac{2d\left[\tr(O_{\ell_2^+}^2 \left[X_{\ell_2}, O_{\ell_2^+}\right]^2) - \tr(O_{\ell_2^+} \left[X_{\ell_2}, O_{\ell_2^+}\right] O_{\ell_2^+} \left[X_{\ell_2}, O_{\ell_2^+}\right])\right]}{d^2-1}\\
    &= \frac{2d}{d^2-1}\left[\frac{d \left(d \tr(O^4)+\tr(O^2)^2-2 \tr(O) \tr(O^3)\right)}{1-d^2} - \frac{2d\left(\tr(O^2)^2-\tr(O)\tr(O^3)\right)}{d^2-1}\right]\\
    &= \frac{2 d^2 \left[4 \tr(O) \tr(O^3)-d \tr(O^4)-3 \tr(O^2)^2\right]}{\left(d^2-1\right)^2},
\end{align}
{where in Eq.~\eqref{eq:eq4} we expand each trace in the first line via an orthonormal basis $\ket{i_1}, \ket{i_2}$ separately, and regroup them together to a single trace operation via projectors $P_{i_2,i_1} = \ketbra{i_2}{i_1}$ and $P_{i_1,i_2} = \ketbra{i_1}{i_2}$ for conveninent calculation with \texttt{RTNI}~\cite{fukuda2019rtni}.}

The fourth one is simply
\begin{align}
    \int dU_{\ell_1\shortto\ell_2}dU_{\ell_2^+}\tr\left(\left[X_{\ell_1}, O_{\ell_1^+}\right]U_{\ell_1\shortto\ell_2}^\dagger O_{\ell_2^+}U_{\ell_1\shortto\ell_2}\right) \tr\left(U_{\ell_1\shortto\ell_2}^\dagger \left[X_{\ell_2}, O_{\ell_2^+}\right]U_{\ell_1\shortto\ell_2}\left[X_{\ell_1},U_{\ell_1\shortto\ell_2}^\dagger \left[X_{\ell_2}, O_{\ell_2^+}\right]U_{\ell_1\shortto\ell_2}\right]\right) = 0
\end{align}
by the cyclic property of trace from expansion.

The fifth term through integration is
\small
\begin{align}
    &\int_{{\calU_{\rm Haar}}} dU_{\ell_1\shortto\ell_2}dU_{\ell_2^+}\tr\left(\left[X_{\ell_1}, O_{\ell_1^+}\right]\left[X_{\ell_1},U_{\ell_1\shortto\ell_2}^\dagger \left[X_{\ell_2}, O_{\ell_2^+}\right]U_{\ell_1\shortto\ell_2}\right] U_{\ell_1\shortto\ell_2}^\dagger \left[X_{\ell_2}, O_{\ell_2^+}\right]O_{\ell_2^+}U_{\ell_1\shortto\ell_2}\right)\nonumber\\
    &=\int_{{\calU_{\rm Haar}}} dU_{\ell_1\shortto\ell_2}dU_{\ell_2^+}\left[\tr\left(X_{\ell_1}U_{\ell_1\shortto\ell_2}^\dagger O_{\ell_2^+} U_{\ell_1\shortto\ell_2} X_{\ell_1} U_{\ell_1\shortto\ell_2}^\dagger \left[X_{\ell_2}, O_{\ell_2^+}\right]^2 O_{\ell_2^+} U_{\ell_1\shortto\ell_2}\right)  - \tr\left(\left[X_{\ell_2}, O_{\ell_2^+}\right]^2 O_{\ell_2^+}^2\right)\right.\nonumber\\
    &\quad \quad \quad \quad \quad \quad \quad \quad  \quad \quad \left.+\tr\left(X_{\ell_1}U_{\ell_1\shortto\ell_2}^\dagger \left[X_{\ell_2}, O_{\ell_2^+}\right]U_{\ell_1\shortto\ell_2} X_{\ell_1} U_{\ell_1\shortto\ell_2}^\dagger \left[X_{\ell_2}, O_{\ell_2^+}\right] O_{\ell_2^+}^2 U_{\ell_1\shortto\ell_2}\right)\right.\nonumber\\
    & \quad \quad \quad \quad \quad \quad \quad \quad \quad \quad \left. -\tr\left(X_{\ell_1}U_{\ell_1\shortto\ell_2}^\dagger O_{\ell_2^+} \left[X_{\ell_2}, O_{\ell_2^+}\right] U_{\ell_1\shortto\ell_2} X_{\ell_1} U_{\ell_1\shortto\ell_2}^\dagger \left[X_{\ell_2}, O_{\ell_2^+}\right] O_{\ell_2^+} U_{\ell_1\shortto\ell_2}\right)\right]\\
    &= \int_{{\calU_{\rm Haar}}} dU_{\ell_2^+}\left[\frac{d\tr(O)\tr(\left[X_{\ell_2}, O_{\ell_2^+}\right]^2 O_{\ell_2^+}) - \tr(\left[X_{\ell_2}, O_{\ell_2^+}\right]^2 O_{\ell_2^+}^2)}{d^2-1} + \frac{\tr(\left[X_{\ell_2}, O_{\ell_2^+}\right]^2 O_{\ell_2^+}^2)}{1-d^2} -\frac{\tr(\left[X_{\ell_2}, O_{\ell_2^+}\right]^2 O_{\ell_2^+}^2)}{1-d^2}\right.\nonumber\\
    & \quad \quad \quad \quad \quad \quad \quad \left. - \tr(\left[X_{\ell_2}, O_{\ell_2^+}\right]^2 O_{\ell_2^+}^2)\right]\\
    &= \int_{{\calU_{\rm Haar}}} dU_{\ell_2^+}\frac{d}{d^2-1}\left[\tr(O)\tr(\left[X_{\ell_2}, O_{\ell_2^+}\right]^2 O_{\ell_2^+}) - d\tr(\left[X_{\ell_2}, O_{\ell_2^+}\right]^2 O_{\ell_2^+}^2)\right]\\
    &= \int_{{\calU_{\rm Haar}}} dU_{\ell_2^+} \left\{\frac{d}{d^2-1}\tr(O)\left[\tr\left(X_{\ell_2}O_{\ell_2^+} X_{\ell_2}O_{\ell_2^+}^2\right) - \tr(O^3)\right] 
     - \frac{d^2}{d^2-1}\left[2\tr(X_{\ell_2}O_{\ell_2^+}^3 X_{\ell_2}O_{\ell_2^+}) - \tr(X_{\ell_2}O_{\ell_2^+}^2 X_{\ell_2}O_{\ell_2^+}^2)-\tr(O^4)\right]\right\} \\
    &= \frac{d}{d^2-1}\tr(O)\frac{d\left[\tr(O)\tr(O^2)-d\tr(O^3)\right]}{d^2-1} + \frac{d^2}{d^2-1}\frac{d}{d^2-1}\left[\tr(O^2)^2-2\tr(O)\tr(O^3)+d\tr(O^4)\right]\\
    &= \frac{d^2}{(d^2-1)^2}\left[\tr(O)^2\tr(O^2)-3d\tr(O)\tr(O^3)+d\tr(O^2)^2+d^2\tr(O^4)\right].
\end{align}
\normalsize

The sixth item is
\small
\begin{align}
    &\int_{{\calU_{\rm Haar}}} dU_{\ell_1\shortto\ell_2}dU_{\ell_2^+}\tr\left(\left[X_{\ell_1}, O_{\ell_1^+}\right]\left[X_{\ell_1},U_{\ell_1\shortto\ell_2}^\dagger \left[X_{\ell_2}, O_{\ell_2^+}\right]U_{\ell_1\shortto\ell_2}\right]U_{\ell_1\shortto\ell_2}^\dagger O_{\ell_2^+}\left[X_{\ell_2}, O_{\ell_2^+}\right]U_{\ell_1\shortto\ell_2}\right)\nonumber\\
    &= \int_{{\calU_{\rm Haar}}} dU_{\ell_1\shortto\ell_2}dU_{\ell_2^+}\left[\tr\left(X_{\ell_1}U_{\ell_1\shortto\ell_2}^\dagger O_{\ell_2^+} U_{\ell_1\shortto\ell_2} X_{\ell_1} U_{\ell_1\shortto\ell_2}^\dagger \left[X_{\ell_2}, O_{\ell_2^+}\right] O_{\ell_2^+} \left[X_{\ell_2}, O_{\ell_2^+}\right] U_{\ell_1\shortto\ell_2}\right) - \tr\left(\left[X_{\ell_2}, O_{\ell_2^+}\right] O_{\ell_2^+}\left[X_{\ell_2}, O_{\ell_2^+}\right] O_{\ell_2^+}\right) \right.\nonumber\\
    & \quad \quad \quad \quad \quad \quad \quad \quad \quad \left. +\tr\left(X_{\ell_1}U_{\ell_1\shortto\ell_2}^\dagger \left[X_{\ell_2}, O_{\ell_2^+}\right]U_{\ell_1\shortto\ell_2} X_{\ell_1} U_{\ell_1\shortto\ell_2}^\dagger  O_{\ell_2^+}\left[X_{\ell_2}, O_{\ell_2^+}\right]O_{\ell_2^+} U_{\ell_1\shortto\ell_2}\right)\right.\nonumber\\
    & \quad \quad \quad \quad \quad \quad \quad \quad \quad \left.-\tr\left(X_{\ell_1}U_{\ell_1\shortto\ell_2}^\dagger O_{\ell_2^+} \left[X_{\ell_2}, O_{\ell_2^+}\right] U_{\ell_1\shortto\ell_2} X_{\ell_1} U_{\ell_1\shortto\ell_2}^\dagger O_{\ell_2^+} \left[X_{\ell_2}, O_{\ell_2^+}\right] U_{\ell_1\shortto\ell_2}\right) \right]\\
    &= \int_{{\calU_{\rm Haar}}} dU_{\ell_2^+}\left[\frac{d\tr(O)\tr(\left[X_{\ell_2}, O_{\ell_2^+}\right]^2 O_{\ell_2^+}) - \tr(\left[X_{\ell_2}, O_{\ell_2^+}\right] O_{\ell_2^+} \left[X_{\ell_2}, O_{\ell_2^+}\right] O_{\ell_2^+})}{d^2-1} + \frac{\tr(\left[X_{\ell_2}, O_{\ell_2^+}\right] O_{\ell_2^+}\left[X_{\ell_2}, O_{\ell_2^+}\right] O_{\ell_2^+})}{1-d^2}\right.\nonumber\\
    & \quad \quad \quad \quad \quad \quad \quad \left.-\frac{\tr(\left[X_{\ell_2}, O_{\ell_2^+}\right] O_{\ell_2^+} \left[X_{\ell_2}, O_{\ell_2^+}\right] O_{\ell_2^+})}{1-d^2} - \tr(\left[X_{\ell_2}, O_{\ell_2^+}\right] O_{\ell_2^+} \left[X_{\ell_2}, O_{\ell_2^+}\right] O_{\ell_2^+})\right]\\
    &= \int_{{\calU_{\rm Haar}}} dU_{\ell_2^+}\frac{d}{d^2-1}\left[\tr(O)\tr(\left[X_{\ell_2}, O_{\ell_2^+}\right]^2 O_{\ell_2^+}) - d\tr(\left[X_{\ell_2}, O_{\ell_2^+}\right] O_{\ell_2^+} \left[X_{\ell_2}, O_{\ell_2^+}\right] O_{\ell_2^+})\right]\\
    &= \int_{{\calU_{\rm Haar}}} dU_{\ell_2^+} \left\{\frac{d}{d^2-1}\tr(O)\left[\tr\left(X_{\ell_2}O_{\ell_2^+} X_{\ell_2}O_{\ell_2^+}^2\right) - \tr(O^3)\right] - \frac{2d^2}{d^2-1}\left[\tr(X_{\ell_2}O_{\ell_2^+}^2 X_{\ell_2}O_{\ell_2^+}^2) - \tr(X_{\ell_2}O_{\ell_2^+}^3 X_{\ell_2}O_{\ell_2^+})\right] \right\}\\
    &= \frac{d}{d^2-1}\tr(O)\frac{d\left[\tr(O)\tr(O^2)-d\tr(O^3)\right]}{d^2-1} - \frac{d^2}{d^2-1}\frac{2d}{d^2-1}\left[\tr(O^2)^2-\tr(O)\tr(O^3)\right]\\
    &= \frac{d^2}{(d^2-1)^2}\left[\tr(O)^2 \tr(O^2) - 2d\tr(O^2)^2 +d\tr(O)\tr(O^3)\right].
\end{align}
\normalsize

The seventh item is
\small
\begin{align}
    &\int_{{\calU_{\rm Haar}}} dU_{\ell_1\shortto\ell_2}dU_{\ell_2^+}\tr\left(\left[X_{\ell_1},U_{\ell_1\shortto\ell_2}^\dagger \left[X_{\ell_2}, O_{\ell_2^+}\right]U_{\ell_1\shortto\ell_2}\right]\left[X_{\ell_1}, O_{\ell_1^+}\right]U_{\ell_1\shortto\ell_2}^\dagger \left[X_{\ell_2}, O_{\ell_2^+}\right]O_{\ell_2^+}U_{\ell_1\shortto\ell_2}\right)\nonumber\\
    &= \int_{{\calU_{\rm Haar}}} dU_{\ell_1\shortto\ell_2}dU_{\ell_2^+}\left[\tr\left(X_{\ell_1}U_{\ell_1\shortto\ell_2}^\dagger \left[X_{\ell_2}, O_{\ell_2^+}\right] U_{\ell_1\shortto\ell_2} X_{\ell_1} U_{\ell_1\shortto\ell_2}^\dagger O_{\ell_2^+} \left[X_{\ell_2}, O_{\ell_2^+}\right] O_{\ell_2^+} U_{\ell_1\shortto\ell_2}\right) - \tr\left(\left[X_{\ell_2}, O_{\ell_2^+}\right] O_{\ell_2^+}\left[X_{\ell_2}, O_{\ell_2^+}\right] O_{\ell_2^+}\right)\right. \nonumber\\
    & \quad \quad \quad \quad \quad \quad \quad \quad \quad \left. +\tr\left(X_{\ell_1}U_{\ell_1\shortto\ell_2}^\dagger O_{\ell_2^+} U_{\ell_1\shortto\ell_2} X_{\ell_1} U_{\ell_1\shortto\ell_2}^\dagger \left[X_{\ell_2}, O_{\ell_2^+}\right] O_{\ell_2^+}\left[X_{\ell_2}, O_{\ell_2^+}\right] U_{\ell_1\shortto\ell_2}\right)\right.\nonumber\\
    & \quad \quad \quad \quad \quad \quad \quad \quad \quad \left.-\tr\left(X_{\ell_1}U_{\ell_1\shortto\ell_2}^\dagger \left[X_{\ell_2}, O_{\ell_2^+}\right] O_{\ell_2^+} U_{\ell_1\shortto\ell_2} X_{\ell_1} U_{\ell_1\shortto\ell_2}^\dagger  \left[X_{\ell_2}, O_{\ell_2^+}\right]O_{\ell_2^+} U_{\ell_1\shortto\ell_2}\right) \right]\\
    &=  \frac{d^2}{(d^2-1)^2}\left[\tr(O)^2 \tr(O^2) - 2d\tr(O^2)^2 +d\tr(O)\tr(O^3)\right].
\end{align}
\normalsize

The eighth item is
\begin{align}
    &\int_{{\calU_{\rm Haar}}} dU_{\ell_1\shortto\ell_2}dU_{\ell_2^+}\tr\left(\left[X_{\ell_1},U_{\ell_1\shortto\ell_2}^\dagger \left[X_{\ell_2}, O_{\ell_2^+}\right]U_{\ell_1\shortto\ell_2}\right]\left[X_{\ell_1}, O_{\ell_1^+}\right]U_{\ell_1\shortto\ell_2}^\dagger O_{\ell_2^+}\left[X_{\ell_2}, O_{\ell_2^+}\right]U_{\ell_1\shortto\ell_2}\right)\nonumber\\
    &= \int_{{\calU_{\rm Haar}}} dU_{\ell_1\shortto\ell_2}dU_{\ell_2^+}\left[\tr\left(X_{\ell_1}U_{\ell_1\shortto\ell_2}^\dagger \left[X_{\ell_2}, O_{\ell_2^+}\right] U_{\ell_1\shortto\ell_2} X_{\ell_1} U_{\ell_1\shortto\ell_2}^\dagger O_{\ell_2^+}^2 \left[X_{\ell_2}, O_{\ell_2^+}\right] U_{\ell_1\shortto\ell_2}\right) - \tr\left(\left[X_{\ell_2}, O_{\ell_2^+}\right]^2 O_{\ell_2^+}^2\right)\right.\nonumber\\
    &\quad \quad \quad \quad \quad \quad \quad \quad \quad \left. +\tr\left(X_{\ell_1}U_{\ell_1\shortto\ell_2}^\dagger O_{\ell_2^+} U_{\ell_1\shortto\ell_2} X_{\ell_1} U_{\ell_1\shortto\ell_2}^\dagger O_{\ell_2^+}\left[X_{\ell_2}, O_{\ell_2^+}\right]^2 U_{\ell_1\shortto\ell_2}\right)\right.\nonumber\\
    &\quad \quad \quad \quad \quad \quad \quad \quad \quad  \left.-\tr\left(X_{\ell_1}U_{\ell_1\shortto\ell_2}^\dagger \left[X_{\ell_2}, O_{\ell_2^+}\right] O_{\ell_2^+} U_{\ell_1\shortto\ell_2} X_{\ell_1} U_{\ell_1\shortto\ell_2}^\dagger  O_{\ell_2^+}\left[X_{\ell_2}, O_{\ell_2^+}\right] U_{\ell_1\shortto\ell_2}\right) \right]\\
    &= \int_{{\calU_{\rm Haar}}} dU_{\ell_2^+}\left[\frac{\tr(\left[X_{\ell_2}, O_{\ell_2^+}\right]^2 O_{\ell_2^+}^2)}{1-d^2}+ \frac{d\tr(O)\tr(\left[X_{\ell_2}, O_{\ell_2^+}\right]^2O_{\ell_2^+}) - \tr(\left[X_{\ell_2}, O_{\ell_2^+}\right]^2 O_{\ell_2^+}^2)}{d^2-1} - \frac{\tr(\left[X_{\ell_2}, O_{\ell_2^+}\right]^2 O_{\ell_2^+}^2)}{1-d^2}\right.\nonumber\\
    &\quad \quad \quad \quad \quad \quad \left.- \tr(\left[X_{\ell_2}, O_{\ell_2^+}\right]^2 O_{\ell_2^+}^2)\right]\\
    &= \int_{{\calU_{\rm Haar}}} dU_{\ell_2^+}\frac{d}{d^2-1}\left[\tr(O)\tr(\left[X_{\ell_2}, O_{\ell_2^+}\right]^2 O_{\ell_2^+}) - d\tr(\left[X_{\ell_2}, O_{\ell_2^+}\right]^2 O_{\ell_2^+}^2)\right]\\
    &= \frac{d^2}{(d^2-1)^2}\left[\tr(O)^2\tr(O^2)-3d\tr(O)\tr(O^3)+d\tr(O^2)^2+d^2\tr(O^4)\right].
\end{align}

The ninth item is
\begin{align}
    &\int_{{\calU_{\rm Haar}}} dU_{\ell_1\shortto\ell_2}dU_{\ell_2^+}\tr\left(\left[X_{\ell_1}, O_{\ell_1^+}\right]U_{\ell_1\shortto\ell_2}^\dagger \left[X_{\ell_2}, O_{\ell_2^+}\right]U_{\ell_1\shortto\ell_2}[X_{\ell_1},U_{\ell_1\shortto\ell_2}^\dagger \left[X_{\ell_2}, O_{\ell_2^+}\right]U_{\ell_1\shortto\ell_2}]U_{\ell_1\shortto\ell_2}^\dagger O_{\ell_2^+}U_{\ell_1\shortto\ell_2}\right)\nonumber\\
    &= \int_{{\calU_{\rm Haar}}} dU_{\ell_1\shortto\ell_2}dU_{\ell_2^+}\left[\tr\left(X_{\ell_1}U_{\ell_1\shortto\ell_2}^\dagger O_{\ell_2^+}\left[X_{\ell_2}, O_{\ell_2^+}\right]U_{\ell_1\shortto\ell_2}X_{\ell_1}U_{\ell_1\shortto\ell_2}^\dagger \left[X_{\ell_2}, O_{\ell_2^+}\right]O_{\ell_2^+}U_{\ell_1\shortto\ell_2}\right) \right.\nonumber\\
    & \quad \quad \quad \quad \quad \quad \quad \quad \quad  + \tr\left(X_{\ell_1}U_{\ell_1\shortto\ell_2}^\dagger \left[X_{\ell_2}, O_{\ell_2^+}\right]^2 U_{\ell_1\shortto\ell_2}X_{\ell_1}U_{\ell_1\shortto\ell_2}^\dagger O_{\ell_2^+}^2 U_{\ell_1\shortto\ell_2}\right)\nonumber\\
    &\quad \quad \quad \quad \quad \quad \quad \quad \quad - \tr\left(X_{\ell_1}U_{\ell_1\shortto\ell_2}^\dagger O_{\ell_2^+}\left[X_{\ell_2}, O_{\ell_2^+}\right]^2 U_{\ell_1\shortto\ell_2}X_{\ell_1}U_{\ell_1\shortto\ell_2}^\dagger O_{\ell_2^+}U_{\ell_1\shortto\ell_2}\right)\nonumber\\
    & \quad \quad \quad \quad \quad \quad \quad \quad \quad \left.- \tr\left(X_{\ell_1}U_{\ell_1\shortto\ell_2}^\dagger \left[X_{\ell_2}, O_{\ell_2^+}\right]U_{\ell_1\shortto\ell_2}X_{\ell_1}U_{\ell_1\shortto\ell_2}^\dagger \left[X_{\ell_2}, O_{\ell_2^+}\right]O_{\ell_2^+}^2 U_{\ell_1\shortto\ell_2}\right)\right]\\
    &= \int_{{\calU_{\rm Haar}}} dU_{\ell_2^+}\left[\frac{\tr(\left[X_{\ell_2}, O_{\ell_2^+}\right]^2 O_{\ell_2^+}^2)}{1-d^2} + \frac{d\tr(O^2)\tr(\left[X_{\ell_2}, O_{\ell_2^+}\right]^2) - \tr(\left[X_{\ell_2}, O_{\ell_2^+}\right]^2 O_{\ell_2^+}^2)}{d^2-1}\right.\nonumber\\
    & \quad \quad \quad \quad \quad \quad \left.- \frac{d\tr(O)\tr(\left[X_{\ell_2}, O_{\ell_2^+}\right]^2 O_{\ell_2^+}) - \tr(\left[X_{\ell_2}, O_{\ell_2^+}\right]^2 O_{\ell_2^+}^2)}{d^2-1} - \frac{\tr(\left[X_{\ell_2}, O_{\ell_2^+}\right]^2 O_{\ell_2^+}^2)}{1-d^2}\right]\\
    &= \int_{{\calU_{\rm Haar}}} dU_{\ell_2^+}\frac{d}{d^2-1}\left[\tr(O^2) \tr(\left[X_{\ell_2}, O_{\ell_2^+}\right]^2) - \tr(O)\tr(\left[X_{\ell_2}, O_{\ell_2^+}\right]^2 O_{\ell_2^+})\right]\\
    &= \frac{d}{d^2-1}\tr(O^2)\frac{2d[\tr(O)^2 - d\tr(O^2)]}{d^2-1} - \frac{d}{d^2-1}\tr(O)\frac{d[\tr(O)\tr(O^2)-d\tr(O^3)]}{d^2-1}\\
    &= \frac{d^2}{(d^2-1)^2}\left[\tr(O^2)\tr(O)^2 - 2d\tr(O^2)^2 + d\tr(O)\tr(O^3)\right].
\end{align}

The tenth item is
\begin{align}
    &\int_{{\calU_{\rm Haar}}} dU_{\ell_1\shortto\ell_2}dU_{\ell_2^+}\tr\left([X_{\ell_1},U_{\ell_1\shortto\ell_2}^\dagger \left[X_{\ell_2}, O_{\ell_2^+}\right]U_{\ell_1\shortto\ell_2}]U_{\ell_1\shortto\ell_2}^\dagger \left[X_{\ell_2}, O_{\ell_2^+}\right]U_{\ell_1\shortto\ell_2}\left[X_{\ell_1}, O_{\ell_1^+}\right]U_{\ell_1\shortto\ell_2}^\dagger O_{\ell_2^+}U_{\ell_1\shortto\ell_2}\right)\nonumber\\
    &= \int_{{\calU_{\rm Haar}}} dU_{\ell_1\shortto\ell_2}dU_{\ell_2^+}\left[\tr\left(X_{\ell_1}U_{\ell_1\shortto\ell_2}^\dagger \left[X_{\ell_2}, O_{\ell_2^+}\right]^2 U_{\ell_1\shortto\ell_2}X_{\ell_1}U_{\ell_1\shortto\ell_2}^\dagger O_{\ell_2^+}^2 U_{\ell_1\shortto\ell_2}\right)\right.\nonumber\\
    & \quad \quad \quad \quad \quad \quad \quad \quad \quad + \tr\left(X_{\ell_1}U_{\ell_1\shortto\ell_2}^\dagger \left[X_{\ell_2}, O_{\ell_2^+}\right] O_{\ell_2^+} U_{\ell_1\shortto\ell_2}X_{\ell_1}U_{\ell_1\shortto\ell_2}^\dagger O_{\ell_2^+} \left[X_{\ell_2}, O_{\ell_2^+}\right] U_{\ell_1\shortto\ell_2}\right)\nonumber\\
    & \quad \quad \quad \quad \quad \quad \quad \quad \quad - \tr\left(X_{\ell_1}U_{\ell_1\shortto\ell_2}^\dagger \left[X_{\ell_2}, O_{\ell_2^+}\right]^2 O_{\ell_2^+} U_{\ell_1\shortto\ell_2}X_{\ell_1}U_{\ell_1\shortto\ell_2}^\dagger O_{\ell_2^+}U_{\ell_1\shortto\ell_2}\right)\nonumber\\
    & \quad \quad \quad \quad \quad \quad \quad \quad \quad \left. - \tr\left(X_{\ell_1}U_{\ell_1\shortto\ell_2}^\dagger \left[X_{\ell_2}, O_{\ell_2^+}\right]U_{\ell_1\shortto\ell_2}X_{\ell_1}U_{\ell_1\shortto\ell_2}^\dagger O_{\ell_2^+}^2 \left[X_{\ell_2}, O_{\ell_2^+}\right]U_{\ell_1\shortto\ell_2}\right)\right]\\
    &= \int_{{\calU_{\rm Haar}}} dU_{\ell_2^+}\left[ \frac{d\tr(O^2)\tr(\left[X_{\ell_2}, O_{\ell_2^+}\right]^2) - \tr(\left[X_{\ell_2}, O_{\ell_2^+}\right]^2 O_{\ell_2^+}^2)}{d^2-1} + \frac{\tr(\left[X_{\ell_2}, O_{\ell_2^+}\right]^2 O_{\ell_2^+}^2)}{1-d^2}\right.\nonumber\\
    & \quad \quad \quad \quad \quad  \quad \left.- \frac{d\tr(O)\tr(\left[X_{\ell_2}, O_{\ell_2^+}\right]^2 O_{\ell_2^+}) - \tr(\left[X_{\ell_2}, O_{\ell_2^+}\right]^2 O_{\ell_2^+}^2)}{d^2-1} - \frac{\tr(\left[X_{\ell_2}, O_{\ell_2^+}\right]^2 O_{\ell_2^+}^2)}{1-d^2}\right]\\
    &=  \frac{d^2}{(d^2-1)^2}\left[\tr(O^2)\tr(O)^2 - 2d\tr(O^2)^2 + d\tr(O)\tr(O^3)\right].
\end{align}

The eleventh (last) item is
\begin{align}
    &\int_{{\calU_{\rm Haar}}} dU_{\ell_1\shortto\ell_2}dU_{\ell_2^+}\tr(\left[X_{\ell_1}, O_{\ell_1^+}\right][X_{\ell_1},U_{\ell_1\shortto\ell_2}^\dagger \left[X_{\ell_2}, O_{\ell_2^+}\right]U_{\ell_1\shortto\ell_2}]) \tr(\left[X_{\ell_2}, O_{\ell_2^+}\right]O_{\ell_2^+}) = 0
\end{align}
{due to} the cyclic property of trace {by expansion of commutators}.

Combing the above eleven terms, we have ${\tr(\rho_0U^\dagger O U)} \frac{\partial^2 \epsilon}{\partial \theta_{\ell_1}\partial\theta_{\ell_2}}\frac{\partial \epsilon}{\partial \theta_{\ell_1}} \frac{\partial \epsilon}{\partial \theta_{\ell_2}}$
\begin{align}
    &\mathbb{E}_{{\calU_{\rm Haar}}} \left[{\tr(\rho_0U^\dagger O U)} \frac{\partial^2 \epsilon}{\partial \theta_{\ell_1}\partial\theta_{\ell_2}}\frac{\partial \epsilon}{\partial \theta_{\ell_1}} \frac{\partial \epsilon}{\partial \theta_{\ell_2}}\right]\nonumber\\
    &= \frac{d}{8(d-1)^2(d+1)^3(d+2)(d+3)}\left[2\tr(O)^4 -3(d-1)\tr(O)^2\tr(O^2) - 3(d+1)\tr(O^2)^2 + (d^2-d+4)\tr(O)\tr(O^3)\right.\nonumber\\
    & \quad \quad \quad \quad \quad \quad \quad \quad \quad \quad \quad \quad \quad \quad \quad \left.+ (d^2-d)\tr(O^4)\right].
    \label{eq:Og12g1g2}
\end{align}

\paragraph{Summary of average dynamical index $\overline{\zeta_0}$ under Haar random ensemble \\}

Summarizing from Eq.~\eqref{eq:Ogllglgl} and~\eqref{eq:Og12g1g2}, combined with the result from Eq.~\eqref{eq:mu0} we thus have $\overline{\epsilon_0\mu_0}$ as
\begin{align}
    \overline{\epsilon_0\mu_0} &= L(L-1)\mathbb{E}_{{\calU_{\rm Haar}}}\left[{\tr(\rho_0U^\dagger O U)} \frac{\partial^2 \epsilon}{\partial \theta_{\ell_1}\partial\theta_{\ell_2}}\frac{\partial \epsilon}{\partial \theta_{\ell_1}} \frac{\partial \epsilon}{\partial \theta_{\ell_2}}\right] + L\mathbb{E}_{{\calU_{\rm Haar}}} \left[{\tr(\rho_0U^\dagger O U)} \frac{\partial^2 \epsilon}{\partial \theta_\ell^2 }\left(\frac{\partial \epsilon}{\partial \theta_\ell}\right)^2 \right] - O_0 \overline{\mu_0} \\
    &= -\frac{L}{4d(d-1)(d+1)^2 (d-2)  (d+2)^2 (d-3) (d+3)^2}\left[\left(d^4-2 d^3-9 d^2+8 d+20\right) \tr(O)^4\right.\nonumber\\
    &+2 d \left(-d^4+d^3+10 d^2+d-29\right) \tr(O^2) \tr(O)^2-4 \left(d^5-11 d^3-6 d^2+38 d+14\right) \tr(O^3)\tr(O)\nonumber\\
    &\left.+\left(d^6+3 d^5-11 d^4-41 d^3+18 d^2+96 d+60\right) \tr(O^2)^2+d \left(d^5-13 d^3-4 d^2+56 d-4\right) \tr(O^4)\right]\nonumber\\
    &+ \frac{L(L-1)d}{8 (d-1)^2 (d+1)^3 (d+2) (d+3)}\left[-2 (d+3) O_0 \tr(O)^3+3d(d+3)O_0\tr(O) \tr(O^2) +(d^2-d+4) \tr(O)\tr(O^3)\right.\nonumber\\
    &\left.+d(d-1)\tr(O^4)-d^2 (d+3) O_0 \tr(O^3)-3 (d-1) \tr(O^2) \tr(O)^2-3 (d+1) \tr(O^2)^2+2 \tr(O)^4\right].
    \label{eq:epsmu0}
\end{align}
One can then find 
\begin{align}
    \overline{\zeta_0} = \overline{\epsilon_0\mu_0}/\overline{K_0}^2,
    \label{eq:zeta0}
\end{align}
where $\overline{\epsilon_0\mu_0}$ and $\overline{K_0}$ can be found in Eq.~\eqref{eq:epsmu0} and~\eqref{eq:K0}, {though too complicated to show it completely here.}

In the asymptotic limit of $L\gg 1, d\gg 1$, Eq.~\eqref{eq:epsmu0} can be reduced to
\begin{align}
    \overline{\epsilon_0\mu_0} &\simeq -\frac{L}{4 d^{6}}\left[\tr(O)^4 -2d \tr(O^2) \tr(O)^2 -4 d \tr(O^3) \tr(O) + d^2 \tr(O^2)^2 +d^2 \tr(O^4)\right]\nonumber\\
    & \quad + \frac{L^2 }{8 d^6}\left[-2d O_0 \tr(O)^3+3d^2 O_0\tr(O) \tr(O^2) + d^2 \tr(O)\tr(O^3) + d^2 \tr(O^4)-d^3 O_0 \tr(O^3)\right.\nonumber\\
    &\quad \quad \quad \quad \left.-3d \tr(O^2) \tr(O)^2-3d \tr(O^2)^2+ 2\tr(O)^4\right].
    \label{eq:epsmu0_asym}
\end{align}
We then have the ratio $\overline{\zeta_0}$ as
\begin{align}
    \overline{\zeta_0} &= \frac{ \overline{\epsilon_0\mu_0}}{\overline{K_0}^2}\\
    &\simeq -\frac{1}{L}\frac{\tr(O)^4 -2d \tr(O^2) \tr(O)^2 -4 d \tr(O^3) \tr(O) + d^2 \tr(O^2)^2 +d^2 \tr(O^4)}{\left(d\tr(O^2) - \tr(O)^2\right)^2}\nonumber\\
    & \quad + \frac{1}{2\left(d\tr(O^2) - \tr(O)^2\right)^2}\left[-2d O_0 \tr(O)^3+3d^2 O_0\tr(O) \tr(O^2) + d^2 \tr(O)\tr(O^3) + d^2 \tr(O^4)-d^3 O_0 \tr(O^3)\right.\nonumber\\
    &\quad \quad \quad \quad \quad \quad \quad \quad \quad \quad \quad \quad \quad \left.-3d \tr(O^2) \tr(O)^2-3d \tr(O^2)^2 + 2\tr(O)^4\right].
    \label{eq:zeta0_asymp}
\end{align}
Specifically, for traceless observable, the $\overline{\zeta_0}$ can be further simplified as
\begin{align}
    \overline{\zeta_0} &= \frac{ \overline{\epsilon_0\mu_0}}{\overline{K_0}^2}\\
    &\simeq -\frac{1}{L}\frac{ \tr(O^2)^2 + \tr(O^4)}{\tr(O^2)^2} + \frac{1}{2 d \tr(O^2)^2}\left[d \tr(O^4)-d^2 O_0 \tr(O^3)-3\tr(O^2)^2\right].
    \label{eq:zeta0_traceless}
\end{align}

\subsection{Fluctuations of error and QNTK under Haar random ensemble}
\label{app:fluctuations_haar}

At the end of discussion about Haar ensemble result, we discuss the standard deviation of total error ${\rm SD}\left[\epsilon_0\right]$ and QNTK ${\rm SD}\left[K_0\right]$. To calculate the standard deviation, we first focus on the variance.

\subsubsection{Relative fluctuation of total error under Haar random ensemble}

The variance of total error $\epsilon$ over Haar random ensemble is 
\begin{align}
    {\rm Var}\left[\epsilon_0\right] &= \overline{\epsilon_0^2} - \overline{\epsilon_0}^2\\
    &= \mathbb{E}_{{\calU_{\rm Haar}}}\left[\left(\braket{O} - O_0\right)^2\right] - \left(\mathbb{E}_{{\calU_{\rm Haar}}}\left[\braket{O}\right] - O_0\right)^2\\
    &= \mathbb{E}_{{\calU_{\rm Haar}}}\left[\braket{O}^2\right] - \mathbb{E}_{{\calU_{\rm Haar}}}\left[\braket{O}\right]^2\\
    &= \frac{\tr(O)^2+\tr(O^2)}{d^2+d} - \frac{\tr(O)^2}{d^2}\\
    &= \frac{d\tr(O^2) -\tr(O)^2}{d^2(d+1)}.
\end{align}
Then the standard deviation of error is
\begin{align}
    {\rm SD}\left[\epsilon_0\right] = \sqrt{{\rm Var}\left[\epsilon_0\right]} = \sqrt{\frac{d\tr(O^2) -\tr(O)^2}{d^2(d+1)}},
\end{align}
and the relative fluctuation can be obtained directly as
\begin{align}
    \frac{{\rm SD}\left[\epsilon_0\right] }{\overline{\epsilon_0}} = \frac{1}{\tr(O)/d-O_0}\sqrt{\frac{d\tr(O^2) -\tr(O)^2}{d^2(d+1)}}.
\end{align}

Specifically, for traceless observable we have ${\rm SD}\left[\epsilon_0\right] = \sqrt{\tr(O^2)/d(d+1)}$ and ${\rm SD}\left[\epsilon_0\right]/\overline{\epsilon_0} = -\sqrt{\tr(O^2)/d(d+1)}/O_0$.

\subsubsection{Relative fluctuation of QNTK under Haar random ensemble}

The variance of $K$ is ${\rm Var}[K_0] = \overline{K_{0}^2} - \overline{K_{0}}^2$, which can be written as
\begin{align}
    {\rm Var}[K_0] &= \sum_{\ell_1,\ell_2}\mathbb{E}_{{\calU_{\rm Haar}}} \left[\left(\frac{\partial\epsilon}{\partial \theta_{\ell_1}}\right)^2 \left(\frac{\partial\epsilon}{\partial \theta_{\ell_2}}\right)^2\right] - \overline{K_{0}}^2\\
    &= 2\sum_{\ell_1<\ell_2} \mathbb{E}_{{\calU_{\rm Haar}}} \left[\left(\frac{\partial\epsilon}{\partial \theta_{\ell_1}}\right)^2 \left(\frac{\partial\epsilon}{\partial \theta_{\ell_2}}\right)^2\right] + \sum_\ell \mathbb{E}_{{\calU_{\rm Haar}}} \left[\left(\frac{\partial\epsilon}{\partial \theta_\ell}\right)^4\right]- \overline{K_{0}}^2\\
    &= L(L-1)\mathbb{E}_{{\calU_{\rm Haar}}}\left[\left(\frac{\partial\epsilon}{\partial \theta_{\ell_1}}\right)^2 \left(\frac{\partial\epsilon}{\partial \theta_{\ell_2}}\right)^2\right] + L\mathbb{E}_{{\calU_{\rm Haar}}} \left[\left(\frac{\partial\epsilon}{\partial \theta_\ell}\right)^4\right]- \overline{K_{0}}^2.
\end{align}
Under the random initialization of circuit parameters, we can calculate the variance via Haar integral as following.

\paragraph{Calculation of $\mathbb{E}\left[\left(\frac{\partial\epsilon}{\partial \theta_\ell}\right)^4\right]$ with Haar random ensemble \\}

We first evaluate $\mathbb{E}_{{\calU_{\rm Haar}}}\left[\left(\frac{\partial\epsilon}{\partial \theta_\ell}\right)^4\right]$, which can be expanded as
\begin{align}
    &\mathbb{E}_{{\calU_{\rm Haar}}}\left[\left(\frac{\partial\epsilon}{\partial \theta_\ell}\right)^4\right] = \frac{1}{16}\mathbb{E}_{{\calU_{\rm Haar}}}\left[\left(\braket{\psi_0|U_{\ell^-}^\dagger\left[X_\ell, O_{\ell^+}\right]U_{\ell^-}|\psi_0}\right)^4\right]\\
    &= \frac{1}{16} \int_{{\calU_{\rm Haar}}} dU_{\ell^-}dU_{\ell^+} \tr\left(\rho_0 U_{\ell^-}^\dagger\left[X_\ell, O_{\ell^+}\right]U_{\ell^-}\rho_0 U_{\ell^-}^\dagger\left[X_\ell, O_{\ell^+}\right]U_{\ell^-}\rho_0 U_{\ell^-}^\dagger\left[X_\ell, O_{\ell^+}\right]U_{\ell^-}\rho_0 U_{\ell^-}^\dagger\left[X_\ell, O_{\ell^+}\right]U_{\ell^-}\right)\\
    &= \int_{{\calU_{\rm Haar}}} dU_{\ell^+} \frac{3\tr([X_\ell,O_{\ell^+}]^2)^2+6\tr\left([X_\ell,O_{\ell^+}]^4\right)}{16d(d+2)(d^2+4d+3)},
\end{align}
where we evaluate the integral over $U_{\ell^-}$ by considering $U_{\ell^-}$ forms Haar random ($4$-design) ensemble.

From the expansion of $\tr([X_\ell, O_{\ell^+}]^4)$, we have
\begin{align}
    &\int_{{\calU_{\rm Haar}}} dU_{\ell^+} \tr\left([X_\ell, O_{\ell^+}]^4\right) = \nonumber\\
    &
    \int_{{\calU_{\rm Haar}}} dU_{\ell^+} \left[2\tr\left(X_\ell O_{\ell^+} X_\ell O_{\ell^+} X_\ell O_{\ell^+}X_\ell O_{\ell^+}\right)-8 \tr\left(X_\ell O_{\ell^+}^3 X_\ell O_{\ell^+}\right)+ 4 \tr\left(X_\ell O_{\ell^+}^2 X_\ell U_{\ell^+}^\dagger O_{\ell^+}^2\right)\right] +2 \tr(O^4).
\end{align}
The expansion follows from Supplementary notes of ~\cite{liu2023analytic}. With the help of \texttt{RTNI}~\cite{fukuda2019rtni}, we have the integrals as
\begin{align}
    &\int_{{\calU_{\rm Haar}}} dU_{\ell^+} \tr\left(X_\ell O_{\ell^+} X_\ell O_{\ell^+} X_\ell O_{\ell^+}X_\ell O_{\ell^+}\right)
    \nonumber
    \\
    &\qquad = \frac{2 d^2 \tr(O^2) \tr(O)^2+\left(d^2+9\right) \tr(O^4)-d \tr(O)^4-8d \tr(O^3)\tr(O)-3d \tr(O^2)^2}{d^4-10 d^2+9}\\
    & \int_{{\calU_{\rm Haar}}} dU_{\ell^+} \tr\left(X_\ell O_{\ell^+}^3 X_\ell O_{\ell^+}\right) = \frac{d\tr(O)\tr(O^3)-\tr(O^4)}{d^2-1}\\
    & \int_{{\calU_{\rm Haar}}} dU_{\ell^+} \tr\left(X_\ell O_{\ell^+}^2 X_\ell O_{\ell^+}^2\right) = \frac{d\tr(O^2)^2-\tr(O^4)}{d^2-1},
\end{align}
where $U_{\ell^+}$ {follows Haar random unitary ($4$-design)}.

The ensemble average of $\mathbb{E}_{{\calU_{\rm Haar}}}\left[\tr([X_\ell,O_{\ell^+}]^4)\right]$ is thus
\begin{align}
&\mathbb{E}_{{\calU_{\rm Haar}}}\left[\tr([X_\ell,O_{\ell^+}]^4)\right]
\nonumber
\\
&= \frac{2 d \left[\left(28-4d^2\right) \tr(O^3) \tr(O)+\left(2 d^2-21\right) \tr(O^2)^2+\left(d^3-7d\right) \tr(O^4)+2 d \tr(O^2) \tr(O)^2-\tr(O)^4\right]}{d^4-10 d^2+9}.
\end{align}

On the other hand, $\tr([X_\ell,O_{\ell^+}]^2)^2$ can be expanded as
\begin{align}
    &\int_{{\calU_{\rm Haar}}} dU_{\ell^+}\tr\left([X_\ell,O_{\ell^+}]^2\right)^2 = \sum_{i_1,i_2}\int dU_{\ell^+}\tr\left(P_{i_2,i_1}[X_\ell,O_{\ell^+}]^2 P_{i_1,i_2}[X_\ell,O_{\ell^+}]^2\right)\\
    &= \sum_{i_1,i_2} \int_{{\calU_{\rm Haar}}} dU_{\ell^+} \left[4\tr\left(P_{i_2,i_1}X_\ell O_{\ell^+} X_\ell O_{\ell^+} P_{i_1,i_2}X_\ell O_{\ell^+} X_\ell O_{\ell^+}\right)
    \nonumber \right.
    \\
    &\quad \quad \left.-8 \tr\left(P_{i_2,i_1}X_\ell O_{\ell^+} X_\ell O_{\ell^+} P_{i_1,i_2}O_{\ell^+}^2\right)+4\tr\left(P_{i_2,i_1}O_{\ell^+}^2 P_{i_1,i_2}O_{\ell^+}^2 \right)\right].
\end{align}
where {again in the first equation} we introduce $P_{i_1,i_2} = \ketbra{i_1}{i_2}$ {to combine the two traces together} such that it can be evaluated by \texttt{RTNI}~\cite{fukuda2019rtni}. Similar to above calculation, we then can find
\begin{align}
    &\sum_{i_1,i_2} \int_{{\calU_{\rm Haar}}} dU_{\ell^+} \tr\left(P_{i_2,i_1}X_\ell O_{\ell^+} X_\ell O_{\ell^+} P_{i_1,i_2}X_\ell O_{\ell^+} X_\ell O_{\ell^+}\right)\nonumber\\
    &= \frac{\left(d^2-6\right) \tr(O)^4+\left(2 d^2-9\right) \tr(O^2)^2-6 d \tr(O^2) \tr(O)^2-6 d \tr(O^4)+24 \tr(O^3) \tr(O)}{d^4-10 d^2+9}\\
    &\sum_{i_1,i_2} \int_{{\calU_{\rm Haar}}} dU_{\ell^+} \tr\left(P_{i_2,i_1}X_\ell O_{\ell^+} X_\ell O_{\ell^+} P_{i_1,i_2}O_{\ell^+}^2\right) = \frac{\tr(O^2)\left(d \tr(O)^2-\tr(O^2)\right)}{d^2-1}\\
    &\sum_{i_1,i_2} \int_{{\calU_{\rm Haar}}} dU_{\ell^+} \tr\left(P_{i_2,i_1}O_{\ell^+}^2 P_{i_1,i_2}O_{\ell^+}^2 \right) = \tr(O^2)^2.
\end{align}
Combining the above three equations, the average over $\tr([X_\ell,O_{\ell^+}]^2)^2$ is
\begin{align}
&\mathbb{E}_{{\calU_{\rm Haar}}}\left[\tr([X_\ell,O_{\ell^+}]^2)^2\right] = 
\nonumber
\\
&\qquad
\frac{4 \left[\left(d^2-6\right) \tr(O)^4-2\left(d^3-6d\right) \tr(O^2) \tr(O)^2+\left(d^4-6 d^2-18\right) \tr(O^2)^2-6 d \tr(O^4)+24 \tr(O^3) \tr(O)\right]}{d^4-10 d^2+9}.
\end{align}

Therefore, the ensemble average of $\left(\frac{\partial\epsilon}{\partial \theta_l}\right)^4$ is
\begin{align}
    \mathbb{E}_{{\calU_{\rm Haar}}} \left[\left(\frac{\partial\epsilon}{\partial \theta_l}\right)^4\right] &= \int_{{\calU_{\rm Haar}}} dU_{\ell^+} \frac{3\tr([X_\ell,O_{\ell^+}]^2)^2+6\tr\left([X_\ell,O_{\ell^+}]^4\right)}{16d(d+2)(d^2+4d+3)}\\
    &= \frac{3 \left[(d^2+3d+3) \tr(O^2)^2+d (d+1) \tr(O^4)+\tr(O)^4-2 d \tr(O^2) \tr(O)^2-4 (d+1) \tr(O^3) \tr(O)\right]}{4 (d-1) d (d+1)^2 (d+3)^2}\\
    &\simeq \frac{3\tr(O^2)^2}{4d^4} + \frac{3\tr(O^4)}{4d^4} + \frac{3\tr(O)^4}{4d^6} - \frac{3\tr(O^2)\tr(O)^2}{2d^5} - \frac{3\tr(O^3)\tr(O)}{d^5},
    \label{eq:gl_4}
\end{align}
where we still approximate by $d\gg 1$.

\paragraph{Calculation of $\mathbb{E}_{{\calU_{\rm Haar}}}\left[\left(\frac{\partial\epsilon}{\partial \theta_{\ell_1}}\right)^2 \left(\frac{\partial\epsilon}{\partial \theta_{\ell_2}}\right)^2\right]$ with Haar random ensemble \\}

For the evaluation of $\mathbb{E}_{{\calU_{\rm Haar}}}\left[\left(\frac{\partial\epsilon}{\partial \theta_{\ell_1}}\right)^2 \left(\frac{\partial\epsilon}{\partial \theta_{\ell_2}}\right)^2\right]$, there are four different unitaries $U_{\ell_1^-},U_{\ell_2^-}, U_{\ell_1^+},U_{\ell_2^+}$ assuming $\ell_1<\ell_2$. 
Note that $U_{\ell_2^-} = U_{\ell_1\shortto\ell_2}U_{\ell_1^-}$ and $U_{\ell_1^+} = U_{\ell_2^+}U_{\ell_1\shortto\ell_2}$ are not fully independent, so the Haar average has to be performed in with respect to $U_{\ell^-},U_{\ell_1\shortto\ell_2},U_{\ell_2^+}$ individually only.
\begin{align}
    &\mathbb{E}_{{\calU_{\rm Haar}}} \left[\left(\frac{\partial\epsilon}{\partial \theta_{\ell_1}}\right)^2\left(\frac{\partial\epsilon}{\partial \theta_{\ell_2}}\right)^2\right]\nonumber\\
    &= \frac{1}{16}\int_{{\calU_{\rm Haar}}} dU_{\ell_1^-}dU_{\ell_1\shortto\ell_2}dU_{\ell_2^+}\braket{\psi_0|U^\dagger_{\ell_1^-}\left[X_{\ell_1}, O_{\ell_1^+}\right]U_{\ell_1^-}|\psi_0}^2\braket{\psi_0|U^\dagger_{\ell_1^-}U_{\ell_1\shortto\ell_2}^\dagger\left[X_{\ell_2}, O_{\ell_2^+}\right]U_{\ell_1 \shortto \ell_2}U_{\ell_1^-}|\psi_0}^2\\
    &= \frac{1}{16}\int_{{\calU_{\rm Haar}}} dU_{\ell_1^-}dU_{\ell_1\shortto\ell_2}dU_{\ell_2^+}\tr\left(\rho_0 U^\dagger_{\ell_1^-}\left[X_{\ell_1}, O_{\ell_1^+}\right]U_{\ell_1^-}\rho_0 U^\dagger_{\ell_1^-}\left[X_{\ell_1}, O_{\ell_1^+}\right]U_{\ell_1^-} \rho_0 U^\dagger_{\ell_1^-}U_{\ell_1\shortto\ell_2}^\dagger \left[X_{\ell_2}, O_{\ell_2^+}\right] U_{\ell_1\shortto\ell_2}U_{\ell_1^-}\right.\nonumber\\
    &\left. \cdot \rho_0 U^\dagger_{\ell_1^-}U_{\ell_1\shortto\ell_2}^\dagger \left[X_{\ell_2}, O_{\ell_2^+}\right] U_{\ell_1\shortto\ell_2}U_{\ell_1^-}\right)\\
    &= \frac{1}{16 d (d+2) \left(d^2+4 d+3\right)} 
    \nonumber
    \\
    &\quad \times\int_{{\calU_{\rm Haar}}} dU_{\ell_1\shortto\ell_2}dU_{\ell_2^+}\left[\tr\left(\left[X_{\ell_1}, O_{\ell_1^+}\right]^2\right) \tr\left(\left[X_{\ell_1}, O_{\ell_2^+}\right]^2\right)+2\tr\left(U_{\ell_1\shortto\ell_2}^\dagger \left[X_{\ell_1}, O_{\ell_2^+}\right] U_{\ell_1\shortto\ell_2}\left[X_{\ell_1}, O_{\ell_1^+}\right]\right)^2\right.\nonumber\\
    &\qquad +4 \tr\left(\left[X_{\ell_1}, O_{\ell_1^+}\right]^2 U_{\ell_1\shortto\ell_2}^\dagger \left[X_{\ell_1}, O_{\ell_2^+}\right]^2 U_{\ell_1\shortto\ell_2}\right)
    \nonumber
    \\
    &\qquad \left.
    +\tr\left(U_{\ell_1\shortto\ell_2}^\dagger \left[X_{\ell_1}, O_{\ell_2^+}\right] U_{\ell_1\shortto\ell_2}\left[X_{\ell_1}, O_{\ell_1^+}\right] U_{\ell_1\shortto\ell_2}^\dagger \left[X_{\ell_1}, O_{\ell_2^+}\right] U_{\ell_1\shortto\ell_2}\left[X_{\ell_1}, O_{\ell_1^+}\right]\right)\right],
\end{align}
{where in the first equation we expand the derivative and write $U_{\ell_2^-}$ and $U_{\ell_1^+}$ in terms of $U_{\ell_1\shortto \ell_2}, U_{\ell_1^-}, U_{\ell_2^+}$.}
Next we evaluate the average over $U_{\ell_1\shortto\ell_2}$ and $U_{\ell_2^+}$ of each term separately.

The average of first term, $\tr(\left[X_{\ell_1}, O_{\ell_1^+}\right]^2) \tr(\left[X_{\ell_2}, O_{\ell_2^+}\right]^2)$, becomes
\begin{align}
    &\int_{{\calU_{\rm Haar}}} dU_{\ell_1\shortto\ell_2} dU_{\ell_2^+}\tr(\left[X_{\ell_1}, O_{\ell_1^+}\right]^2) \tr(\left[X_{\ell_2}, O_{\ell_2^+}\right]^2)\\
    &= \int_{{\calU_{\rm Haar}}} dU_{\ell_2^+}2\tr(\left[X_{\ell_2}, O_{\ell_2^+}\right]^2)\left[\frac{d \tr(O)^2-\tr(O^2)}{d^2-1} - \tr(O^2)\right]\\
    &=4\left[\frac{d \tr(O)^2-\tr(O^2)}{d^2-1} - \tr(O^2)\right]\left[\frac{d \tr(O)^2-\tr(O^2)}{d^2-1} - \tr(O^2)\right]\\
    &= \frac{4d^2\left[\tr(O)^2-d\tr(O^2)\right]^2}{(d^2-1)^2}.
\end{align}

The average of second term, $\tr(U_{\ell_1\shortto\ell_2}^\dagger \left[X_{\ell_2}, O_{\ell_2^+}\right] U_{\ell_1\shortto\ell_2}\left[X_{\ell_1}, O_{\ell_1^+}\right])^2$ is
\small
\begin{align}
    &\int_{{\calU_{\rm Haar}}} dU_{\ell_1\shortto\ell_2} d U_{\ell_2^+} \tr(U_{\ell_1\shortto\ell_2}^\dagger \left[X_{\ell_2}, O_{\ell_2^+}\right] U_{\ell_1\shortto\ell_2}\left[X_{\ell_1}, O_{\ell_1^+}\right])^2\nonumber\\
    &= \sum_{i_1,i_2}\int_{{\calU_{\rm Haar}}} dU_{\ell_1\shortto\ell_2}dU_{\ell_2^+} \tr\left(P_{i_2,i_1}U_{\ell_1\shortto\ell_2}^\dagger \left[X_{\ell_2}, O_{\ell_2^+}\right] U_{\ell_1\shortto\ell_2}\left[X_{\ell_1}, O_{\ell_1^+}\right] P_{i_1,i_2}U_{\ell_1\shortto\ell_2}^\dagger \left[X_{\ell_2}, O_{\ell_2^+}\right] U_{\ell_1\shortto\ell_2}\left[X_{\ell_1}, O_{\ell_1^+}\right] \right)\\
    &= \sum_{i_1,i_2}\int_{{\calU_{\rm Haar}}} dU_{\ell_1\shortto\ell_2}dU_{\ell_2^+} \left[\tr\left(O_{\ell_2^+}\left[X_{\ell_2}, O_{\ell_2^+}\right] U_{\ell_1\shortto\ell_2}X_{\ell_1}U_{\ell_1\shortto\ell_2}^\dagger\right)^2 + \tr\left(\left[X_{\ell_2}, O_{\ell_2^+}\right] O_{\ell_2^+}U_{\ell_1\shortto\ell_2}X_{\ell_1}U_{\ell_1\shortto\ell_2}^\dagger\right)^2\right.\nonumber\\
    &\qquad \qquad \qquad \qquad \qquad \quad \left.- 2 \tr\left(O_{\ell_2^+}\left[X_{\ell_2}, O_{\ell_2^+}\right]U_{\ell_1\shortto\ell_2}X_{\ell_1}U_{\ell_1\shortto\ell_2}^\dagger\right)\tr\left(\left[X_{\ell_2}, O_{\ell_2^+}\right] O_{\ell_2^+}U_{\ell_1\shortto\ell_2}X_{\ell_1}U_{\ell_1\shortto\ell_2}^\dagger\right)\right]\\
    &= \int_{{\calU_{\rm Haar}}} dU_{\ell_2^+} 2\left[\frac{d \tr(O_{\ell_2^+}\left[X_{\ell_2}, O_{\ell_2^+}\right] O_{\ell_2^+}\left[X_{\ell_2}, O_{\ell_2^+}\right])-\tr(O_{\ell_2^+}\left[X_{\ell_2}, O_{\ell_2^+}\right])^2}{d^2-1} + \frac{\tr(O_{\ell_2^+}\left[X_{\ell_2}, O_{\ell_2^+}\right])^2-d \tr(\left[X_{\ell_2}, O_{\ell_2^+}\right]^2 O_{\ell_2^+}^2)}{d^2-1}\right]\\
    &= \int_{{\calU_{\rm Haar}}} dU_{\ell_2^+} 2d\frac{\tr(O_{\ell_2^+}\left[X_{\ell_2}, O_{\ell_2^+}\right] O_{\ell_2^+}\left[X_{\ell_2}, O_{\ell_2^+}\right])-\tr(\left[X_{\ell_2}, O_{\ell_2^+}\right]^2 O_{\ell_2^+}^2)}{d^2-1}\\
    &= \frac{2d}{d^2-1}\int_{{\calU_{\rm Haar}}} dU_{\ell_2^+} \left[2\tr\left(O_{\ell_2^+}^2 X_{\ell_2}O_{\ell_2^+}^2 X_{\ell_2}\right)-2\tr\left(O_{\ell_2^+}X_{\ell_2}O_{\ell_2^+}^3 X_{\ell_2}\right)
    - 2\tr\left(X_{\ell_2}O_{\ell_2^+} X_{\ell_2}O_{\ell_2^+}^3 \right)+\tr\left(X_{\ell_2}O_{\ell_2^+}^2 X_{\ell_2}O_{\ell_2^+}^2\right)+\tr(O^4) \right]\\
    &= \frac{2d}{d^2-1}\int_{{\calU_{\rm Haar}}} dU_{\ell_2^+} \left[3\tr\left(O_{\ell_2^+}^2 X_{\ell_2}O_{\ell_2^+}^2 X_{\ell_2}\right)-4\tr\left(O_{\ell_2^+}X_{\ell_2}O_{\ell_2^+}^3 X_{\ell_2}\right)
    +\tr(O^4) \right]\\
    &= \frac{2 d^2 \left[d \tr(O^4)+3 \tr(O^2)^2-4 \tr(O) \tr(O^3)\right]}{\left(d^2-1\right)^2}.
\end{align}
\normalsize

The average of third term, $\tr(\left[X_{\ell_1}, O_{\ell_1^+}\right]^2 U_{\ell_1\shortto\ell_2}^\dagger \left[X_{\ell_2}, O_{\ell_2^+}\right]^2 U_{\ell_1\shortto\ell_2})$ is
\small
\begin{align}
    &\int_{{\calU_{\rm Haar}}} dU_{\ell_1\shortto\ell_2}dU_{\ell_2^+} \tr(\left[X_{\ell_1}, O_{\ell_1^+}\right]^2 U_{\ell_1\shortto\ell_2}^\dagger \left[X_{\ell_2}, O_{\ell_2^+}\right]^2 U_{\ell_1\shortto\ell_2})\nonumber\\
    &= \int_{{\calU_{\rm Haar}}} dU_{\ell_1\shortto\ell_2}dU_{\ell_2^+}\left[\tr(X_{\ell_1}O_{\ell_1^+}X_{\ell_1}U_{\ell_1\shortto\ell_2}^\dagger O_{\ell_2^+} \left[X_{\ell_2}, O_{\ell_2^+}\right]^2 U_{\ell_1\shortto\ell_2})+\tr(X_{\ell_1} O_{\ell_1^+}X_{\ell_1}U_{\ell_1\shortto\ell_2}^\dagger \left[X_{\ell_2}, O_{\ell_2^+}\right]^2 O_{\ell_2^+}  U_{\ell_1\shortto\ell_2})\right.\nonumber\\
    &\qquad \qquad \qquad \qquad \qquad \left.- \tr(X_{\ell_1}O_{\ell_1^+}^2 X_{\ell_1}U_{\ell_1\shortto\ell_2}^\dagger \left[X_{\ell_2}, O_{\ell_2^+}\right]^2 U_{\ell_1\shortto\ell_2})-\tr(O_{\ell_2^+}^2\left[X_{\ell_2}, O_{\ell_2^+}\right]^2)\right]\\
    &= \int_{{\calU_{\rm Haar}}} dU_{\ell_2^+}\left[2\frac{d \tr(O) \tr(\left[X_{\ell_2}, O_{\ell_2^+}\right]^2 O_{\ell_2^+})-\tr(\left[X_{\ell_2}, O_{\ell_2^+}\right]^2O_{\ell_2^+}^2)}{d^2-1} - \frac{d \tr(\left[X_{\ell_2}, O_{\ell_2^+}\right]^2) \tr(O^2)-\tr(\left[X_{\ell_2}, O_{\ell_2^+}\right]^2 O_{\ell_2^+}^2)}{d^2-1}\right.\nonumber\\
    &\qquad \qquad \qquad \quad \left. -\tr(O_{\ell_2^+}^2 \left[X_{\ell_2}, O_{\ell_2^+}\right]^2)\right]\\
    &= \int_{{\calU_{\rm Haar}}} dU_{\ell_2^+}\frac{d \left[-d \tr(\left[X_{\ell_2}, O_{\ell_2^+}\right]^2 O_{\ell_2^+}^2) + 2 \tr(O) \tr(\left[X_{\ell_2}, O_{\ell_2^+}\right]^2 O_{\ell_2^+})-\tr(\left[X_{\ell_2}, O_{\ell_2^+}\right]^2) \tr(O^2)\right]}{d^2-1}\\
    &= \frac{d}{d^2-1}\left[d\frac{d \left[d \tr(O^4)+\tr(O^2)^2-2 \tr(O) \tr(O^3)\right]}{d^2-1} + 2\tr(O)\frac{d [\tr(O) \tr(O^2)-d \tr(O^3)]}{d^2-1} -\tr(O^2) \frac{2 d \left[\tr(O)^2-d \tr(O^2)\right]}{d^2-1}\right]\\
    &= \frac{d^3 \left[d \tr(O^4)+3 \tr(O^2)^2-4 \tr(O)\tr(O^3)\right]}{\left(d^2-1\right)^2}
\end{align}
\normalsize

The average over $U_{\ell_1\shortto\ell_2}$ in the last term, $\tr(U_{\ell_1\shortto\ell_2}^\dagger \left[X_{\ell_2}, O_{\ell_2^+}\right] U_{\ell_1\shortto\ell_2} \left[X_{\ell_1}, O_{\ell_1^+}\right] U_{\ell_1\shortto\ell_2}^\dagger \left[X_{\ell_2}, O_{\ell_2^+}\right] U_{\ell_1\shortto\ell_2}\left[X_{\ell_1}, O_{\ell_1^+}\right])$ is
\begin{align}
    &\int_{{\calU_{\rm Haar}}} dU_{\ell_1\shortto\ell_2}dU_{\ell_2^+}\tr(U_{\ell_1\shortto\ell_2}^\dagger \left[X_{\ell_2}, O_{\ell_2^+}\right] U_{\ell_1\shortto\ell_2} \left[X_{\ell_1}, O_{\ell_1^+}\right] U_{\ell_1\shortto\ell_2}^\dagger \left[X_{\ell_2}, O_{\ell_2^+}\right] U_{\ell_1\shortto\ell_2}\left[X_{\ell_1}, O_{\ell_1^+}\right])\nonumber\\
    &= \int_{{\calU_{\rm Haar}}} dU_{\ell_1\shortto\ell_2}dU_{\ell_2^+}\left[\tr\left(X_{\ell_1}U_{\ell_1\shortto\ell_2}^\dagger O_{\ell_2^+}\left[X_{\ell_2}, O_{\ell_2^+}\right] U_{\ell_1\shortto\ell_2}X_{\ell_1}U_{\ell_1\shortto\ell_2}^\dagger O_{\ell_2^+} \left[X_{\ell_2}, O_{\ell_2^+}\right] U_{\ell_1\shortto\ell_2} \right) \right.\nonumber\\
    & \qquad \qquad \qquad \qquad +\tr\left(X_{\ell_1}U_{\ell_1\shortto\ell_2}^\dagger \left[X_{\ell_2}, O_{\ell_2^+}\right] O_{\ell_2^+}U_{\ell_1\shortto\ell_2}X_{\ell_1}U_{\ell_1\shortto\ell_2}^\dagger \left[X_{\ell_2}, O_{\ell_2^+}\right]O_{\ell_2^+}  U_{\ell_1\shortto\ell_2} \right)\nonumber\\
    & \qquad \qquad \qquad \qquad \left.- 2\tr\left(O_{\ell_2^+}\left[X_{\ell_2}, O_{\ell_2^+}\right] O_{\ell_2^+} U_{\ell_1\shortto\ell_2}X_{\ell_1}U_{\ell_1\shortto\ell_2}^\dagger  \left[X_{\ell_2}, O_{\ell_2^+}\right]U_{\ell_1\shortto\ell_2}X_{\ell_1}U_{\ell_1\shortto\ell_2}^\dagger \right)\right]\\
    &= \int_{{\calU_{\rm Haar}}} dU_{\ell_2^+}2\left[\frac{d \tr(\left[X_{\ell_2}, O_{\ell_2^+}\right]O_{\ell_2^+})^2-\tr(\left[X_{\ell_2}, O_{\ell_2^+}\right]O_{\ell_2^+}\left[X_{\ell_2}, O_{\ell_2^+}\right]O_{\ell_2^+})}{d^2-1} + \frac{\tr(O_{\ell_2^+}\left[X_{\ell_2}, O_{\ell_2^+}\right]O_{\ell_2^+}\left[X_{\ell_2}, O_{\ell_2^+}\right])}{d^2-1}\right]\\
    &= \int_{{\calU_{\rm Haar}}} dU_{\ell_2^+}\frac{2 d \tr(\left[X_{\ell_2}, O_{\ell_2^+}\right]O_{\ell_2^+})^2}{d^2-1}\\
    &= 0,
\end{align}
where the last line can be found by the cyclic property of trace directly.

Therefore, we can conclude on $\mathbb{E}_{{\calU_{\rm Haar}}} \left[\left(\frac{\partial\epsilon}{\partial \theta_{\ell_1}}\right)^2\left(\frac{\partial\epsilon}{\partial \theta_{\ell_2}}\right)^2\right]$ is
\begin{align}
    &\mathbb{E}_{{\calU_{\rm Haar}}} \left[\left(\frac{\partial\epsilon}{\partial \theta_{\ell_1}}\right)^2\left(\frac{\partial\epsilon}{\partial \theta_{\ell_2}}\right)^2\right]\nonumber\\
    &= \frac{1}{16 d (d+2) \left(d^2+4 d+3\right)} \left(\frac{4d^2\left[\tr(O)^2-d\tr(O^2)\right]^2}{(d^2-1)^2}+2\frac{2 d^2 \left[d \tr(O^4)+3 \tr(O^2)^2-4 \tr(O) \tr(O^3)\right]}{\left(d^2-1\right)^2}\right.\nonumber\\
    & \qquad \qquad \qquad \qquad \qquad \qquad \qquad \left.+4 \frac{d^3 \left[d \tr(O^4)+3 \tr(O^2)^2-4 \tr(O)\tr(O^3)\right]}{\left(d^2-1\right)^2}+0\right)\\
    &= \frac{d \left[(d^2+3d+3) \tr(O^2)^2+d (d+1) \tr(O^4)+\tr(O)^4-2 d \tr(O^2) \tr(O)^2-4 (d+1) \tr(O^3) \tr(O)\right]}{4 (d-1)^2 (d+1)^3 (d+2) (d+3)}\\
    &\simeq \frac{\tr(O^2)^2}{4d^4} +\frac{\tr(O^4)}{4d^4} + \frac{\tr(O)^4}{4d^6} - \frac{\tr(O^2)\tr(O)^2}{2d^5} - \frac{\tr(O^3)\tr(O)}{d^5}.
    \label{eq:g1g1g2g2}
\end{align}

\paragraph{Summary of relative fluctuation ${\rm SD}[K_0]/\overline{K_0}$ under random initialization \\}

To summarize from Eq.~\eqref{eq:gl_4} and~\eqref{eq:g1g1g2g2}, the ensemble average of ${\rm Var}[K_0]$ is
\begin{align}
    {\rm Var}[K_0] &= L(L-1)\mathbb{E}_{{\calU_{\rm Haar}}} \left[\left(\frac{\partial\epsilon}{\partial \theta_{\ell_1}}\right)^2 \left(\frac{\partial\epsilon}{\partial \theta_{\ell_2}}\right)^2\right] + L\mathbb{E}_{{\calU_{\rm Haar}}} \left[\left(\frac{\partial\epsilon}{\partial \theta_\ell}\right)^4\right] - \overline{K_0}^2 \\
    &= L(L-1)\frac{d \left[(d^2+3d+3) \tr(O^2)^2+d (d+1) \tr(O^4)+\tr(O)^4-2 d \tr(O^2) \tr(O)^2-4 (d+1) \tr(O^3) \tr(O)\right]}{4 (d-1)^2 (d+1)^3 (d+2) (d+3)}\nonumber\\
    &+ L\frac{3 \left[(d^2+3d+3) \tr(O^2)^2+d (d+1) \tr(O^4)+\tr(O)^4-2 d \tr(O^2) \tr(O)^2-4 (d+1) \tr(O^3) \tr(O)\right]}{4 (d-1) d (d+1)^2 (d+3)^2}\nonumber\\
    &- \left(L\frac{d\tr(O^2)-\tr(O)^2}{2 (d-1) (d+1)^2}\right)^2.
    \label{eq:Kvar0}
\end{align}
The relative fluctuation is then
\begin{align}
    {\rm SD}[K_0]/\overline{K_0} = \sqrt{{\rm Var}[K_0]}/\overline{K_0},
    \label{eq:Kfluc0}
\end{align}
where ${\rm Var}[K_0]$ and $\overline{K_0}$ can be found in Eq.~\eqref{eq:Kvar0} and~\eqref{eq:K0}.

In the asymptotic limit of $L, d \gg 1$, we have
\begin{align}
    {\rm Var}[K_0] &\simeq \frac{L^2+3L}{4d^6}\left[d^2 \tr(O^2)^2+d^2 \tr(O^4)+\tr(O)^4-2 d \tr(O^2) \tr(O)^2-4 d \tr(O^3) \tr(O)\right]
    - L^2\frac{\left(d\tr(O^2)-\tr(O)^2\right)^2}{4 d^6}.
\end{align}
Thus we have the standard deviation of QNTK as
\begin{align}
    &{\rm SD}[K_0] \simeq \nonumber
    \\
    &\frac{1}{2d^3}\left((L^2+3L)\left[d^2 \tr(O^2)^2+d^2 \tr(O^4)+\tr(O)^4-2 d \tr(O^2) \tr(O)^2-4 d \tr(O^3) \tr(O)\right]
    - L^2\left(d\tr(O^2)-\tr(O)^2\right)^2\right)^{\frac{1}{2}}.
    \label{eq:Kstd0_asymp}
\end{align}
and the relative fluctuation is
\begin{align}
    &\frac{{\rm SD}[K_0]}{\overline{K_0}} \simeq \nonumber
    \\
    &\left((L^2+3L)\left[d^2 \tr(O^2)^2+d^2 \tr(O^4)+\tr(O)^4-2 d \tr(O^2) \tr(O)^2-4 d \tr(O^3) \tr(O)\right]
    - L^2\left(d\tr(O^2)-\tr(O)^2\right)^2\right)^{1/2}\nonumber\\
    &\quad  \times \frac{1}{L\left(d\tr(O^2)-\tr(O)^2\right)}.
\end{align}

For traceless observable $O$, the standard deviation and relative fluctuation are reduced to 
\begin{align}
    {\rm SD}[K_0] &\simeq \frac{1}{2d^2}\sqrt{L^2 \tr(O^4) + 3L \tr(O^2)^2} \label{eq:Kstd0_traceless},\\
    \frac{{\rm SD}[K_0]}{\overline{K_0}} 
    &\simeq \frac{1}{\sqrt{L}} \sqrt{ L \frac{\tr(O^4)}{\tr(O^2)^2} + 3}.
\end{align}


\textit{Remark}

A further simplification is considered in~\cite{liu2023analytic} by treating the four unitaries $U_{\ell_1^-}, U_{\ell_1,L}, U_{1,\ell_2-1}, U_{\ell_2^+}$ that appears in $\mathbb{E}\left[\left(\frac{\partial\epsilon}{\partial \theta_{\ell_1}}\right)^2\left(\frac{\partial\epsilon}{\partial \theta_{\ell_2}}\right)^2\right]$ are independent sampled from Haar random, which leads to the scaling of ${\rm SD}[K_0] \sim \sqrt{L}$ only.

\section{Results with restricted Haar ensemble}
\label{app:rh_result}


We first review the restricted Haar random ensemble here. Recall the unitary in the restricted Haar ensemble {$\calU_{\rm RH}$} is defined as (see App.~\ref{app:rst_def})
\begin{align}
    U_{\rm RH} = 
    \begin{pmatrix}
    1 & {\bm 0} \\
    {\bm 0} & V
\end{pmatrix},
\end{align}
where $V$ is a unitary of dimension $d-1$, and in restricted ensemble, we assume $V$ follows Haar random ensemble.
As the overall circuit unitary $U_{\rm RH}=U_{\ell^-}U_{\ell ^+}=U_{{\ell_1}^-}U_{\ell_1 \shortto \ell_2}U_{\ell_2^+}$ (see definitions around Eqs.~\eqref{eq:U_ell_def},~\eqref{eq:U_1to2_def}), the form of Eq.~\eqref{eq:toyU_app} will lead to constraint on the unitaries $U_{\ell^-}, U_{\ell ^+}, U_{{\ell_1}^-}, U_{\ell_1 \shortto \ell_2}, U_{\ell_2^+}$ utilized in evaluation. It turns out that the specific distribution of $V$ for the overall unitary ensemble $\{U_{\rm RH}\}$ does not affect the ensemble averages and thus we do not specify the distribution of $V$. To implement the constraint, we can assume that the unitaries of a part of circuit including $U_{\ell^-}, U_{\ell_1^-}, U_{\ell_1\shortto \ell_2} \sim \calU_{\rm Haar}(d)$ follow independent Haar random distribution; while $U_{\ell^+}, U_{\ell_2^+}$ are directly determined by $U_{\ell ^+} = U_{\rm RH} U_{\ell^-}^\dagger$ and $U_{\ell_2^+} = U_{\rm RH} U_{\ell^-}^\dagger U_{\ell_1 \shortto \ell_2}^\dagger$ due to the constraint.

To prepare a target state $O = \ketbra{\Phi}{\Phi}$, we can keep the loss function in Eq.~\eqref{loss_function} in the main text with a target value $O_0 \ge 0$, and the total error as
\begin{align}
    \epsilon = |\braket{\Phi|U|\psi_0}|^2 - O_0.
\end{align}
As fidelity between arbitrary two quantum states is bounded from zero to unity, when $0 < O_0 <1$, we expect the error can be decreased to zero with sufficiently large $L$, leading to the \textit{frozen kernel dynamics}; On the other hand if $O_0 >1$, the error can only be optimized to a negative constant leading to the \textit{frozen error dynamics}; $O_0=1$ will become the critical point. To capture late-time dynamics, we have the fidelity
\begin{align}
    F_0 = |\braket{\Phi|U|\psi_0}|^2 ={ O_0 + R},
    \label{eq:F0_relation}
\end{align}
{where $R = \lim_{t\to \infty} \epsilon(t) = \min\{1-O_0,0\}$ which is consistent with definition in the main text}.

\subsection{Average QNTK under restricted Haar ensemble}

We can evaluate average of  QNTK $\overline{K_\infty}$ under restricted Haar ensemble. Recall that the QNTK is defined as $\overline{K_{\infty}} = \sum_{\ell} \mathbb{E}_{{\calU_{\rm RH}}}\left[\left(\frac{\partial \epsilon}{\partial \theta_{\ell}}\right)^2\right]$, thus we have
\begin{align}
    &\mathbb{E}_{{\calU_{\rm RH}}}\left[\left(\frac{\partial\epsilon}{\partial \theta_\ell}\right)^2\right]
    = -\frac{1}{4}\int dU_{\ell^-}dU_{\ell^+} \tr\left(\rho_0 U_{\ell^-}^\dagger\left[X_\ell, O_{\ell^+}\right]U_{\ell^-}\rho_0 U_{\ell^-}^\dagger\left[X_\ell, O_{\ell^+}\right]U_{\ell^-}\right)\\
    &= -\frac{1}{4}\int_{{\calU_{\rm RH}}}dU_{\rm RH}\int_{{\calU_{\rm Haar}}} dU_{\ell^-} \left[\tr\left(\rho_0 U_{\ell^-}^\dagger X_\ell U_{\ell^-} O_{\rm RH} \rho_0 U_{\ell^-}^\dagger X_\ell U_{\ell^-} O_{\rm RH}\right) + \tr\left(\rho_0 O_{\rm RH} U_{\ell^-}^\dagger X_\ell U_{\ell^-}  \rho_0 O_{\rm RH} U_{\ell^-}^\dagger X_\ell U_{\ell^-} \right)\right.\nonumber\\
    &\left. \qquad  \qquad \qquad \qquad \qquad \qquad - \tr\left(\rho_0 U_{\ell^-}^\dagger X_\ell U_{\ell^-} O_{\rm RH}   \rho_0 O_{\rm RH} U_{\ell^-}^\dagger X_\ell U_{\ell^-} \right) - \tr\left(\rho_0 O_{\rm RH} U_{\ell^-}^\dagger X_\ell U_{\ell^-} \rho_0 U_{\ell^-}^\dagger X_\ell U_{\ell^-} O_{\rm RH} \right)\right]\\
    &= -\frac{2}{4}\int_{{\calU_{\rm RH}}} dU_{\rm RH} \left[\frac{d\tr(\rho_0 O_{\rm RH})^2 - \tr(\rho_0 O_{\rm RH} \rho_0 O_{\rm RH})}{d^2 - 1} - \frac{d\tr(O_{\rm RH} \rho_0 O_{\rm RH}) - \tr(\rho_0 O_{\rm RH} \rho_0 O_{\rm RH})}{d^2 - 1}\right]\\
    &= \int_{{\calU_{\rm RH}}} dU_{\rm RH}  \frac{2d\left[\tr(\rho_0 O_{\rm RH})^2 - \tr(O_{\rm RH} \rho_0 O_{\rm RH})\right]}{4(d^2 - 1)}\\
    &= \frac{d F_0 (1-F_0)}{2(d^2 - 1)},
    \label{eq:glgl_rh}
\end{align}
where $O_{\rm RH} = U_{\rm RH}^\dagger O U_{\rm RH}$ is defined for simplicity. In the last line, we utilize the fact that $\tr(\rho_0 O_{\rm RH})^2 = F_0^2$ and $\tr(O_{\rm RH} \rho_0 O_{\rm RH}) = F_0$. Thus the QNTK is
\begin{align}
    \overline{K_\infty} &= L\mathbb{E}_{{\calU_{\rm RH}}} \left[\left(\frac{\partial\epsilon}{\partial \theta_\ell}\right)^2\right] = \frac{L d F_0 (1-F_0)}{2(d^2 - 1)}\\
    &= \frac{Ld}{2(d^2 - 1)} \left(O_0 {+ R}\right)\left(1-O_0 {-R}\right) \simeq \frac{L}{2d}\left(O_0 {+ R}\right)\left(1-O_0 {-R}\right),
    \label{eq:K_rh}
\end{align}
where in the last equation we utilize the relation between $F_0$ and $O_0, R$ in Eq.~\eqref{eq:F0_relation}. {The approximation in the last line is taken for $d\gg 1$ for direct identification on its scaling.} When $O_0 > 1$ with $R = 1-O_0$, we directly have $\overline{K_\infty} = 0$; when $O_0 = 1$ with $R = 0$, we also have $\overline{K_\infty} = 0$; however when $O_0 <1$ with $R =0$, we have a finite nonzero QNTK as $\overline{K_\infty} \propto O_0 (1-O_0)$ and specifically, $\overline{K_\infty} \propto O_0$ in the limit of $O_0$ close to unity.

\subsection{Average relative dQNTK under restricted Haar ensemble}

In this section, we evaluate the factor $\overline{\lambda_\infty} = \overline{\mu_{{\infty}}}/\overline{K_\infty}$ under restricted Haar ensemble. As $\overline{K_\infty}$ is already calculated above, we focus on dQNTK $\overline{\mu_{{\infty}}}$ in the following.
Recall that $\overline{\mu_{\infty}} = L(L-1) \mathbb{E}_{{\calU_{\rm RH}}} \left[\frac{\partial^2 \epsilon}{\partial \theta_{\ell_1}\partial\theta_{\ell_2}}\frac{\partial \epsilon}{\partial \theta_{\ell_1}} \frac{\partial \epsilon}{\partial \theta_{\ell_2}}\right] + L\mathbb{E}_{{\calU_{\rm RH}}} \left[\frac{\partial^2 \epsilon}{\partial\theta_\ell^2}\left(\frac{\partial \epsilon}{\partial \theta_\ell}\right)^2 \right]$, we evaluate the two ensemble average separately. 

\paragraph{$\mathbb{E}_{{\calU_{\rm RH}}}\left[\frac{\partial^2 \epsilon}{\partial\theta_\ell^2}\left(\frac{\partial \epsilon}{\partial \theta_\ell}\right)^2 \right]$ under restricted Haar ensemble \\}

We first consider $\mathbb{E}_{{\calU_{\rm RH}}} \left[\frac{\partial^2 \epsilon}{\partial\theta_\ell^2}\left(\frac{\partial \epsilon}{\partial \theta_\ell}\right)^2 \right]$.
\small
\begin{align}
    &\mathbb{E}_{{\calU_{\rm RH}}} \left[\frac{\partial^2 \epsilon}{\partial\theta_\ell^2}\left(\frac{\partial \epsilon}{\partial \theta_\ell}\right)^2 \right] =  \frac{1}{16} \int dU_{\ell^-}dU_{\ell^+} \tr\left(\rho_0 U_{\ell^-}^\dagger O_{\ell^+} U_{\ell^-}\rho_0 U_{\ell^-}^\dagger [X_\ell,[X_\ell, O_{\ell^+}]]U_{\ell^-}\rho_0 U_{\ell^-}^\dagger[X_\ell, O_{\ell^+}]U_{\ell^-}\rho_0 U_{\ell^-}^\dagger[X_\ell, O_{\ell^+}] U_{\ell^-}\right)\\
    &= \frac{2}{16}\int_{{\calU_{\rm RH}}} dU_{\rm RH} \int_{{\calU_{\rm Haar}}} dU_{\ell^-} \left[\tr\left(O_{\rm RH} \rho_0 O_{\rm RH} \rho_0 U_{\ell^+}^\dagger X_\ell U_{\ell^+} O_{\rm RH} \rho_0 U_{\ell^+}^\dagger X_\ell U_{\ell^+} \right) + \tr\left(\rho_0 O_{\rm RH} \rho_0 O_{\rm RH}   U_{\ell^+}^\dagger X_\ell U_{\ell^+}  \rho_0 O_{\rm RH} U_{\ell^+}^\dagger X_\ell U_{\ell^+} \right) 
    \right.\nonumber\\
    & - \tr\left( \rho_0 O_{\rm RH} \rho_0 U_{\ell^+}^\dagger X_\ell U_{\ell^+}  O_{\rm RH} \rho_0 O_{\rm RH} U_{\ell^+}^\dagger X_\ell U_{\ell^+} \right) - \tr\left(O_{\rm RH} \rho_0 O_{\rm RH} \rho_0 O_{\rm RH} U_{\ell^+}^\dagger X_\ell U_{\ell^+}  \rho_0 U_{\ell^+}^\dagger X_\ell U_{\ell^+} \right) \nonumber\\
    & + \tr\left(\rho_0 U_{\ell^+}^\dagger X_\ell U_{\ell^+} O_{\rm RH} U_{\ell^+}^\dagger X_\ell U_{\ell^+} \rho_0 U_{\ell^+}^\dagger X_\ell U_{\ell^+} O_{\rm RH} \rho_0 O_{\rm RH} U_{\ell^+}^\dagger X_\ell U_{\ell^+}\right)\nonumber\\
    &+ \tr\left( O_{\rm RH}  \rho_0 U_{\ell^+}^\dagger X_\ell U_{\ell^+} O_{\rm RH} U_{\ell^+}^\dagger X_\ell U_{\ell^+} \rho_0 O_{\rm RH} U_{\ell^+}^\dagger X_\ell U_{\ell^+} \rho_0 U_{\ell^+}^\dagger X_\ell U_{\ell^+}\right)\nonumber\\
    & - \tr\left( O_{\rm RH} \rho_0 U_{\ell^+}^\dagger X_\ell U_{\ell^+} O_{\rm RH} U_{\ell^+}^\dagger X_\ell U_{\ell^+} \rho_0 U_{\ell^+}^\dagger X_\ell U_{\ell^+} O_{\rm RH} \rho_0 U_{\ell^+}^\dagger X_\ell U_{\ell^+}\right)\nonumber\\
    &\left. - \tr\left( \rho_0 U_{\ell^+}^\dagger X_\ell U_{\ell^+} O_{\rm RH} U_{\ell^+}^\dagger X_\ell U_{\ell^+} \rho_0 O_{\rm RH} U_{\ell^+}^\dagger X_\ell U_{\ell^+} \rho_0 O_{\rm RH} U_{\ell^+}^\dagger X_\ell U_{\ell^+}\right)\right].
\end{align}
\normalsize

{One can see that the first two terms equals and they are}
\begin{align}
    &\int_{{\calU_{\rm RH}}} dU_{\rm RH} \int_{{\calU_{\rm Haar}}} dU_{\ell^-} \left[\tr\left(O_{\rm RH} \rho_0 O_{\rm RH} \rho_0 U_{\ell^+}^\dagger X_\ell U_{\ell^+} O_{\rm RH} \rho_0 U_{\ell^+}^\dagger X_\ell U_{\ell^+} \right) + \tr\left(\rho_0 O_{\rm RH} \rho_0 O_{\rm RH}   U_{\ell^+}^\dagger X_\ell U_{\ell^+}  \rho_0 O_{\rm RH} U_{\ell^+}^\dagger X_\ell U_{\ell^+} \right)\right]\nonumber\\
    &= \int_{{\calU_{\rm RH}}} dU_{\rm RH} \frac{2}{d^2 - 1}\left[d\tr(\rho_0 O_{\rm RH}) \tr(\rho_0 O_{\rm RH} \rho_0 O_{\rm RH}) - \tr(\rho_0 O_{\rm RH} \rho_0 O_{\rm RH} \rho_0 O_{\rm RH})\right]\nonumber\\
    &= \frac{2(d-1) F_0^3}{d^2 - 1}.
\end{align}
The third and {fourth term also equals and are}
\begin{align}
    &\int_{{\calU_{\rm RH}}} dU_{\rm RH} \int_{{\calU_{\rm Haar}}} dU_{\ell^-} \left[\tr\left(\rho_0 O_{\rm RH} \rho_0 U_{\ell^+}^\dagger X_\ell U_{\ell^+}  O_{\rm RH} \rho_0 O_{\rm RH} U_{\ell^+}^\dagger X_\ell U_{\ell^+} \right) + \tr\left(O_{\rm RH} \rho_0 O_{\rm RH} \rho_0 O_{\rm RH} U_{\ell^+}^\dagger X_\ell U_{\ell^+}  \rho_0 U_{\ell^+}^\dagger X_\ell U_{\ell^+} \right)\right] \nonumber\\
    &= \int_{{\calU_{\rm RH}}} dU_{\rm RH} \frac{1}{d^2 - 1}\left[d\tr(\rho_0 O_{\rm RH})^2 - \tr(\rho_0 O_{\rm RH} \rho_0 O_{\rm RH} \rho_0 O_{\rm RH})+ d\tr(\rho_0 O_{\rm RH} \rho_0 O_{\rm RH})- \tr(\rho_0 O_{\rm RH} \rho_0 O_{\rm RH} \rho_0 O_{\rm RH}) \right]\\
    &= \frac{2(d F_0^2 - F_0^3)}{d^2-1}.
\end{align}
The fifth term is
\begin{align}
    & \int_{{\calU_{\rm RH}}} dU_{\rm RH} \int_{{\calU_{\rm Haar}}} dU_{\ell^-} \tr\left(\rho_0 U_{\ell^+}^\dagger X_\ell U_{\ell^+} O_{\rm RH} U_{\ell^+}^\dagger X_\ell U_{\ell^+} \rho_0 U_{\ell^+}^\dagger X_\ell U_{\ell^+} O_{\rm RH} \rho_0 O_{\rm RH} U_{\ell^+}^\dagger X_\ell U_{\ell^+}\right)\nonumber\\
    &= \int_{{\calU_{\rm RH}}} dU_{\rm RH} \frac{1}{(d^2-9)(d^2-1)} \left[\left(d^2-3\right) \tr(\rho_0 O_{\rm RH} \rho_0 O_{\rm RH} \rho_0 O_{\rm RH})+2 (d-3) \tr(\rho_0 O_{\rm RH}) (-\tr(\rho_0 O_{\rm RH})+d+2)\right.\nonumber\\
    & \qquad\qquad\qquad\qquad\qquad\qquad\qquad \left.-2 \tr(\rho_0 O_{\rm RH} \rho_0 O_{\rm RH}) (d \tr(\rho_0 O_{\rm RH})+3 d-9)\right]\\
    &= \frac{F_0 \left(d \left(F_0 ^2+2\right)+F_0 ^2-8 F_0 +4\right)}{d^3+3 d^2-d-3}.
\end{align}
The sixth term can be found to be equal to the fifth one above. The seventh term is
\begin{align}
    &\int_{{\calU_{\rm RH}}} dU_{\rm RH} \int_{{\calU_{\rm Haar}}} dU_{\ell^-} \tr\left( O_{\rm RH} \rho_0 U_{\ell^+}^\dagger X_\ell U_{\ell^+} O_{\rm RH} U_{\ell^+}^\dagger X_\ell U_{\ell^+} \rho_0 U_{\ell^+}^\dagger X_\ell U_{\ell^+} O_{\rm RH} \rho_0 U_{\ell^+}^\dagger X_\ell U_{\ell^+}\right)\nonumber\\
    &= \frac{1}{(d^2 - 9) (d^2 - 1)}\left[\tr(\rho_0 O_{\rm RH})^2 (3 \tr(\rho_0 O_{\rm RH})+d (2 d-5)-6)+d \tr(\rho_0 O_{\rm RH} \rho_0 O_{\rm RH}) (-3 \tr(\rho_0 O_{\rm RH})+d-4)\right.\nonumber\\
    &\qquad\qquad\qquad\qquad \left. +6 (\tr(\rho_0 O_{\rm RH} \rho_0 O_{\rm RH})+\tr(\rho_0 O_{\rm RH} \rho_0 O_{\rm RH} \rho_0 O_{\rm RH}))\right]\\
    &= \frac{3 F_0^2 (d-F_0 )}{d^3+3 d^2-d-3}.
\end{align}
The eighth term is also equal to the seventh one above.

Conclude from the above calculation, we have
\begin{align}
    \mathbb{E}_{{\calU_{\rm RH}}}\left[\frac{\partial^2 \epsilon}{\partial\theta_\ell^2}\left(\frac{\partial \epsilon}{\partial \theta_\ell}\right)^2 \right] &=\frac{1}{16} \int dU_{\ell^-}dU_{\ell^+} \tr\left(\rho_0 U_{\ell^-}^\dagger O_{\ell^+} U_{\ell^-}\rho_0 U_{\ell^-}^\dagger [X_\ell,[X_\ell, O_{\ell^+}]]U_{\ell^-}\rho_0 U_{\ell^-}^\dagger[X_\ell, O_{\ell^+}]U_{\ell^-}\rho_0 U_{\ell^-}^\dagger[X_\ell, O_{\ell^+}] U_{\ell^-}\right)\nonumber\\
    &= \frac{2}{8}\left[\frac{(d-1) F_0^3}{d^2 - 1} - \frac{(d F_0^2 - F_0^3)}{d^2 - 1} + \frac{F_0 \left[d \left(F_0 ^2+2\right)+F_0 ^2-8 F_0 +4\right]}{d^3+3 d^2-d-3} - \frac{3 F_0 ^2 (d-F_0 )}{d^3+3 d^2-d-3}\right]\\
    &= \frac{(d+2) (F_0 -1) F_0 \left[(d+2) F_0 -2\right]}{4 (d-1) (d+1) (d+3)}.
    \label{eq:gllglgl_rh}
\end{align}

\paragraph{$\mathbb{E}_{{\calU_{\rm RH}}}\left[\frac{\partial^2 \epsilon}{\partial \theta_{\ell_1}\partial\theta_{\ell_2}}\frac{\partial \epsilon}{\partial \theta_{\ell_1}} \frac{\partial \epsilon}{\partial \theta_{\ell_2}}\right]$ under restricted Haar ensemble\\}

The other part $\mathbb{E}_{{\calU_{\rm RH}}}\left[\frac{\partial^2 \epsilon}{\partial \theta_{\ell_1}\partial\theta_{\ell_2}}\frac{\partial \epsilon}{\partial \theta_{\ell_1}} \frac{\partial \epsilon}{\partial \theta_{\ell_2}}\right]$ is
\begin{align}
    &\mathbb{E}_{{\calU_{\rm RH}}}\left[\frac{\partial^2 \epsilon}{\partial \theta_{\ell_1}\partial\theta_{\ell_2}}\frac{\partial \epsilon}{\partial \theta_{\ell_1}} \frac{\partial \epsilon}{\partial \theta_{\ell_2}}\right]\nonumber\\
    &=\frac{1}{16}\int dU_{\ell_1^-}dU_{\ell_1\shortto\ell_2}dU_{\ell_2^+}
    \nonumber
    \\
    &\qquad \qquad \tr\left(\rho_0 U_{\ell_1^-}^\dagger \left[X_{\ell_1}, U_{\ell_1\shortto\ell_2}^\dagger \left[X_{\ell_2}, O_{\ell_2^+}\right]U_{\ell_1\shortto\ell_2}\right] U_{\ell_1^-}\rho_0U_{\ell_1^-}^\dagger \left[X_{\ell_1}, O_{\ell_1^+}\right] U_{\ell_1^-} \rho_0 U_{\ell_1^-}^\dagger U_{\ell_1\shortto\ell_2}^\dagger \left[X_{\ell_2}, O_{\ell_2^+}\right] U_{\ell_1\shortto\ell_2} U_{\ell_1^-}\right)\nonumber\\
    &= \frac{1}{16}\int_{{\calU_{\rm RH}}} dU_{\rm RH} \int_{{\calU_{\rm Haar}}} dU_{\ell_1^-}dU_{\ell_1\shortto\ell_2} \left[\tr(O_{\rm RH} \rho_0 U_{\ell_1^-}^\dagger X_{\ell_1} U_{\ell_1\shortto \ell_2}^\dagger X_{\ell_2} U_{\ell_1\shortto \ell_2} U_{\ell_1^-} O_{\rm RH} \rho_0  U_{\ell_1^-}^\dagger X_{\ell_1}U_{\ell_1^-} O_{\rm RH} \rho_0 U_{\ell_1^-}^\dagger U_{\ell_1\shortto \ell_2}^\dagger X_{\ell_2} U_{\ell_1\shortto \ell_2} U_{\ell_1^-})\right.\nonumber\\
    &+ \tr(\rho_0 U_{\ell_1^-}^\dagger X_{\ell_1} U_{\ell_1\shortto \ell_2}^\dagger X_{\ell_2} U_{\ell_1\shortto \ell_2} U_{\ell_1^-} O_{\rm RH} \rho_0 O_{\rm RH}  U_{\ell_1^-}^\dagger X_{\ell_1}U_{\ell_1^-}  \rho_0 O_{\rm RH} U_{\ell_1^-}^\dagger U_{\ell_1\shortto \ell_2}^\dagger X_{\ell_2} U_{\ell_1\shortto \ell_2} U_{\ell_1^-})\nonumber\\
    & - \tr( \rho_0 U_{\ell_1^-}^\dagger X_{\ell_1} U_{\ell_1\shortto \ell_2}^\dagger X_{\ell_2} U_{\ell_1\shortto \ell_2} U_{\ell_1^-} O_{\rm RH} \rho_0  U_{\ell_1^-}^\dagger X_{\ell_1}U_{\ell_1^-}  O_{\rm RH} \rho_0 O_{\rm RH} U_{\ell_1^-}^\dagger U_{\ell_1\shortto \ell_2}^\dagger X_{\ell_2} U_{\ell_1\shortto \ell_2} U_{\ell_1^-})\nonumber\\
    & - \tr(O_{\rm RH} \rho_0 U_{\ell_1^-}^\dagger X_{\ell_1} U_{\ell_1\shortto \ell_2}^\dagger X_{\ell_2} U_{\ell_1\shortto \ell_2} U_{\ell_1^-} O_{\rm RH} \rho_0 O_{\rm RH}  U_{\ell_1^-}^\dagger X_{\ell_1}U_{\ell_1^-}  \rho_0 U_{\ell_1^-}^\dagger U_{\ell_1\shortto \ell_2}^\dagger X_{\ell_2} U_{\ell_1\shortto \ell_2} U_{\ell_1^-})\nonumber\\
    & + \tr(O_{\rm RH} \rho_0 U_{\ell_1^-}^\dagger X_{\ell_1} U_{\ell_1^-} O_{\rm RH}  U_{\ell_1^-}^\dagger U_{\ell_1\shortto \ell_2}^\dagger X_{\ell_2} U_{\ell_1\shortto \ell_2} U_{\ell_1^-}  \rho_0 O_{\rm RH} U_{\ell_1^-}^\dagger X_{\ell_1} U_{\ell_1^-} \rho_0 U_{\ell_1^-}^\dagger U_{\ell_1\shortto \ell_2}^\dagger X_{\ell_2} U_{\ell_1\shortto \ell_2} U_{\ell_1^-})\nonumber\\
    & + \tr( \rho_0 U_{\ell_1^-}^\dagger X_{\ell_1} U_{\ell_1^-} O_{\rm RH}  U_{\ell_1^-}^\dagger U_{\ell_1\shortto \ell_2}^\dagger X_{\ell_2} U_{\ell_1\shortto \ell_2} U_{\ell_1^-}  \rho_0 U_{\ell_1^-}^\dagger X_{\ell_1} U_{\ell_1^-} O_{\rm RH} \rho_0 O_{\rm RH} U_{\ell_1^-}^\dagger U_{\ell_1\shortto \ell_2}^\dagger X_{\ell_2} U_{\ell_1\shortto \ell_2} U_{\ell_1^-})\nonumber\\
    & - \tr(O_{\rm RH} \rho_0 U_{\ell_1^-}^\dagger X_{\ell_1} U_{\ell_1^-} O_{\rm RH}  U_{\ell_1^-}^\dagger U_{\ell_1\shortto \ell_2}^\dagger X_{\ell_2} U_{\ell_1\shortto \ell_2} U_{\ell_1^-}  \rho_0 U_{\ell_1^-}^\dagger X_{\ell_1} U_{\ell_1^-} O_{\rm RH} \rho_0 U_{\ell_1^-}^\dagger U_{\ell_1\shortto \ell_2}^\dagger X_{\ell_2} U_{\ell_1\shortto \ell_2} U_{\ell_1^-})\nonumber\\
    & - \tr( \rho_0 U_{\ell_1^-}^\dagger X_{\ell_1} U_{\ell_1^-} O_{\rm RH}  U_{\ell_1^-}^\dagger U_{\ell_1\shortto \ell_2}^\dagger X_{\ell_2} U_{\ell_1\shortto \ell_2} U_{\ell_1^-}  \rho_0 O_{\rm RH} U_{\ell_1^-}^\dagger X_{\ell_1} U_{\ell_1^-} \rho_0 O_{\rm RH}  U_{\ell_1^-}^\dagger U_{\ell_1\shortto \ell_2}^\dagger X_{\ell_2} U_{\ell_1\shortto \ell_2} U_{\ell_1^-})\nonumber\\
    & + \tr(\rho_0 U_{\ell_1^-}^\dagger U_{\ell_1\shortto \ell_2}^\dagger X_{\ell_2} U_{\ell_1\shortto \ell_2} U_{\ell_1^-} O_{\rm RH}  U_{\ell_1^-}^\dagger X_{\ell_1}  U_{\ell_1^-}  \rho_0 U_{\ell_1^-}^\dagger X_{\ell_1} U_{\ell_1^-} O_{\rm RH} \rho_0 O_{\rm RH} U_{\ell_1^-}^\dagger U_{\ell_1\shortto \ell_2}^\dagger X_{\ell_2} U_{\ell_1\shortto \ell_2} U_{\ell_1^-})\nonumber\\
    & + \tr(O_{\rm RH} \rho_0 U_{\ell_1^-}^\dagger U_{\ell_1\shortto \ell_2}^\dagger X_{\ell_2} U_{\ell_1\shortto \ell_2} U_{\ell_1^-} O_{\rm RH}  U_{\ell_1^-}^\dagger X_{\ell_1}  U_{\ell_1^-}  \rho_0 O_{\rm RH} U_{\ell_1^-}^\dagger X_{\ell_1} U_{\ell_1^-} \rho_0  U_{\ell_1^-}^\dagger U_{\ell_1\shortto \ell_2}^\dagger X_{\ell_2} U_{\ell_1\shortto \ell_2} U_{\ell_1^-})\nonumber\\
    & - \tr(O_{\rm RH} \rho_0 U_{\ell_1^-}^\dagger U_{\ell_1\shortto \ell_2}^\dagger X_{\ell_2} U_{\ell_1\shortto \ell_2} U_{\ell_1^-} O_{\rm RH}  U_{\ell_1^-}^\dagger X_{\ell_1}  U_{\ell_1^-}  \rho_0 U_{\ell_1^-}^\dagger X_{\ell_1} U_{\ell_1^-} O_{\rm RH} \rho_0  U_{\ell_1^-}^\dagger U_{\ell_1\shortto \ell_2}^\dagger X_{\ell_2} U_{\ell_1\shortto \ell_2} U_{\ell_1^-})\nonumber\\
    & - \tr( \rho_0 U_{\ell_1^-}^\dagger U_{\ell_1\shortto \ell_2}^\dagger X_{\ell_2} U_{\ell_1\shortto \ell_2} U_{\ell_1^-} O_{\rm RH}  U_{\ell_1^-}^\dagger X_{\ell_1}  U_{\ell_1^-}  \rho_0 O_{\rm RH} U_{\ell_1^-}^\dagger X_{\ell_1} U_{\ell_1^-} \rho_0 O_{\rm RH} U_{\ell_1^-}^\dagger U_{\ell_1\shortto \ell_2}^\dagger X_{\ell_2} U_{\ell_1\shortto \ell_2} U_{\ell_1^-})\nonumber\\
    & + \tr(O_{\rm RH} \rho_0 O_{\rm RH} U_{\ell_1^-}^\dagger U_{\ell_1\shortto \ell_2}^\dagger X_{\ell_2} U_{\ell_1\shortto \ell_2} X_{\ell_1} U_{\ell_1^-} \rho_0 U_{\ell_1^-}^\dagger X_{\ell_1} U_{\ell_1^-} O_{\rm RH} \rho_0  U_{\ell_1^-}^\dagger U_{\ell_1\shortto \ell_2}^\dagger X_{\ell_2} U_{\ell_1\shortto \ell_2} U_{\ell_1^-})\nonumber\\
    & + \tr( \rho_0 O_{\rm RH} U_{\ell_1^-}^\dagger U_{\ell_1\shortto \ell_2}^\dagger X_{\ell_2} U_{\ell_1\shortto \ell_2} X_{\ell_1} U_{\ell_1^-} \rho_0 O_{\rm RH} U_{\ell_1^-}^\dagger X_{\ell_1} U_{\ell_1^-} \rho_0 O_{\rm RH} U_{\ell_1^-}^\dagger U_{\ell_1\shortto \ell_2}^\dagger X_{\ell_2} U_{\ell_1\shortto \ell_2} U_{\ell_1^-})\nonumber\\
    & - \tr( \rho_0 O_{\rm RH} U_{\ell_1^-}^\dagger U_{\ell_1\shortto \ell_2}^\dagger X_{\ell_2} U_{\ell_1\shortto \ell_2} X_{\ell_1} U_{\ell_1^-} \rho_0 U_{\ell_1^-}^\dagger X_{\ell_1} U_{\ell_1^-} O_{\rm RH} \rho_0  O_{\rm RH} U_{\ell_1^-}^\dagger U_{\ell_1\shortto \ell_2}^\dagger X_{\ell_2} U_{\ell_1\shortto \ell_2} U_{\ell_1^-})\nonumber\\
    &\left. + \tr(O_{\rm RH} \rho_0 O_{\rm RH} U_{\ell_1^-}^\dagger U_{\ell_1\shortto \ell_2}^\dagger X_{\ell_2} U_{\ell_1\shortto \ell_2} X_{\ell_1} U_{\ell_1^-} \rho_0 O_{\rm RH} U_{\ell_1^-}^\dagger X_{\ell_1} U_{\ell_1^-} \rho_0  U_{\ell_1^-}^\dagger U_{\ell_1\shortto \ell_2}^\dagger X_{\ell_2} U_{\ell_1\shortto \ell_2} U_{\ell_1^-})\right].
\end{align}

The integral over $U_{\ell_1^-}, U_{\ell_1\shortto \ell_2}$ of first term is
\begin{align}
    &\int_{{\calU_{\rm RH}}} dU_{\rm RH} \int_{{\calU_{\rm Haar}}} dU_{\ell_1^-}dU_{\ell_1\shortto\ell_2} \tr(O_{\rm RH}  \rho_0 U_{\ell_1^-}^\dagger X_{\ell_1} U_{\ell_1\shortto \ell_2}^\dagger X_{\ell_2} U_{\ell_1\shortto \ell_2} U_{\ell_1^-} O_{\rm RH} \rho_0  U_{\ell_1^-}^\dagger X_{\ell_1}U_{\ell_1^-} O_{\rm RH} \rho_0 U_{\ell_1^-}^\dagger U_{\ell_1\shortto \ell_2}^\dagger X_{\ell_2} U_{\ell_1\shortto \ell_2} U_{\ell_1^-})\nonumber\\
    &= \int_{{\calU_{\rm RH}}} dU_{\rm RH} \frac{1}{(d^2 - 1)^2} \left[(d^2+1)\tr(\rho_0 O_{\rm RH} \rho_0 O_{\rm RH} \rho_0 O_{\rm RH}) - 2d\tr(\rho_0 O_{\rm RH}) \tr(\rho_0 O_{\rm RH}\rho_0 O_{\rm RH}) \right]\\
    &= \frac{F_0^3}{(d+1)^2}.
\end{align}

The second term is
\begin{align}
    &\int_{{\calU_{\rm RH}}} dU_{\rm RH} \int_{{\calU_{\rm Haar}}} dU_{\ell_1^-}dU_{\ell_1\shortto\ell_2} \tr(\rho_0 U_{\ell_1^-}^\dagger X_{\ell_1} U_{\ell_1\shortto \ell_2}^\dagger X_{\ell_2} U_{\ell_1\shortto \ell_2} U_{\ell_1^-} O_{\rm RH} \rho_0 O_{\rm RH}  U_{\ell_1^-}^\dagger X_{\ell_1}U_{\ell_1^-}  \rho_0 O_{\rm RH} U_{\ell_1^-}^\dagger U_{\ell_1\shortto \ell_2}^\dagger X_{\ell_2} U_{\ell_1\shortto \ell_2} U_{\ell_1^-})\nonumber\\
    &= \frac{F_0^2 (F_0 + d^2 -2d)}{(d^2 - 1)^2}.
\end{align}

The third term is
\begin{align}
    &\int_{{\calU_{\rm RH}}} dU_{\rm RH} \int_{{\calU_{\rm Haar}}} dU_{\ell_1^-}dU_{\ell_1\shortto\ell_2} \tr(\rho_0 U_{\ell_1^-}^\dagger X_{\ell_1} U_{\ell_1\shortto \ell_2}^\dagger X_{\ell_2} U_{\ell_1\shortto \ell_2} U_{\ell_1^-} O_{\rm RH} \rho_0  U_{\ell_1^-}^\dagger X_{\ell_1}U_{\ell_1^-}  O_{\rm RH} \rho_0 O_{\rm RH} U_{\ell_1^-}^\dagger U_{\ell_1\shortto \ell_2}^\dagger X_{\ell_2} U_{\ell_1\shortto \ell_2} U_{\ell_1^-})\nonumber\\
    &= \frac{F_0^2 (d-F_0)}{(d+1)^2 (d-1)}.
\end{align}
The fourth one equals the third one above. 

The fifth one is
\begin{align}
    &\int_{{\calU_{\rm RH}}} dU_{\rm RH} \int_{{\calU_{\rm Haar}}} dU_{\ell_1^-}dU_{\ell_1\shortto\ell_2}  \tr(O_{\rm RH} \rho_0 U_{\ell_1^-}^\dagger X_{\ell_1} U_{\ell_1^-} O_{\rm RH}  U_{\ell_1^-}^\dagger U_{\ell_1\shortto \ell_2}^\dagger X_{\ell_2} U_{\ell_1\shortto \ell_2} U_{\ell_1^-}  \rho_0 O_{\rm RH} U_{\ell_1^-}^\dagger X_{\ell_1} U_{\ell_1^-} \rho_0 U_{\ell_1^-}^\dagger U_{\ell_1\shortto \ell_2}^\dagger X_{\ell_2} U_{\ell_1\shortto \ell_2} U_{\ell_1^-})\nonumber\\
    &= \frac{F_0 (F_0-d)^2}{(d^2-1)^2}.
\end{align}

The sixth term equals the first; the seventh and eighth equals the third and fourth; the ninth equals the fifth; the tenth equals the sixth; the eleventh and twelfth equals the seventh and eighth; the thirteenth equals the second; the fourteenth equals the first; the fifteenth and sixteenth equals the third and fourth. 

Conclude from the above calculation, we have
\begin{align}
    &\mathbb{E}_{{\calU_{\rm RH}}}\left[\frac{\partial^2 \epsilon}{\partial \theta_{\ell_1}\partial\theta_{\ell_2}}\frac{\partial \epsilon}{\partial \theta_{\ell_1}} \frac{\partial \epsilon}{\partial \theta_{\ell_2}}\right]\nonumber\\
    &=\frac{1}{16}\int dU_{\ell_1^-}dU_{\ell_1\shortto\ell_2}dU_{\ell_2^+}
    \nonumber
    \\
    &\qquad \tr\left(\rho_0 U_{\ell_1^-}^\dagger \left[X_{\ell_1}, U_{\ell_1\shortto\ell_2}^\dagger \left[X_{\ell_2}, O_{\ell_2^+}\right]U_{\ell_1\shortto\ell_2}\right] U_{\ell_1^-}\rho_0U_{\ell_1^-}^\dagger \left[X_{\ell_1}, O_{\ell_1^+}\right] U_{\ell_1^-} \rho_0 U_{\ell_1^-}^\dagger U_{\ell_1\shortto\ell_2}^\dagger \left[X_{\ell_2}, O_{\ell_2^+}\right] U_{\ell_1\shortto\ell_2} U_{\ell_1^-}\right)\nonumber\\
    &= \frac{1}{16}\left[\frac{4F_0^3}{(d+1)^2} + \frac{2F_0^2(F_0 + d^2 - 2d)}{(d^2 - 1)^2} - \frac{8F_0^2(d-F_0)}{(d+1)^2 (d-1)} + \frac{2F_0(F_0-d)^2}{(d^2 - 1)^2}\right]\\
    &= \frac{d^2 F_0(F_0-1) (2F_0-1)}{8(d^2 - 1)^2}.
    \label{eq:g12g1g2_rh}
\end{align}

\paragraph{Summary of average relative dQNTK $\overline{\lambda_\infty}$ with restricted Haar ensemble \\}

Combining Eq.~\eqref{eq:gllglgl_rh} and~\eqref{eq:g12g1g2_rh}, {restricted Haar ensemble averaged} dQNTK $\overline{\mu_\infty}$ is
\begin{align}
    \overline{\mu_{{\infty}}} &= L(L-1) \mathbb{E}_{{\calU_{\rm RH}}}\left[\frac{\partial^2 \epsilon}{\partial \theta_{\ell_1}\partial\theta_{\ell_2}}\frac{\partial \epsilon}{\partial \theta_{\ell_1}} \frac{\partial \epsilon}{\partial \theta_{\ell_2}}\right] + L  \mathbb{E}_{{\calU_{\rm RH}}}\left[\frac{\partial^2 \epsilon}{\partial\theta_\ell^2}\left(\frac{\partial \epsilon}{\partial \theta_\ell}\right)^2 \right] \nonumber \\
    &= L(L-1) \frac{d^2 F_0(F_0-1) (2F_0-1)}{8(d^2 - 1)^2} + L \frac{(d+2) (F_0 -1) F_0 \left[(d+2) F_0 -2\right]}{4 (d-1) (d+1) (d+3)}.
    \label{eq:mu_rh}
\end{align}

Combining with the average QNTK calculated above, we have
\begin{align}
    \overline{\lambda_\infty} &= \frac{ \overline{\mu_{{\infty}}} }{\overline{K_\infty}}\\
    &= (L-1) \frac{d (1-2F_0)}{4(d^2 - 1)} - \frac{(d+2)\left[(d+2) F_0 -2\right]}{2 d (d+3)}\\
    &= \frac{(L-1)d}{4(d^2-1)}(1-2O_0{-2R}) - \frac{(d+2)}{2d(d+3)}\left[(d+2)(O_0{+R})-2\right] \label{eq:lambda_rh}\\
    &\simeq \frac{L}{4d}\left(1-2O_0 {-2R}\right) - \frac{O_0 {+R}}{2}, \label{eq:lambda_rh_asymp}
\end{align}
{where in the last line we still make approximations under $d\gg 1$ for direct understanding on its scaling, and we can see that $\overline{\lambda_\infty}$}
is a constant $\propto L/d$ regardless of $O_0 \lesseqgtr 1$.

\subsection{Average dynamical index under restricted Haar ensemble}

{We evaluate the restricted Haar ensemble averaged $\overline{\zeta_\infty} = \overline{\epsilon_\infty\mu_\infty}/\overline{K_\infty}^2$ in the following.}
\begin{align}
    \overline{\epsilon_{{\infty}}\mu_{{\infty}}} &= \mathbb{E}_{{\calU_{\rm RH}}} \left[BZ{\tr(\rho_0 U_{\rm RH}^\dagger O U_{\rm RH})}\mu \right] - O_0 \overline{\mu_{{\infty}}}\nonumber\\
    &= L(L-1) \mathbb{E}_{{\calU_{\rm RH}}} \left[{\tr(\rho_0 U_{\rm RH}^\dagger O U_{\rm RH})} \frac{\partial^2 \epsilon}{\partial \theta_{\ell_1}\partial\theta_{\ell_2}}\frac{\partial \epsilon}{\partial \theta_{\ell_1}} \frac{\partial \epsilon}{\partial \theta_{\ell_2}}\right] + L\mathbb{E}\left[{\tr(\rho_0 U_{\rm RH}^\dagger O U_{\rm RH})}\frac{\partial^2 \epsilon}{\partial\theta_\ell^2}\left(\frac{\partial \epsilon}{\partial \theta_\ell}\right)^2 \right]- O_0 \overline{\mu_{{\infty}}}.
\end{align}

\paragraph{$\mathbb{E}_{{\calU_{\rm RH}}}\left[{\tr(\rho_0 U_{\rm RH}^\dagger O U_{\rm RH})}\frac{\partial^2 \epsilon}{\partial\theta_\ell^2}\left(\frac{\partial \epsilon}{\partial \theta_\ell}\right)^2\right]$ under restricted Haar ensemble \\}

We first consider $\mathbb{E}_{{\calU_{\rm RH}}}\left[{\tr(\rho_0 U_{\rm RH}^\dagger O U_{\rm RH})}\frac{\partial^2 \epsilon}{\partial\theta_\ell^2}\left(\frac{\partial \epsilon}{\partial \theta_\ell}\right)^2\right]$.
\begin{align}
    &\mathbb{E}_{{\calU_{\rm RH}}}\left[{\tr(\rho_0 U_{\rm RH}^\dagger O U_{\rm RH})} \frac{\partial^2 \epsilon}{\partial\theta_\ell^2}\left(\frac{\partial \epsilon}{\partial \theta_\ell}\right)^2\right]
    \nonumber\\
    &= \frac{1}{16}\int dU_{\ell^-}dU_{\ell^+} \tr\left(\rho_0 U_{\ell^-}^\dagger O_{\ell^+} U_{\ell^-}\rho_0 U_{\ell^-}^\dagger [X_\ell,[X_\ell, O_{\ell^+}]]U_{\ell^-}\rho_0 U_{\ell^-}^\dagger[X_\ell, O_{\ell^+}]U_{\ell^-}\rho_0 U_{\ell^-}^\dagger[X_\ell, O_{\ell^+}] U_{\ell^-}\right)
    \nonumber\\
    &= \frac{2}{16}\int_{{\calU_{\rm RH}}} dU_{\rm RH} \int_{{\calU_{\rm Haar}}} dU_{\ell^-} \left[\tr\left(O_{\rm RH} \rho_0 O_{\rm RH} \rho_0 O_{\rm RH} \rho_0 U_{\ell^+}^\dagger X_\ell U_{\ell^+} O_{\rm RH} \rho_0 U_{\ell^+}^\dagger X_\ell U_{\ell^+} \right) 
    \right.
    \nonumber
    \\
    &+ \tr\left(\rho_0 O_{\rm RH} \rho_0 O_{\rm RH} \rho_0 O_{\rm RH}  U_{\ell^+}^\dagger X_\ell U_{\ell^+}  \rho_0 O_{\rm RH} U_{\ell^+}^\dagger X_\ell U_{\ell^+} \right) - \tr\left(\rho_0 O_{\rm RH} \rho_0 O_{\rm RH} \rho_0 U_{\ell^+}^\dagger X_\ell U_{\ell^+}  O_{\rm RH} \rho_0 O_{\rm RH} U_{\ell^+}^\dagger X_\ell U_{\ell^+} \right)
    \nonumber\\
    &  - \tr\left(O_{\rm RH} \rho_0 O_{\rm RH} \rho_0 O_{\rm RH} \rho_0 O_{\rm RH} U_{\ell^+}^\dagger X_\ell U_{\ell^+}  \rho_0 U_{\ell^+}^\dagger X_\ell U_{\ell^+} \right) \nonumber\\
    & + \tr\left(\rho_0 O_{\rm RH} \rho_0 U_{\ell^+}^\dagger X_\ell U_{\ell^+} O_{\rm RH} U_{\ell^+}^\dagger X_\ell U_{\ell^+} \rho_0 U_{\ell^+}^\dagger X_\ell U_{\ell^+} O_{\rm RH} \rho_0 O_{\rm RH} U_{\ell^+}^\dagger X_\ell U_{\ell^+}\right)\nonumber\\
    &+ \tr\left( O_{\rm RH} \rho_0 O_{\rm RH} \rho_0 U_{\ell^+}^\dagger X_\ell U_{\ell^+} O_{\rm RH} U_{\ell^+}^\dagger X_\ell U_{\ell^+} \rho_0 O_{\rm RH} U_{\ell^+}^\dagger X_\ell U_{\ell^+} \rho_0 U_{\ell^+}^\dagger X_\ell U_{\ell^+}\right)\nonumber\\
    & - \tr\left( O_{\rm RH} \rho_0 O_{\rm RH} \rho_0 U_{\ell^+}^\dagger X_\ell U_{\ell^+} O_{\rm RH} U_{\ell^+}^\dagger X_\ell U_{\ell^+} \rho_0 U_{\ell^+}^\dagger X_\ell U_{\ell^+} O_{\rm RH} \rho_0 U_{\ell^+}^\dagger X_\ell U_{\ell^+}\right)\nonumber\\
    &\left. - \tr\left( \rho_0 O_{\rm RH} \rho_0 U_{\ell^+}^\dagger X_\ell U_{\ell^+} O_{\rm RH} U_{\ell^+}^\dagger X_\ell U_{\ell^+} \rho_0 O_{\rm RH} U_{\ell^+}^\dagger X_\ell U_{\ell^+} \rho_0 O_{\rm RH} U_{\ell^+}^\dagger X_\ell U_{\ell^+}\right)\right].
\end{align}

The first and second terms {equal and} are
\begin{align}
    &\int_{{\calU_{\rm RH}}} dU_{\rm RH} \int_{{\calU_{\rm Haar}}} dU_{\ell^-} \left[\tr\left(O_{\rm RH} \rho_0 O_{\rm RH} \rho_0 O_{\rm RH} \rho_0 U_{\ell^+}^\dagger X_\ell U_{\ell^+} O_{\rm RH} \rho_0 U_{\ell^+}^\dagger X_\ell U_{\ell^+} \right) 
    \nonumber \right.
    \\
    &\qquad \qquad \qquad \left.+ \tr\left(\rho_0 O_{\rm RH} \rho_0 O_{\rm RH} \rho_0 O_{\rm RH}  U_{\ell^+}^\dagger X_\ell U_{\ell^+}  \rho_0 O_{\rm RH} U_{\ell^+}^\dagger X_\ell U_{\ell^+} \right)\right] \nonumber\\
    &= \int_{{\calU_{\rm RH}}} dU_{\rm RH} \frac{2}{d^2 - 1}\left[d\tr(\rho_0 O_{\rm RH}) \tr(\rho_0 O_{\rm RH} \rho_0 O_{\rm RH} \rho_0 O_{\rm RH}) - \tr(\rho_0 O_{\rm RH} \rho_0 O_{\rm RH} \rho_0 O_{\rm RH} \rho_0 O_{\rm RH})\right]\nonumber\\
    &= \frac{2(d-1) F_0^4}{d^2 - 1}.
\end{align}

The third and fourth terms {equal and} are
\begin{align}
    &\int_{{\calU_{\rm RH}}} dU_{\rm RH} \int_{{\calU_{\rm Haar}}} dU_{\ell^-} \left[\tr\left(\rho_0 O_{\rm RH} \rho_0 O_{\rm RH} \rho_0 U_{\ell^+}^\dagger X_\ell U_{\ell^+}  O_{\rm RH} \rho_0 O_{\rm RH} U_{\ell^+}^\dagger X_\ell U_{\ell^+} \right) 
    \nonumber \right.
    \\
    &\qquad \qquad \left.+ \tr\left(O_{\rm RH} \rho_0 O_{\rm RH} \rho_0 O_{\rm RH} \rho_0 O_{\rm RH} U_{\ell^+}^\dagger X_\ell U_{\ell^+}  \rho_0 U_{\ell^+}^\dagger X_\ell U_{\ell^+} \right)\right] \nonumber\\
    &= \int_{{\calU_{\rm RH}}} dU_{\rm RH}  \frac{1}{d^2 - 1}\left[d\tr(\rho_0 O_{\rm RH}) \tr(\rho_0 O_{\rm RH} \rho_0 O_{\rm RH}) - \tr(\rho_0 O_{\rm RH} \rho_0 O_{\rm RH} \rho_0 O_{\rm RH} \rho_0 O_{\rm RH}) + d\tr(\rho_0 O_{\rm RH} \rho_0 O_{\rm RH} \rho_0 O_{\rm RH})\right.\\
    &\left.- \tr(\rho_0 O_{\rm RH} \rho_0 O_{\rm RH} \rho_0 O_{\rm RH} \rho_0 O_{\rm RH}) \right]\\
    &= \frac{2(d F_0^3 - F_0^4)}{d^2-1}.
\end{align}

The fifth term is
\begin{align}
    & \int_{{\calU_{\rm RH}}} dU_{\rm RH} \int_{{\calU_{\rm Haar}}} dU_{\ell^-} \tr\left(\rho_0 O_{\rm RH} \rho_0 U_{\ell^+}^\dagger X_\ell U_{\ell^+} O_{\rm RH} U_{\ell^+}^\dagger X_\ell U_{\ell^+} \rho_0 U_{\ell^+}^\dagger X_\ell U_{\ell^+} O_{\rm RH} \rho_0 O_{\rm RH} U_{\ell^+}^\dagger X_\ell U_{\ell^+}\right)\nonumber\\
    &= \int_{{\calU_{\rm RH}}} dU_{\rm RH} \frac{1}{(d^2-9)(d^2-1)} \left\{\left(d^2-3\right) \tr(\rho_0 O_{\rm RH} \rho_0 O_{\rm RH} \rho_0 O_{\rm RH} \rho_0 O_{\rm RH})+2 (d-3) (d+2) \tr(\rho_0 O_{\rm RH})^2 +3 \tr(\rho_0 O_{\rm RH})^3\right.\nonumber\\
    &-\tr(\rho_0 O_{\rm RH}) \left[(4 d-6) \tr(\rho_0 O_{\rm RH} \rho_0 O_{\rm RH})+d \tr(\rho_0 O_{\rm RH} \rho_0 O_{\rm RH} \rho_0 O_{\rm RH})\right] -d \tr(\rho_0 O_{\rm RH} \rho_0 O_{\rm RH})^2
    \nonumber
    \\
    &\left.+(15-4 d) \tr(\rho_0 O_{\rm RH} \rho_0 O_{\rm RH} \rho_0 O_{\rm RH})\right\}\\
    &= \frac{F_0 ^2 \left[d \left(F_0 ^2+2\right)+F_0 ^2-8 F_0 +4\right]}{d^3+3 d^2-d-3}.
\end{align}
The sixth term can be found to be equal to the fifth one above. 

The seventh term is
\begin{align}
    &\int_{{\calU_{\rm RH}}} dU_{\rm RH} \int_{{\calU_{\rm Haar}}} dU_{\ell^-} \tr\left( O_{\rm RH} \rho_0 O_{\rm RH} \rho_0 U_{\ell^+}^\dagger X_\ell U_{\ell^+} O_{\rm RH} U_{\ell^+}^\dagger X_\ell U_{\ell^+} \rho_0 U_{\ell^+}^\dagger X_\ell U_{\ell^+} O_{\rm RH} \rho_0 U_{\ell^+}^\dagger X_\ell U_{\ell^+}\right)\nonumber\\
    &= \frac{1}{(d^2 - 9) (d^2 - 1)}\left[\tr(\rho_0 O_{\rm RH} \rho_0 O_{\rm RH}) \left(\tr(\rho_0 O_{\rm RH}) \left[3 \tr(\rho_0 O_{\rm RH})+d (2 d-5)-6\right]-d \tr(\rho_0 O_{\rm RH} \rho_0 O_{\rm RH})\right)\right.\nonumber\\
    &\left.+d \tr(\rho_0 O_{\rm RH} \rho_0 O_{\rm RH} \rho_0 O_{\rm RH}) (-2 \tr(\rho_0 O_{\rm RH})+d-4)+6 (\tr(\rho_0 O_{\rm RH} \rho_0 O_{\rm RH} \rho_0 O_{\rm RH})+\tr(\rho_0 O_{\rm RH} \rho_0 O_{\rm RH} \rho_0 O_{\rm RH} \rho_0 O_{\rm RH}))\right]\\
    &= \frac{3 F_0 ^3 (d-F_0 )}{d^3+3 d^2-d-3}.
\end{align}
The eighth term also {equals} to the seventh one above.

Summarizing from above, we have
\begin{align}
    &\mathbb{E}_{{\calU_{\rm RH}}}\left[{\tr(\rho_0 U_{\rm RH}^\dagger O U_{\rm RH})} \frac{\partial^2 \epsilon}{\partial\theta_\ell^2}\left(\frac{\partial \epsilon}{\partial \theta_\ell}\right)^2 \right]\nonumber\\
    &= \frac{2}{8}\left[\frac{(d-1) F_0^4}{d^2 - 1} - \frac{(d F_0^3 - F_0^4)}{d^2 - 1} + \frac{F_0 ^2 \left[d \left(F_0 ^2+2\right)+F_0 ^2-8 F_0 +4\right]}{d^3+3 d^2-d-3} - \frac{3 F_0 ^3 (d-F_0 )}{d^3+3 d^2-d-3}\right]\\
    &= \frac{(d+2) (F_0 -1) F_0 ^2 \left[(d+2) F_0 -2\right]}{4 (d-1) (d+1) (d+3)}.
    \label{eq:Ogllglgl_rh}
\end{align}

\paragraph{$\mathbb{E}_{{\calU_{\rm RH}}}\left[{\tr(\rho_0 U_{\rm RH}^\dagger O U_{\rm RH})} \frac{\partial^2 \epsilon}{\partial \theta_{\ell_1}\partial\theta_{\ell_2}}\frac{\partial \epsilon}{\partial \theta_{\ell_1}} \frac{\partial \epsilon}{\partial \theta_{\ell_2}}\right]$ under restricted Haar ensemble \\}

The other item $\mathbb{E}_{{\calU_{\rm RH}}}\left[{\tr(\rho_0 U_{\rm RH}^\dagger O U_{\rm RH})} \frac{\partial^2 \epsilon}{\partial \theta_{\ell_1}\partial\theta_{\ell_2}}\frac{\partial \epsilon}{\partial \theta_{\ell_1}} \frac{\partial \epsilon}{\partial \theta_{\ell_2}}\right]$ can be expanded as
\small
\begin{align}
    &\mathbb{E}_{{\calU_{\rm RH}}}\left[{\tr(\rho_0 U_{\rm RH}^\dagger O U_{\rm RH})} \frac{\partial^2 \epsilon}{\partial \theta_{\ell_1}\partial\theta_{\ell_2}}\frac{\partial \epsilon}{\partial \theta_{\ell_1}} \frac{\partial \epsilon}{\partial \theta_{\ell_2}}\right]\nonumber\\
    &= \frac{1}{16}\int dU_{\ell_1^-}dU_{\ell_1\shortto\ell_2}dU_{\ell_2^+}\tr\left(\rho_0 U_{\ell_1^-}^\dagger U_{\ell_1\shortto\ell_2}^\dagger O_{\ell_2^+}U_{\ell_1\shortto\ell_2}U_{\ell_1^-}\rho_0 U_{\ell_1^-}^\dagger \left[X_{\ell_1}, U_{\ell_1\shortto\ell_2}^\dagger \left[X_{\ell_2}, O_{\ell_2^+}\right]U_{\ell_1\shortto\ell_2}\right] U_{\ell_1^-}\rho_0U_{\ell_1^-}^\dagger \left[X_{\ell_1}, O_{\ell_1^+}\right] U_{\ell_1^-} \right.\nonumber\\
    &\left.\cdot \rho_0 U_{\ell_1^-}^\dagger U_{\ell_1\shortto\ell_2}^\dagger \left[X_{\ell_2}, O_{\ell_2^+}\right] U_{\ell_1\shortto\ell_2} U_{\ell_1^-}\right)\\
    &=  \frac{1}{16}\int_{{\calU_{\rm RH}}} dU_{\rm RH} \int_{{\calU_{\rm Haar}}} dU_{\ell_1^-}dU_{\ell_1\shortto\ell_2} \left[\tr(O_{\rm RH} \rho_0 O_{\rm RH} \rho_0 U_{\ell_1^-}^\dagger X_{\ell_1} U_{\ell_1\shortto \ell_2}^\dagger X_{\ell_2} U_{\ell_1\shortto \ell_2} U_{\ell_1^-} O_{\rm RH} \rho_0  U_{\ell_1^-}^\dagger X_{\ell_1}U_{\ell_1^-} O_{\rm RH} \rho_0 U_{\ell_1^-}^\dagger U_{\ell_1\shortto \ell_2}^\dagger X_{\ell_2} U_{\ell_1\shortto \ell_2} U_{\ell_1^-})\right.\nonumber\\
    &+ \tr(\rho_0 O_{\rm RH} \rho_0 U_{\ell_1^-}^\dagger X_{\ell_1} U_{\ell_1\shortto \ell_2}^\dagger X_{\ell_2} U_{\ell_1\shortto \ell_2} U_{\ell_1^-} O_{\rm RH} \rho_0 O_{\rm RH}  U_{\ell_1^-}^\dagger X_{\ell_1}U_{\ell_1^-}  \rho_0 O_{\rm RH} U_{\ell_1^-}^\dagger U_{\ell_1\shortto \ell_2}^\dagger X_{\ell_2} U_{\ell_1\shortto \ell_2} U_{\ell_1^-})\nonumber\\
    & - \tr(\rho_0 O_{\rm RH} \rho_0 U_{\ell_1^-}^\dagger X_{\ell_1} U_{\ell_1\shortto \ell_2}^\dagger X_{\ell_2} U_{\ell_1\shortto \ell_2} U_{\ell_1^-} O_{\rm RH} \rho_0  U_{\ell_1^-}^\dagger X_{\ell_1}U_{\ell_1^-}  O_{\rm RH} \rho_0 O_{\rm RH} U_{\ell_1^-}^\dagger U_{\ell_1\shortto \ell_2}^\dagger X_{\ell_2} U_{\ell_1\shortto \ell_2} U_{\ell_1^-})\nonumber\\
    & - \tr(O_{\rm RH} \rho_0 O_{\rm RH} \rho_0 U_{\ell_1^-}^\dagger X_{\ell_1} U_{\ell_1\shortto \ell_2}^\dagger X_{\ell_2} U_{\ell_1\shortto \ell_2} U_{\ell_1^-} O_{\rm RH} \rho_0 O_{\rm RH}  U_{\ell_1^-}^\dagger X_{\ell_1}U_{\ell_1^-}  \rho_0 U_{\ell_1^-}^\dagger U_{\ell_1\shortto \ell_2}^\dagger X_{\ell_2} U_{\ell_1\shortto \ell_2} U_{\ell_1^-})\nonumber\\
    & + \tr(O_{\rm RH} \rho_0 O_{\rm RH} \rho_0 U_{\ell_1^-}^\dagger X_{\ell_1} U_{\ell_1^-} O_{\rm RH}  U_{\ell_1^-}^\dagger U_{\ell_1\shortto \ell_2}^\dagger X_{\ell_2} U_{\ell_1\shortto \ell_2} U_{\ell_1^-}  \rho_0 O_{\rm RH} U_{\ell_1^-}^\dagger X_{\ell_1} U_{\ell_1^-} \rho_0 U_{\ell_1^-}^\dagger U_{\ell_1\shortto \ell_2}^\dagger X_{\ell_2} U_{\ell_1\shortto \ell_2} U_{\ell_1^-})\nonumber\\
    & + \tr(\rho_0 O_{\rm RH} \rho_0 U_{\ell_1^-}^\dagger X_{\ell_1} U_{\ell_1^-} O_{\rm RH}  U_{\ell_1^-}^\dagger U_{\ell_1\shortto \ell_2}^\dagger X_{\ell_2} U_{\ell_1\shortto \ell_2} U_{\ell_1^-}  \rho_0 U_{\ell_1^-}^\dagger X_{\ell_1} U_{\ell_1^-} O_{\rm RH} \rho_0 O_{\rm RH} U_{\ell_1^-}^\dagger U_{\ell_1\shortto \ell_2}^\dagger X_{\ell_2} U_{\ell_1\shortto \ell_2} U_{\ell_1^-})\nonumber\\
    & - \tr(O_{\rm RH} \rho_0 O_{\rm RH} \rho_0 U_{\ell_1^-}^\dagger X_{\ell_1} U_{\ell_1^-} O_{\rm RH}  U_{\ell_1^-}^\dagger U_{\ell_1\shortto \ell_2}^\dagger X_{\ell_2} U_{\ell_1\shortto \ell_2} U_{\ell_1^-}  \rho_0 U_{\ell_1^-}^\dagger X_{\ell_1} U_{\ell_1^-} O_{\rm RH} \rho_0 U_{\ell_1^-}^\dagger U_{\ell_1\shortto \ell_2}^\dagger X_{\ell_2} U_{\ell_1\shortto \ell_2} U_{\ell_1^-})\nonumber\\
    & - \tr(\rho_0 O_{\rm RH} \rho_0 U_{\ell_1^-}^\dagger X_{\ell_1} U_{\ell_1^-} O_{\rm RH}  U_{\ell_1^-}^\dagger U_{\ell_1\shortto \ell_2}^\dagger X_{\ell_2} U_{\ell_1\shortto \ell_2} U_{\ell_1^-}  \rho_0 O_{\rm RH} U_{\ell_1^-}^\dagger X_{\ell_1} U_{\ell_1^-} \rho_0 O_{\rm RH}  U_{\ell_1^-}^\dagger U_{\ell_1\shortto \ell_2}^\dagger X_{\ell_2} U_{\ell_1\shortto \ell_2} U_{\ell_1^-})\nonumber\\
    & + \tr(\rho_0 O_{\rm RH} \rho_0 U_{\ell_1^-}^\dagger U_{\ell_1\shortto \ell_2}^\dagger X_{\ell_2} U_{\ell_1\shortto \ell_2} U_{\ell_1^-} O_{\rm RH}  U_{\ell_1^-}^\dagger X_{\ell_1}  U_{\ell_1^-}  \rho_0 U_{\ell_1^-}^\dagger X_{\ell_1} U_{\ell_1^-} O_{\rm RH} \rho_0 O_{\rm RH} U_{\ell_1^-}^\dagger U_{\ell_1\shortto \ell_2}^\dagger X_{\ell_2} U_{\ell_1\shortto \ell_2} U_{\ell_1^-})\nonumber\\
    & + \tr(O_{\rm RH} \rho_0 O_{\rm RH} \rho_0 U_{\ell_1^-}^\dagger U_{\ell_1\shortto \ell_2}^\dagger X_{\ell_2} U_{\ell_1\shortto \ell_2} U_{\ell_1^-} O_{\rm RH}  U_{\ell_1^-}^\dagger X_{\ell_1}  U_{\ell_1^-}  \rho_0 O_{\rm RH} U_{\ell_1^-}^\dagger X_{\ell_1} U_{\ell_1^-} \rho_0  U_{\ell_1^-}^\dagger U_{\ell_1\shortto \ell_2}^\dagger X_{\ell_2} U_{\ell_1\shortto \ell_2} U_{\ell_1^-})\nonumber\\
    & - \tr(O_{\rm RH} \rho_0 O_{\rm RH} \rho_0 U_{\ell_1^-}^\dagger U_{\ell_1\shortto \ell_2}^\dagger X_{\ell_2} U_{\ell_1\shortto \ell_2} U_{\ell_1^-} O_{\rm RH}  U_{\ell_1^-}^\dagger X_{\ell_1}  U_{\ell_1^-}  \rho_0 U_{\ell_1^-}^\dagger X_{\ell_1} U_{\ell_1^-} O_{\rm RH} \rho_0  U_{\ell_1^-}^\dagger U_{\ell_1\shortto \ell_2}^\dagger X_{\ell_2} U_{\ell_1\shortto \ell_2} U_{\ell_1^-})\nonumber\\
    & - \tr(\rho_0 O_{\rm RH} \rho_0 U_{\ell_1^-}^\dagger U_{\ell_1\shortto \ell_2}^\dagger X_{\ell_2} U_{\ell_1\shortto \ell_2} U_{\ell_1^-} O_{\rm RH}  U_{\ell_1^-}^\dagger X_{\ell_1}  U_{\ell_1^-}  \rho_0 O_{\rm RH} U_{\ell_1^-}^\dagger X_{\ell_1} U_{\ell_1^-} \rho_0 O_{\rm RH} U_{\ell_1^-}^\dagger U_{\ell_1\shortto \ell_2}^\dagger X_{\ell_2} U_{\ell_1\shortto \ell_2} U_{\ell_1^-})\nonumber\\
    & + \tr(O_{\rm RH} \rho_0 O_{\rm RH} \rho_0 O_{\rm RH} U_{\ell_1^-}^\dagger U_{\ell_1\shortto \ell_2}^\dagger X_{\ell_2} U_{\ell_1\shortto \ell_2} X_{\ell_1} U_{\ell_1^-} \rho_0 U_{\ell_1^-}^\dagger X_{\ell_1} U_{\ell_1^-} O_{\rm RH} \rho_0  U_{\ell_1^-}^\dagger U_{\ell_1\shortto \ell_2}^\dagger X_{\ell_2} U_{\ell_1\shortto \ell_2} U_{\ell_1^-})\nonumber\\
    & + \tr(\rho_0 O_{\rm RH} \rho_0 O_{\rm RH} U_{\ell_1^-}^\dagger U_{\ell_1\shortto \ell_2}^\dagger X_{\ell_2} U_{\ell_1\shortto \ell_2} X_{\ell_1} U_{\ell_1^-} \rho_0 O_{\rm RH} U_{\ell_1^-}^\dagger X_{\ell_1} U_{\ell_1^-} \rho_0 O_{\rm RH} U_{\ell_1^-}^\dagger U_{\ell_1\shortto \ell_2}^\dagger X_{\ell_2} U_{\ell_1\shortto \ell_2} U_{\ell_1^-})\nonumber\\
    & - \tr(\rho_0 O_{\rm RH} \rho_0 O_{\rm RH} U_{\ell_1^-}^\dagger U_{\ell_1\shortto \ell_2}^\dagger X_{\ell_2} U_{\ell_1\shortto \ell_2} X_{\ell_1} U_{\ell_1^-} \rho_0 U_{\ell_1^-}^\dagger X_{\ell_1} U_{\ell_1^-} O_{\rm RH} \rho_0  O_{\rm RH} U_{\ell_1^-}^\dagger U_{\ell_1\shortto \ell_2}^\dagger X_{\ell_2} U_{\ell_1\shortto \ell_2} U_{\ell_1^-})\nonumber\\
    &\left. + \tr(O_{\rm RH} \rho_0 O_{\rm RH} \rho_0 O_{\rm RH} U_{\ell_1^-}^\dagger U_{\ell_1\shortto \ell_2}^\dagger X_{\ell_2} U_{\ell_1\shortto \ell_2} X_{\ell_1} U_{\ell_1^-} \rho_0 O_{\rm RH} U_{\ell_1^-}^\dagger X_{\ell_1} U_{\ell_1^-} \rho_0  U_{\ell_1^-}^\dagger U_{\ell_1\shortto \ell_2}^\dagger X_{\ell_2} U_{\ell_1\shortto \ell_2} U_{\ell_1^-})\right].
\end{align}
\normalsize

The integral over $U_{\ell_1^-}, U_{\ell_1\shortto \ell_2}$ of first term is
\begin{align}
    &\int_{{\calU_{\rm RH}}} dU_{\rm RH} \int_{{\calU_{\rm Haar}}} dU_{\ell_1^-}dU_{\ell_1\shortto\ell_2} \rho_0 O_{\rm RH} \rho_0 U_{\ell_1^-}^\dagger X_{\ell_1} U_{\ell_1\shortto \ell_2}^\dagger X_{\ell_2} U_{\ell_1\shortto \ell_2} U_{\ell_1^-} O_{\rm RH} \rho_0  U_{\ell_1^-}^\dagger X_{\ell_1}U_{\ell_1^-} O_{\rm RH} \rho_0 U_{\ell_1^-}^\dagger U_{\ell_1\shortto \ell_2}^\dagger X_{\ell_2} U_{\ell_1\shortto \ell_2} U_{\ell_1^-})\nonumber\\
    &= \int_{{\calU_{\rm RH}}} dU_{\rm RH} \frac{1}{(d^2 - 1)^2} \left[(d^2+1)\tr(\rho_0 O_{\rm RH} \rho_0 O_{\rm RH} \rho_0 O_{\rm RH} \rho_0 O_{\rm RH}) - d\tr(\rho_0 O_{\rm RH}) \tr(\rho_0 O_{\rm RH}\rho_0 O_{\rm RH}\rho_0 O_{\rm RH})\right.\nonumber\\
    &\left.- d\tr(\rho_0 O_{\rm RH}\rho_0 O_{\rm RH})^2 \right]\\
    &= \frac{F_0^4}{(d+1)^2}.
\end{align}
The second term is
\small
\begin{align}
    &\int_{{\calU_{\rm RH}}} dU_{\rm RH} \int_{{\calU_{\rm Haar}}} dU_{\ell_1^-}dU_{\ell_1\shortto\ell_2}  \tr(\rho_0 O_{\rm RH} \rho_0 U_{\ell_1^-}^\dagger X_{\ell_1} U_{\ell_1\shortto \ell_2}^\dagger X_{\ell_2} U_{\ell_1\shortto \ell_2} U_{\ell_1^-} O_{\rm RH} \rho_0 O_{\rm RH}  U_{\ell_1^-}^\dagger X_{\ell_1}U_{\ell_1^-}  \rho_0 O_{\rm RH} U_{\ell_1^-}^\dagger U_{\ell_1\shortto \ell_2}^\dagger X_{\ell_2} U_{\ell_1\shortto \ell_2} U_{\ell_1^-})\nonumber\\
    &= \frac{1}{(d^2 - 1)^2}\left[\tr(\rho_0 O_{\rm RH} \rho_0 O_{\rm RH} \rho_0 O_{\rm RH} \rho_0 O_{\rm RH}) + d^2 \tr(\rho_0 O_{\rm RH} \rho_0 O_{\rm RH} \rho_0 O_{\rm RH}) - 2d\tr(\rho_0 O_{\rm RH})\tr(\rho_0 O_{\rm RH}\rho_0 O_{\rm RH})\right]\\
    &= \frac{F_0^3 (F_0 + d^2 -2d)}{(d^2 - 1)^2}.
\end{align}
\normalsize

The third term is
\small
\begin{align}
    &\int_{{\calU_{\rm RH}}} dU_{\rm RH} \int_{{\calU_{\rm Haar}}} dU_{\ell_1^-}dU_{\ell_1\shortto\ell_2}  \tr(\rho_0 O_{\rm RH} \rho_0 U_{\ell_1^-}^\dagger X_{\ell_1} U_{\ell_1\shortto \ell_2}^\dagger X_{\ell_2} U_{\ell_1\shortto \ell_2} U_{\ell_1^-} O_{\rm RH} \rho_0  U_{\ell_1^-}^\dagger X_{\ell_1}U_{\ell_1^-}  O_{\rm RH} \rho_0 O_{\rm RH} U_{\ell_1^-}^\dagger U_{\ell_1\shortto \ell_2}^\dagger X_{\ell_2} U_{\ell_1\shortto \ell_2} U_{\ell_1^-})\nonumber\\
    &= \frac{1}{(d^2 - 1)^2}\left[\tr(\rho_0 O_{\rm RH} \rho_0 O_{\rm RH} \rho_0 O_{\rm RH} \rho_0 O_{\rm RH}) + d^2 \tr(\rho_0 O_{\rm RH} \rho_0 O_{\rm RH} \rho_0 O_{\rm RH})  \right. \nonumber\\
    &\quad \quad  \quad  \quad \quad \quad \left. - d\tr(\rho_0 O_{\rm RH})\left(\tr(\rho_0 O_{\rm RH}\rho_0 O_{\rm RH}) +\tr(\rho_0 O_{\rm RH}\rho_0 O_{\rm RH} \rho_0 O_{\rm RH})\right)\right]\\
    &= \frac{F_0^3 (d-F_0)}{(d+1)^2 (d-1)}.
\end{align}
\normalsize
The fourth one equals the third one above. 

The fifth one is
\small
\begin{align}
    &\int_{{\calU_{\rm RH}}} dU_{\rm RH} \int_{{\calU_{\rm Haar}}} dU_{\ell_1^-}dU_{\ell_1\shortto\ell_2}   \tr(O_{\rm RH} \rho_0 O_{\rm RH} \rho_0 U_{\ell_1^-}^\dagger X_{\ell_1} U_{\ell_1^-} O_{\rm RH}  U_{\ell_1^-}^\dagger U_{\ell_1\shortto \ell_2}^\dagger X_{\ell_2} U_{\ell_1\shortto \ell_2} U_{\ell_1^-}  \rho_0 O_{\rm RH} U_{\ell_1^-}^\dagger X_{\ell_1} U_{\ell_1^-} \rho_0 U_{\ell_1^-}^\dagger U_{\ell_1\shortto \ell_2}^\dagger X_{\ell_2} U_{\ell_1\shortto \ell_2} U_{\ell_1^-})\nonumber\\
    &= \frac{1}{(d^2 - 1)^2}\left[\tr(\rho_0 O_{\rm RH} \rho_0 O_{\rm RH} \rho_0 O_{\rm RH} \rho_0 O_{\rm RH}) + d^2 \tr(\rho_0 O_{\rm RH}  \rho_0 O_{\rm RH}) - 2d\tr(\rho_0 O_{\rm RH})\tr(\rho_0 O_{\rm RH}\rho_0 O_{\rm RH})\right]\\
    &= \frac{F_0^2 (F_0-d)^2}{(d^2-1)^2}.
\end{align}
\normalsize
The sixth term equals the first; the seventh and eighth equals the third and fourth; the ninth equals the fifth; the tenth equals the sixth; the eleventh and twelfth equals the seventh and eighth; the thirteenth equals the second; the fourteenth equals the first; the fifteenth and sixteenth equals the third and fourth. 

Concluding from the above sixteenth terms, we have
\begin{align}
    \mathbb{E}_{{\calU_{\rm RH}}}\left[{\tr(\rho_0 U_{\rm RH}^\dagger O U_{\rm RH})} \frac{\partial^2 \epsilon}{\partial \theta_{\ell_1}\partial\theta_{\ell_2}}\frac{\partial \epsilon}{\partial \theta_{\ell_1}} \frac{\partial \epsilon}{\partial \theta_{\ell_2}}\right]
    &= \frac{1}{16}\left[\frac{4F_0^4}{(d+1)^2} + \frac{2F_0^3(F_0 + d^2 - 2d)}{(d^2 - 1)^2} - \frac{8F_0^3(d-F_0)}{(d+1)^2 (d-1)} + \frac{2F_0^2(F_0-d)^2}{(d^2 - 1)^2}\right]\\
    &= \frac{d^2 F_0^2(F_0-1) (2F_0-1)}{8(d^2 - 1)^2}.
    \label{eq:Og12g1g2_rh}
\end{align}

Summarizing from Eq.~\eqref{eq:Ogllglgl_rh} and~\eqref{eq:Og12g1g2_rh}, $\mathbb{E}\left[{\tr(\rho_0 U_{\rm RH}^\dagger O U_{\rm RH})}\mu\right]$ becomes
\begin{align}
    &\mathbb{E}_{{\calU_{\rm RH}}}\left[{\tr(\rho_0 U_{\rm RH}^\dagger O U_{\rm RH})} \mu\right]\nonumber\\
    &= L(L-1)\mathbb{E}_{{\calU_{\rm RH}}} \left[{\tr(\rho_0 U_{\rm RH}^\dagger O U_{\rm RH})} \frac{\partial^2 \epsilon}{\partial \theta_{\ell_1}\partial\theta_{\ell_2}}\frac{\partial \epsilon}{\partial \theta_{\ell_1}} \frac{\partial \epsilon}{\partial \theta_{\ell_2}}\right] + L \mathbb{E}_{{\calU_{\rm RH}}} \left[{\tr(\rho_0 U_{\rm RH}^\dagger O U_{\rm RH})} \frac{\partial^2 \epsilon}{\partial\theta_\ell^2}\left(\frac{\partial \epsilon}{\partial \theta_\ell}\right)^2 \right]\\
    &= L(L-1) \frac{d^2 F_0^2(F_0-1) (2F_0-1)}{8(d^2 - 1)^2} + L \frac{(d+2) (F_0 -1) F_0 ^2 \left[(d+2) F_0 -2\right]}{4 (d-1) (d+1) (d+3)}.
    \label{eq:Omu_rh}
\end{align}

\paragraph{Summary of average dynamical index $\overline{\zeta_\infty}$ with restricted Haar ensemble \\}

By subtracting $O_0\overline{\mu_{\infty}}$ solved in Eq.~\eqref{eq:mu_rh}, the ensemble average of $\epsilon\mu$ under restricted Haar ensemble is
\begin{align}
    \overline{\epsilon_{{\infty}}\mu_{{\infty}}} &= \mathbb{E}\left[{\tr(\rho_0 U_{\rm RH}^\dagger O U_{\rm RH})}\mu\right] - O_0 \overline{\mu_{\infty}}\\
     &=
     L(L-1) \frac{d^2 F_0^2(F_0-1) (2F_0-1)}{8(d^2 - 1)^2} + L \frac{(d+2) (F_0 -1) F_0 ^2 \left[(d+2) F_0 -2\right]}{4 (d-1) (d+1) (d+3)}\nonumber\\
     & \quad - O_0 \left[L(L-1) \frac{d^2 F_0(F_0-1) (2F_0-1)}{8(d^2 - 1)^2} + L \frac{(d+2) (F_0 -1) F_0 \left[(d+2) F_0 -2\right]}{4 (d-1) (d+1) (d+3)}\right]\\
    & = L(L-1)\frac{d^2 F_0 (F_0-1) (2F_0-1)}{8(d^2 - 1)^2}\left(F_0 - O_0\right) + L \frac{(d+2) (F_0 -1) F_0 \left[(d+2) F_0 -2\right]}{4 (d-1) (d+1) (d+3)}\left(F_0 - O_0\right)\\
    &\simeq L^2 \frac{F_0 (F_0-1) (2F_0-1)}{8 d^2 }\left(F_0 - O_0\right) + L \frac{(F_0 -1) F_0^2}{4 d}\left(F_0 - O_0\right).
    \label{eq:epsmu_rh}
\end{align}

The ratio $\overline{\zeta_{\infty}}$ becomes
\begin{align}
    \overline{\zeta_\infty} &= \frac{ \overline{\epsilon_{\infty} \mu_{\infty}} }{\overline{K_\infty}^2}\\
    &= (L-1)\frac{(2F_0-1)}{2 L F_0(F_0-1) }\left(F_0 - O_0\right) + \frac{(d+2)(d^2-1) \left[(d+2) F_0 -2\right]}{ L d^2 (d+3) F_0 (F_0 -1) }\left(F_0 - O_0\right)\\
    &= \frac{L-1}{2L}\frac{(2O_0 {+ 2R}-1){R}}{(O_0 {+R})({O_0+R-1}) } + \frac{(d+2)(d^2-1)}{Ld^2(d+3)} \frac{\left[(d+2) (O_0 {+ R} ) -2\right]{R}}{ (O_0 {+ R}) ({O_0 + R-1}) } \label{eq:zeta_rh}\\
    &\simeq {\frac{R}{O_0 + R-1}}\left(1 - \frac{1}{2(O_0 {+R})} + \frac{d}{ L }\right). \label{eq:zeta_rh_asymp}
\end{align}
{where in the last line approximate it with $L,d \gg 1$ to simplify the formula.}
When $O_0 <1$ with $R=0$, we directly have {$\overline{\zeta_\infty} = 0$}.
At the critical point of $O_0 = 1$ with $R=0$, we have {$\overline{\zeta_\infty} = 1/2 + d/L$}, and in the large limit $L \gg d$, we have $\overline{\zeta_\infty} = 1/2$. However for $O_0 >1$ with $R={1-O_0}$, we have {$\lim_{R \to (1-O_0)^-}\overline{\zeta_\infty} \to +\infty$} which diverges to infinity.

\end{widetext}



\end{document}